\documentclass[aps,prb,twocolumn]{revtex4}
\usepackage{amssymb,macros2erev4,graphicx}
\usepackage{times}
\newcommand{\rechecked}{}
\newcommand{\ind}[1]{{\mathrm{#1}}}
\newcommand{\be}{\begin{equation}}
\newcommand{\ee}{\end{equation}}
\newcommand{\bea}{\begin{eqnarray}}
\newcommand{\eea}{\end{eqnarray}}
\newcommand{\half}{\frac{1}{2}}
\newcommand{\ts}{\hskip0.1ex\raisebox{-1ex}[0ex][0.8ex]{\rule{0.1ex}{2.75ex}\hskip0.2ex}}
\newif\ifpdf
	\ifx\pdfoutput\undefined
	\pdffalse 
	\newcommand{\fig}[2]{\includegraphics[width=#1]{./figures/#2.eps}}
	\newcommand{\Fig}[1]{\includegraphics[width=\columnwidth]{./figures/#1.eps}}
         \newlength{\bilderlength} 
	\newcommand{\bilderscale}{0.35}
	
	\newcommand{\bilderskip}{\hspace*{0.8ex}}
	
	\newcommand{\diagram}[1]{%
	\settowidth{\bilderlength}{\bilderskip%
	\includegraphics[scale=\bilderscale]{./figures/#1.eps}\bilderskip}%
	\parbox{\bilderlength}{\bilderskip%
	\includegraphics[scale=\bilderscale]{./figures/#1.eps}\bilderskip}}
	
\else
	\pdfoutput=1 
	\newcommand{\fig}[2]{\includegraphics[width=#1]{./figures/#2.pdf}}
	\newcommand{\Fig}[1]{\includegraphics[width=\columnwidth]{./figures/#1.pdf}}
	\newlength{\bilderlength} 
	\newcommand{\bilderscale}{0.35}
	
	\newcommand{\bilderskip}{\hspace*{0.8ex}}
	
	\newcommand{\diagram}[1]{%
	\settowidth{\bilderlength}{\bilderskip%
	\includegraphics[scale=\bilderscale]{./figures/#1.pdf}\bilderskip}%
	\parbox{\bilderlength}{\bilderskip%
	\includegraphics[scale=\bilderscale]{./figures/#1.pdf}\bilderskip}}
	
	\pdftrue
\fi
\newcommand{\sgn}{{\mathrm{sgn}}}
\newcommand{\rme}{{\mathrm{e}}}
\newcommand{\rmd}{{\mathrm{d}}}
\newcommand{\nn}{\nonumber}

\newcommand{\E}{\epsilon}

\begin{document}
%
\title{\sffamily\bfseries\large 2-Loop Functional Renormalization Group
Theory of the Depinning Transition}
\author{\sffamily\bfseries\normalsize Pierre Le Doussal{$^1$}, Kay
J\"org Wiese{$^2$} and  Pascal Chauve{$^3$} \vspace*{3mm}}
\affiliation{{$^1$} CNRS-Laboratoire de Physique Th{\'e}orique de
l'Ecole Normale Sup{\'e}rieure,
24 rue Lhomond, 75005 Paris, France\\
{$^2$} Institute of Theoretical Physics, University of
California at
Santa Barbara, Santa Barbara, CA 93106-4030, USA\\
{$^3$} CNRS-Laboratoire de Physique des Solides, Universit{\'e} de
Paris-Sud, B{\^a}t.\ 510, 91405 Orsay, France\medskip }
\date{\small May 6, 2002}
\begin{abstract}
We construct the field theory which describes the universal properties of
the quasi-static isotropic depinning transition for
interfaces and elastic periodic systems at zero temperature,
taking properly into account the non-analytic form of the
dynamical action. This cures the inability of the 1-loop flow-equations
to distinguish between statics and quasi-static depinning, and thus to
account for the irreversibility of the latter. We prove two-loop
renormalizability, 
obtain the 2-loop $\beta$-function and show the generation of
``irreversible'' anomalous terms, originating from the non-analytic
nature of the theory, which cause the statics and driven dynamics to differ
at 2-loop order. We obtain the roughness exponent $\zeta$ and dynamical
exponent $z$ to order $\epsilon^2$. This allows to test several
previous conjectures made on the basis of the 1-loop 
result. First it demonstrates that   random-field disorder does
indeed attract all  disorder of shorter range. It also shows that the
conjecture $\zeta=\epsilon/3$ is incorrect, and allows to compute the
violations, as $\zeta=\frac{\epsilon}{3}(1 + 0.14331 \epsilon)$,
$\epsilon=4-d$. This solves a longstanding discrepancy with simulations.
For long-range elasticity it yields $\zeta=\frac{\epsilon}{3}(1 +
0.39735 \epsilon)$, $\epsilon=2-d$ (vs.\ the standard prediction $
\zeta=1/3$
for $d=1$), in reasonable agreement with the most  recent
simulations. The high value of $\zeta\approx 0.5$ found in experiments
both on the   
contact line depinning of liquid Helium and on slow crack fronts is discussed.
\end{abstract}
\maketitle

\section{Introduction} \label{intro}
\subsection{Overview} \label{overview}
Pinning of coherent structures by quenched disorder, and one of its
most striking manifestations, the depinning transition, are important,
ubiquitous and not fully understood phenomena
\cite{Kardar1997,DSFisher1998,BlatterFeigelmanGeshkenbeinLarkinVinokur1994}.
Even a single particle in a quenched random potential exhibits a
depinning threshold at zero temperature: Unbounded motion occurs only
when the additional external applied force $f$ exceeds a critical
force $f_c$.  Depinning also occurs for systems with many interacting
particles, and depending on the degree of order in the structure, it
ranges from the so-called plastic depinning
\cite{ChandranScalettarZimanyi2002} to  elastic depinning.  Here we
focus on elastic depinning  where the particles form a lattice
or more generally a well ordered structure. The  depinning
transition is then a rather non-trivial collective phenomenon,
intrinsically out of equilibrium and irreversible: It is well known
for instance to be a source of hysteresis in magnets and
superconductors \cite{Bertotti1998}.

For many experimental systems which exhibit a depinning transition a
modelization in terms of an elastic object pinned by random impurities
is a good starting point.  The type of disorder, which they
experience, depends on their symmetries and their local environment.
Domain walls in magnets \cite{NattermannBookYoung}, whose study is of
importance to information storage technology, behave as elastic
interfaces and can experience either random-bond disorder (RB), which
is short range (SR), or random-field disorder (RF), which
has long range (LR) spatial correlations.  Dislocation lines in metals
exhibit a depinning threshold as the stress in increased
\cite{ZapperiZaiser2000}.  Charge density waves (CDW) in solids
exhibit a similiar conduction threshold. If the applied electric field
becomes large enough, the CDW starts to slide \cite{Gruner1988}. Being
periodic objects the disorder they feel is also periodic
\cite{MiddletonFisher1993}.  This is also the case for
superconductors, where vortex lines form, in presence of weak
disorder, a quasi ordered periodic Bragg glass phase
\cite{GiamarchiLeDoussal1995,GiamarchiBookYoung}. These systems have
similarities with (vortex free) continuous XY spins in presence of
random fields, and generally constitute the random-periodic (RP) universality
class. 

The contact line of a liquid helium meniscus on a rough substrate can
be thought of as an interface, but is governed by long range
elasticity and so are slowly propagating cracks
\cite{PrevostThese,PrevostRolleyGuthmann1999,ErtasKardar1996,%
DSFisher1998,SchmittbuhlMaloy1997}.  Solid friction is another example
of a depinning phenomenon.  Of course, in each of these systems it
must be checked separately whether the elastic description holds for
depinning. It is far from obvious that this is true for all relevant
scales. In any case, in order to be capable to confirm or rule out
such a description, it is necessary to first obtain precise theoretical
predictions for the  expected behavior in the case of elastic
depinning, what we aim to achieve here.

It was proposed some time ago, starting from the study of a fully
connected mean-field-type model \cite{DSFisher1985}, that the elastic
depinning transition can be viewed in the framework of standard
critical phenomena. The ordered phase is then the moving phase with
force $f>f_c$, and the order parameter the velocity $v$, which
vanishes as $v \sim (f-f_c)^\beta$ at the critical point $f=f_c$.  The
analogy with standard critical phenomena in a pure system however has
some limits: Additional fluctuation exponents were later identified
\cite{NarayanDSFisher1992b,MiddletonFisher1993}, and some
non-universality was noticed in the fully connected model
\cite{DSFisher1985,VannimenusDerrida2001}.

It is thus important to develop a renormalization-group description of
depinning.  An important step in that direction was performed within
the framework of the so-called Functional Renormalization Group (FRG),
to 1-loop order using the Wilson scheme
\cite{NattermannStepanowTangLeschhorn1992,LeschhornNattermannStepanowTang1997,%
NarayanDSFisher1992b,NarayanDSFisher1993a}.  The upper critical
dimension was identified as $d_{\mathrm{uc}} = 4$, $d$ being the
internal dimension of the elastic manifold. The peculiarity of the
problem is that for $d< d_{\mathrm{uc}} = 4$ an infinite set of
operators becomes relevant, parameterized by a full function
$\Delta(u)$, the second cumulant of the random pinning force.  This
problem turns out to be closely related to the statics, i.e.\
describing the pinned state with minimal energy in the absence of an
applied force $f=0$ for which the FRG was initially developed
\cite{DSFisher1986} (there, the flowing function is the second
cumulant $R(u)$ of the random potential).  Both problems are notably
difficult due to so-called dimensional reduction (DR) which renders
the naive $T=0$ perturbation theory useless
\cite{EfetovLarkin1977,NattermannBookYoung,AharonyImryShangkeng1976,Grinstein1976,ParisiSourlas1979,Cardy1983}.
Indeed to {\em any} order in the disorder at zero temperature $T=0$,
{\em any} physical observable is found to be {\em identical} to its
(trivial) average in a Gaussian random force (Larkin) model. This
phenomenon is not restricted to elastic manifolds in disorder, but
occurs in a broad class of disordered systems as e.g.\ random field
spin models and solving it here may open the way to a solution in
other models as well. The FRG at depinning and in the statics seems to
provide a way out of the DR puzzle: the key feature is that the coarse
grained disorder correlator becomes {\em non-analytic} beyond the
Larkin scale $L_c$, yielding large-scale results distinct from naive
perturbation theory, which assumes an analytic disorder correlator.
Explicit solution of the 1-loop Functional RG equation (FRG) for the
disorder correlators $R(u)$ and $\Delta(u)$ gives several non-trivial
attractive fixed points (FP)
\cite{DSFisher1986,GiamarchiLeDoussal1995} and critical exponents for
statics and depinning
\cite{DSFisher1986,GiamarchiLeDoussal1995,NattermannStepanowTangLeschhorn1992,NarayanDSFisher1992b,NarayanDSFisher1993a}
to lowest order in $\epsilon=4-d$. All these fixed points exhibit a
``cusp'' singularity, which has the form $\Delta^{*}(u) -
\Delta^{*}(0) \sim |u|$ at small $|u|$. The existence of the cusp
nicely accounts for the existence of a critical threshold
force\cite{NattermannStepanowTangLeschhorn1992}, as it is found that
$f_c \sim \frac{\rmd }{\rmd u}\ts _{u=0^{+}}\Delta^{*}(u)$.

There are however several highly unsatisfactory and puzzling
features within the 1-loop treatment, which  prompted the
present and related works. First it was found that the FRG flow
equation for the statics and depinning are {\em identical} to one loop
(with $\Delta(u) = - R''(u)$).  This implies for instance that, within a
given universality class (RB,RF and RP), the 1-loop RG is a priori
unable to distinguish static observables, such as the roughness
exponent $\zeta$ at zero applied force $f=0$ from those at depinning
$f=f_c$. This is a rather surprising and unphysical result since one
knows that depinning is an irreversible out of equilibrium 
process, quite different from the statics. In an attempt to recover
the expected physics, and to extend conclusions from the 1-loop study
to higher orders, three {\em conjectures} were put forward
\cite{NattermannStepanowTangLeschhorn1992,LeschhornNattermannStepanowTang1997,NarayanDSFisher1992b,NarayanDSFisher1993a}:

\begin{enumerate} 
\item  At more than 1-loop order depinning should differ from statics.
\item  At depinning the RB universality class should flow to the RF
universality class: Indeed, since for $f \to f_c^+$ the manifold does
not move backward it cannot feel the ``potential'' character of RB
disorder.
\item The roughness exponent of the RF universality class at depinning
is $\zeta = \epsilon/3$ to all orders (the Narayan Fisher (NF) conjecture
\cite{NarayanDSFisher1992b,NarayanDSFisher1993a}), with $\epsilon=4-d$ for
standard manifold elasticity  and $\epsilon=2-d$
for LR elasticity. 
\end{enumerate}

While conjectures 1 and 2 seem reasonable on physical grounds, we
emphasize that they were based on qualitative arguments: In the
absence of any (renormalizable) theory beyond one loop, they appear
putative. A 1-loop study including the effect of a finite velocity
\cite{ChauveGiamarchiLeDoussal2000} indeed indicated that 2 is
correct. It strongly relies on a finite velocity, and the
behavior in the limit $v=0^+$ was found to be subtle and difficult to
fully control within that approach.

The NF conjecture 3 is based on a study of the structure of higher
orders, but it lacks a controlled field theory argument. With the
time, it
got more and more in disagreement with numerical simulations and
experiments, as we discuss below. In addition, if one considers
that this value $\zeta = \epsilon/3$ is expected instead for the {\it
statics} RF class, the NF conjecture seems rather unnatural. 

There are also more fundamental reasons to study the FRG beyond one
loop.  In the last fifteen years since
\cite{DSFisher1985,DSFisher1986}, no study has addressed whether the
FRG yields, beyond one loop, a renormalizable field theory able to
predict universal results.  There have been 2-loop studies
previously but they assumed an analytic correlator and thus they only
applied below the Larkin length
\cite{BucheliWagnerGeshkenbeinLarkinBlatter1998,%
WagnerGeshkenbeinLarkinBlatter1999,Stepanow1995}.  Doubts were even raised
\cite{BalentsDSFisher1993} about the validity of the
$\epsilon$-expansion beyond  order $\epsilon$. 

The aim of the present paper is to develop a more systematic field
theoretic description of depinning which extends beyond one loop. A
short summary of our study was already published
\cite{ChauveLeDoussalWiese2000a} together with a companion study on
the statics. The main and highly non-trivial difficulty is the
non-analytic nature of the theory (i.e.\ of the fixed-point action) at
$T=0$, which makes it a priori quite different from conventional
critical phenomena. It is not even obvious whether this is a
legitimate field theory and how to construct it.  For the depinning
transition with $f=f_c^+$, which is the focus of the present paper, we
are able to develop a meaningful  perturbation theory in a
non-analytic disorder 
which allows us to show renormalizability at 2-loop order.  Even the
way renormalizability works here is slightly different from the
conventional one.  To handle the non-analyticity in the static problem
is even more challenging, and we propose a solution of the problem to
2-loop\cite{ChauveLeDoussalWiese2000a,LeDoussalWieseChauve2002a} and
3-loop\cite{LeDoussalWiese2002b} order as well as at large-$N\ $
\cite{LeDoussalWiese2001}.

In this paper we focus on the so-called ``isotropic depinning''
universality class. This means that the starting model has sufficient
rotational invariance, as discussed below, which guarantees that
additional Kardar-Parisi-Zhang terms are absent. A general discussion
of the various universality classes can be found in
\cite{TangKardarDhar1995,AlbertBarabasiCarleDougherty1998} and an
application of our non-analytic field theory (NAFT) methods to the
case of "anisotropic depinning" will be presented in
\cite{LeDoussalWiese2002c}.

Before we summarize the novel results of the present paper, let us recall some
important features about the model, the scaling and statistical fluctuations
at the depinning threshold.

\subsection{Model, scaling and fluctuations} \label{scaling}

Elastic objects can be parameterized by a $N$-component height or
displacement field $u_x$, where $x$ denotes the $d$-dimensional
internal coordinate of the elastic object (we will use $u_q$ to denote
Fourier components). An interface in the 3D random field Ising model
has $d=2$, $N=1$, a vortex lattice $d=3$, $N=2$, a contact-line $d=1$
and $N=1$. In this paper we
restrict our study to $N=1$. In the presence of a random potential the
equilibrium problem is defined by the Hamiltonian:
\begin{eqnarray}
&& {\cal H}= \int_q \frac{c(q)}2 u_q u_{-q} + \int_x V(u_x,x)
\label{ham}
\end{eqnarray}
with $c(q) = c q^2$ for standard short-range elasticity, $c(q) = c
|q|$ for long-range elasticity and we denote $\int_q = \int \frac {d^d
q}{(2 \pi)^d}$ and $\int_{x}=\int \rmd ^{d}x$.  Long-range elasticity
appears e.g.\ for the contact line by integrating out the bulk-degrees
of freedom \cite{ErtasKardar1994b}.  For periodic systems the
integration is over the first Brillouin zone. More generally a short
scale UV cutoff is implied at $q \sim \Lambda$, and the system size is
denoted by $L$. As will become clear later, the random potential can
without loss of generality be chosen Gaussian with second cumulant
\begin{equation}\label{corrstat}
\overline{V (u, x) V(u',x')} = R(u-u') \delta^d(x-x') \ .
\end{equation}
Periodic systems are described by a periodic function $R(u)$, random
bond disorder by a short range function and random field disorder of
amplitude $\sigma$ by $R(u) \sim - \sigma |u|$ at large $u$.

We study the over-damped dynamics of the manifold in this random
potential, described (in the case of SR-elasticity) by the equation of
motion
\begin{eqnarray}\label{eqn.motion}
\eta \partial_t u_{xt} = c \nabla_x^2 u_{xt}  + F(x, u_{xt} ) + f
\label{eqmo1}
\end{eqnarray}
with friction $\eta$. In presence of an applied force $f$ the
center of mass velocity is $v = L^{-d} \int_x \partial_t u_{xt}$. The
pinning force is $F(u,x) = - \partial_u V(u,x)$ and thus the second
cumulant of the force is
\begin{equation}
 \overline{F(x,u) F(x',u')} = \Delta(u-u') \delta^d(x-x') \ ,
\end{equation}
such that $\Delta(u) = - R''(u)$ in the bare model. As we will see
below it does not remain so in the driven dynamics.  The
``isotropic depinning'' class contains more general equations of motion
than (\ref{eqmo1}). For instance some cellular automaton models are
believed to be in this class \cite{Alava2002}.  They must obey
rotational invariance, as discussed in
Ref. \cite{TangKardarDhar1995,AlbertBarabasiCarleDougherty1998,%
LeDoussalWiese2002c}, which prevents the additional KPZ term $\lambda
(\nabla_x u_{xt})^2$ to be generated at $f=f_c^{+}$. There is always a KPZ
term generated at $v >0$ from the broken symmetry $x \to -x$, but
$\lambda$ can vanish or not as $v\to 0^+$, depending on whether
rotational invariance is broken or not. Here this symmetry is implied
by the  statistical tilt symmetry (STS)
\cite{SchulzVillainBrezinOrland1988,HwaFisher1994b}  $u_{xt} \to
u_{xt} + g_x$. It also holds in the statics and accounts for the
non-renormalization of the elastic coefficient, here set to $c=1$.

A quantity measured in numerical simulations and experiments is the roughness
exponent at the depinning threshold $f=f_c$
\begin{eqnarray}
C_L(x-x') = \overline{|u(x) - u(x')| ^2} \sim |x-x'|^{2 \zeta} \ ,
\label{corr}
\end{eqnarray}
which can be compared to the static one $\zeta_{\mathrm{eq}}$. Other
exponents have been introduced
\cite{DSFisher1985,NattermannStepanowTangLeschhorn1992,LeschhornNattermannStepanowTang1997,NarayanDSFisher1992b,NarayanDSFisher1993a}. The
velocity near the depinning threshold behaves as $v \sim
(f-f_c)^\beta$; the dynamical response scales with the dynamical
exponent $t \sim x^z$ and the local velocity correlation length $\xi$
diverges at threshold with $\xi \sim (f-f_c)^{-\nu}$.  There have also
been some studies below threshold
\cite{MiddletonFisher1991,MiddletonFisher1993}. The following exponent
relations were found to hold
\cite{NattermannStepanowTangLeschhorn1992}:
\begin{eqnarray}
\beta &=& \nu (z - \zeta) \label{Nat1} \\
\nu &=& \frac{1}{2 - \zeta}\label{Nat2}
\end{eqnarray}
the latter using STS. There are various ways to measure the roughness
exponent. In some simulations
\cite{RossoKrauth2001a,RossoKrauth2001b,RossoKrauth2002} it has been
extracted from the critical configuration, i.e.\ as $f$ is increased
to $f_c$ in a given sample it is obtained from the last blocking
configuration. It can also be defined as the limit $v \to 0^+$ of the
roughness in the moving state, which we will refer to as the
``quasi-static'' depinning limit to distinguish it from the previous
one. This is the situation studied in this paper.  Although it is
widely believed that both are the same, the depinning theory has
enough peculiarities that one should be careful. In particular, beyond
scaling arguments and simulations, there is presently no rigorous
method capable to connect the behavior  below and above threshold.

Another peculiarity was noted in \cite{NarayanDSFisher1992b}. It was found that
the finite-size fluctuations of the critical force
can scale with a different exponent:
\begin{eqnarray}
f_c(L) - f_c \sim L^{-1/\nu_{\mathrm{FS}}}
\end{eqnarray}
and it was questioned whether $\nu_{\mathrm{FS}} = \nu$. The  bound
\begin{eqnarray} \label{FCCS}
\nu_{\ind{FS}} \geq 2/(d + \zeta)
\end{eqnarray}
follows from a general argument of
Ref. \cite{ChayesChayesFisherSpencer1986}. For charge density waves
where $\zeta=0$ one sees that $\nu=1/2$ and thus $\nu$ and
$\nu_{\mathrm{FS}}$ must be different for $d<4$. For interfaces it was
noted \cite{NarayanDSFisher1992b} that $\nu = \nu_{\mathrm{FS}}$ is
possible provided $\zeta \geq \epsilon/3$. If one assumes $\nu =
\nu_{\mathrm{FS}}$, the NF-conjecture $\zeta=\epsilon/3$ is then
equivalent to saturating the bound (\ref{FCCS}).  We will address the
question of whether $\nu = \nu_{\mathrm{FS}}$ below.

Finally note that at $f=f_c$ the condition of equilibrium of a piece of
interface expresses that the elastic force, which acts only on the perimeter,
balances the excess force on the bulk, yielding the scaling:
\begin{eqnarray}
L^{d-1} u(a,L) \sim (f_c(L) - f_c) L^{d}
\ ,
\end{eqnarray}
where $u(a,L) \sim \sqrt{C_L(a)}$ is the relative displacement
(\ref{corr}) between two neighbors averaged 
over the perimeter. This shows that
\begin{eqnarray}
u(a,L) \sim L^{1 - \frac{1}{\nu_{\mathrm{FS}}} }
\end{eqnarray}
thus for CDW the displacements between two neighbors grows unboundedly
\cite{CoppersmithMillis1991} with $L$ for $d \leq 2$.  For interfaces
(non-periodic disorder), if one assumes $\nu = \nu_{\mathrm{FS}}$ one
obtains that the displacements between two neighbors grows with $L$
only when $\zeta >1$.

\subsection{Summary of results} \label{results}
Let us now discuss the main results of our study.

First we show that, at depinning, 1- and 2-loop diagrams can be
computed using a non-analytic action in an unambiguous and well
defined way, allowing to escape dimensional reduction. The mechanism
is non-trivial and works because the manifold only moves forward in
the steady state which allows to remove all ambiguities.  We show that
the limit $v \to 0^+$ can be taken safely without additional
unexpected singularities arising in this limit.

Next we identify the divergences in the 2-loop diagrams using
dimensional regularization in $d=4-\epsilon$. We identify the 1-loop
and 2-loop counter-terms and perform the renormalization program. We
find that the $1/\epsilon$ divergences cancel nicely in the $\beta $-%
function for the disorder correlator and in the dynamical
exponent. The theory is finite to 2-loop order and yields universal
results.

The obtained FRG flow equation for the disorder (the $\beta
$-function) contains new ``anomalous'' terms, absent in an analytic
theory (e.g.\ in the flow obtained in
Ref. \cite{BucheliWagnerGeshkenbeinLarkinBlatter1998,%
WagnerGeshkenbeinLarkinBlatter1999}). These terms are different in the
static theory (obtained in \cite{ChauveLeDoussalWiese2000a}) and at
depinning, showing that indeed {\em static and depinning differ to two
loops}. {\bf Thus the minimal consistent theory for depinning requires two
loops.}

Next we study the fixed point solutions of our 2-loop FRG equations
at depinning. For non-periodic disorder (e.g.\ interfaces) with
correlator of range shorter or equal to  random-field, we find that
there is a single universality class, the random-field class. Thus
random-bond disorder does flow to random field. Specifically we find that the
flow of $\int \Delta$ is corrected to two loops and thus $\int \Delta$
cannot remain at its random-bond value, which is zero. This is explained in
more detail in section \ref{sec:fixedpoints}. The problem does not remain
potential and irreversibility is manifest. For short range elasticity,
we find the roughness-exponent at depinning:
\begin{equation}
\zeta = \frac{\epsilon}{3} \Big(1 + 0.143313\, \epsilon\Big)
\label{sr}
\end{equation}
with $\epsilon=4-d$, and for long range elasticity:
\begin{equation}
\zeta = \frac{\epsilon}{3} \Big(1 + 0.39735\, \epsilon\Big)
\label{lr}
\end{equation}
with $\epsilon=2-d$. Thus the NF-conjecture
\cite{NarayanDSFisher1992b,NarayanDSFisher1993a} that $\zeta
=\frac{\epsilon }{3}$  is incorrect. We
also compute the dynamical exponent $z$ and obtain $\beta$ and $\nu$
by the  scaling relations (\ref{Nat1}) and (\ref{Nat2}). We also find that
$\nu_{\mathrm{FS}}=\nu$ holds to two loops.

For periodic disorder, relevant for  charge density waves, we
find a fixed point which leads to a universal logarithmic growth of
displacements. This fixed point is however unstable, as an additional
Larkin random force is generated. The true correlations are the sum of
this logarithmic growth and of a power law growth so that the true
$\zeta = (4-d)/2$. This is similar to
\cite{LeDoussalGiamarchi1997}. Then we find
\begin{eqnarray}
 \nu &=& \frac{1}{2} \\
\nu_{\mathrm{FS}} &=& \frac{2}{d}\ ,
\end{eqnarray}
which holds presumably to all orders.

\subsection{Numerical simulations and experiments} \label{experiments}
Over many years, numerous simulations near depinning
\cite{NattermannStepanowTangLeschhorn1992,%
LeschhornNattermannStepanowTang1997,Leschhorn1993,%
RotersHuchtLubeckNowakUsadel1999,NowakUsadel1998,ThomasPaczuski1996}
accumulated evidence that $\zeta\neq \epsilon/3$. In $d=1$ in
particular often an exponent $\zeta >1$ was observed. Our results show
that $\zeta > \epsilon /3$ and thus
resolve this long standing discrepancy between numerical simulations
and the renormalization group. They are summarized in Tables
\ref{alpha=2 table} and \ref{alpha=1 table} in Section
\ref{sec:fixedpoints}, where we compare them to numerical
simulations. Of course it is not possible to give strict error bars
from the FRG calculation without further knowledge of higher orders,
but one can still give rough estimates, based on different
Pad\'e-approximants.

Let us in the following discuss recent numerical results.
Following shortly our paper \cite{ChauveLeDoussalWiese2000a}, Rosso
and Krauth obtained a set of  precision numerical results using a
powerful algorithm to determine the critical
configuration at depinning (the last blocking configuration) up to
large sizes\cite{RossoKrauth2001a,RossoKrauth2001b,RossoKrauth2002}. 
They obtained results in $d=1$ which, despite being far
from $d=4$, compare well with our results.  For short range elasticity
they find
\begin{eqnarray}
 \zeta = 1.17
\label{rossouno}
\end{eqnarray}
close to our 2-loop result (\ref{sr}). Note that displacement
correlations scaling as
\begin{equation}
 \overline{u_{q} u_{-q}} \sim q^{- ( d + 2 \zeta)}
\end{equation}
with $\zeta >1$ are perfectly legitimate. It simply means that
\begin{equation}
C_L(x) \sim 2 \int_q (1-\cos(q x)) q^{- ( d + 2 \zeta)} \sim L^{2(\zeta -
1)} x^2 \ .
\end{equation}
The size dependent factor comes from the infrared divergence of the
integral. Thus in a simulation neighboring monomers will be spread
further and further apart, which is fine if their attraction is purely
quadratic. Of course in a realistic physical situation their bond will
eventually break, but as a model it is mathematically well defined.
For the anisotropic depinning universality class, not studied here,
they found $\zeta=0.63$ as many other authors using cellular automaton
models
\cite{TangLeschhorn1992,BuldyrevBarabasiCasertaHavlinStanleyVicsek1992,%
GlotzerGyureSciortinoConiglioStanley1994}. 

For isotropic depinning with long range elasticity they obtained:
\begin{eqnarray}
&& \zeta = 0.390 \pm 0.002
\label{rossodue}
\ ,
\end{eqnarray}
which lies roughly at midpoint of the 1-loop and 2-loop prediction
setting $\epsilon=1$ in (\ref{lr}).  So do their most recent estimates
\cite{RossoHartmannKrauthInPreparation} for SR disorder. In $d=2$ this
is $\zeta= 0.753 \pm 0.002$ and for $d=3$ they obtain $0.35 < \zeta
< 0.4$.  These results (\ref{rossouno}) and (\ref{rossodue}) are
close to estimates from the 2-loop expansion and clearly rule
out the NF-conjecture.

\begin{figure}[t]
\centerline{\Fig{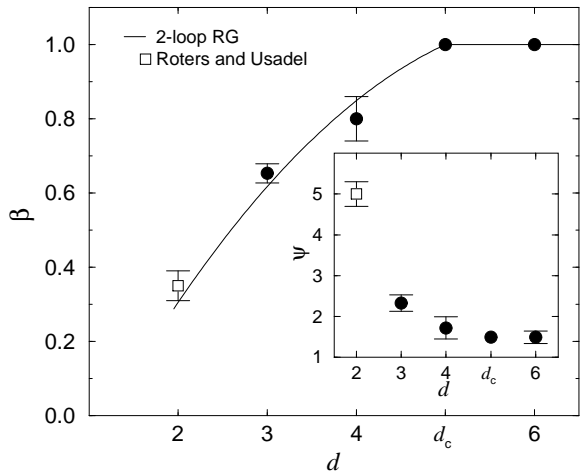}}
\caption{Figure from Ref.~\cite{RotersUsadelSubmitted} which compares
new numerical values (black circles) and a previous one (white square) obtained
for the exponent $\beta$ with  our prediction from the FRG.}
\label{fig:RotersEtAl.}
\end{figure}
Another recent work \cite{RotersUsadelSubmitted} studies an interface
in the random field Ising model in high dimension. The authors confirm that
$d=4$ is the upper critical dimension. They further extract the velocity
exponent $\beta$ and compare their results with our 2-loop FRG
prediction for $\beta$:
\begin{equation}
\beta = 1 - \frac \epsilon 9 + 0.040123 \epsilon^2  
\ .
\end{equation}
The results are shown in Fig.~\ref{fig:RotersEtAl.}. One can see a
clear curvature downwards and that the straight line giving the 
1-loop result is well above the obtained results (the 1-loop
approximation would predict $\beta=0.78$ in $d=2$).

Thus, although there is still some spread   and uncertainty in the results,
it seems that there is now a trend towards a convergence between
theory and numerical simulations.

The situation concerning experiments is presently unclear. Let us
first outline the generic findings before analyzing the details. The
measured exponents corresponding to LR-elasticity and $d=1$ seem to be
consistently in the range $\zeta \approx 0.5-0.55$. This is slightly
above our 2-loop result (\ref{lr}) but not fully incompatible with
it. Our calculation holds for quasi-static depinning, i.e.\ $v>0 \to
0^+$, and most experiments are also performed from the moving side,
hopefully reaching the same quasi-static limit $v \to 0^+$. On the
other hand if one believes that the numerical result \ref{rossodue}
(also compatible with our calculation, from below) obtained for
$f=f_c^-$ also holds for quasi-static depinning (a rather natural, but
as yet unproved assumption) then one must conclude that the elastic
models, in their simplest form at least, may not faithfully represent
the experimental situation. Care must however be exercised before any
such conclusion is reached. One could argue that disorder $\Delta
(u)\sim u^{-\alpha }$ of range longer than RF ($\alpha <1$) could
produce higher exponents $\zeta=\epsilon/(2 + \alpha)$ (see end of
Section \ref{non-period systems}) but that does not seem to apply to
those experiments where disorder is well controlled.  Also, since the
exponent $\zeta =0.5$ is the Larkin DR-exponent, which should hold
below the Larkin length $L_c$ one must make sure that $L_c$ is well
identified and that one is not simply observing a slow crossover to
the asymptotic regime. In some of these experiments $L_c$ has been
identified to be rather small.

Let us now examine the situation in more details.

One much studied experimental system is the contact line of a fluid
\cite{PrevostThese,PrevostRolleyGuthmann2002}. It advances on a rough
substrate and is pushed by adding fluid to the reservoir. The
elasticity of the line is short range at short scale but at larger
scales it is mediated by the elasticity of the two dimensional
meniscus and thus it becomes long range and should be compared with
(\ref{lr}), (\ref{rossodue}). Disorder is random-field, but one should
distinguish between microscopic disorder, which is poorly
characterized, and macroscopic one which is well controlled. The
situation has been studied for a helium meniscus on a macroscopically
disordered substrate where $\zeta=0.55$ was found
\cite{PrevostRolleyGuthmann2002}.  Although there are good indications
that these experiments probe quasi-static depinning (the contact line
jumps from a reproducible pinned configuration to the next one) the
precise nature of the dynamics remains open. Indeed it was found that
propagation of perturbations along the line can be as fast as
avalanches, showing inertial regime for helium\footnote{M. Poujade,
C. Guthmann and E. Rolley, submitted to Europhys. Lett. (2002).
Preliminary report in J. of Low Temp. Phys. 126 (2002) 379.}.
Experiments were repeated for viscous liquids \footnote{S. Moulinet,
C. Guthmann and E. Rolley, in preparation; to be submitted to EPJ B.}
yielding $\zeta = 0.51 \pm 0.03$. There it was checked that the system
is over-damped and near depinning. In both cases there is also
evidence of thermal activation effects
\cite{PrevostRolleyGuthmann1999} characteristic of depinning (not
creep).  It was argued that these may be a signature that a more
complicated dynamics (e.g. plastic) takes place at the very short
scales and produces an effective dynamics at larger scales with
complicated non-linear (e.g.  exponential) velocity and temperature
dependent damping.  Very similar effects have also been shown to occur
in solid friction\footnote{L. Bureau, T. Baumberger, C. Caroli,
cond-mat/0101357.}  were the activation volume was also found to
correspond to microscopic scales.

Another class of much studied experimental systems are crack fronts in
heterogeneous media\footnote{for a review see E. Bouchaud,
J. Phys. Condensed Matter {\bf 9}, 4319 (1997), and Refs.\
\cite{ParisiCaldarelliPiotronero2000,ThomasPaczuski1996}.}.  These are
characterized by two displacement fields, one out-of-plane component
$h$ and an in-plane one $f$. Cracks can either be studied stopped or
slowly advancing. At the simplest level the in-plane displacement $f$
is expected to be described as an elastic line $d=1$, $N=1$ with LR
elasticity $c |q|$, at quasi static depinning \footnote{See
e.g. S. Ramanathan and D.S. Fisher, condmat/9712181; S. Ramanathan,
D. Ertas and D.S. Fisher, condmat/9611196, and references
therein.}. In experiments \cite{SchmittbuhlMaloy1997}$^{,}$%
\footnote{A. Delaplace et al. Phys. Rev. E {\bf 60} 1337 (1999),
P. Daguier et al. Europhys. Lett. 367 (1995).}  the observed roughness
is again $\zeta_f \approx 0.55$. Since the crack propagates in an
elastic medium, elastic waves which can in principle affect the
roughness as the crack front advances producing a more complicated
dynamics than Eq.~(\ref{eqn.motion}). Some proposals have been put
forward on mechanisms to produce higher roughness exponents
\footnote{See e.g. E. Bouchaud et al., condmat/0108261, and references
therein.}  They rely however on a finite velocity and it is unclear
whether they can modify roughness in the quasi-static limit. Even if
instantaneous velocities during avalanches become large enough, a
detailed description on how these could change the line configurations
remains to be understood. Then of course a major issue is whether the
experiment, and in which sense, is in the quasi-static limit.  There
again microscopic dynamics could be more complex as at small scales
the material may be damaged and the notion of a single front may not
apply. Finally, since there are two components to displacement one
should also be careful to understand interactions between them near
depinning \cite{ErtasKardar1994}.

Another interesting experimental system is a domain wall in a
very thin magnetic film
\cite{LemerleFerreChappertMathetGiamarchiLeDoussal1998} which
experiences RB disorder. Up to now however only the thermally activated
motion has been studied, which gives a quite remarkable confirmation
of the creep law
\cite{LemerleFerreChappertMathetGiamarchiLeDoussal1998} with RB
exponents.  It would be interesting to study depinning there and to
check whether it also belongs to the isotropic universality class. In
that case, the crossover from RB to RF resulting in overhangs beyond
some scale at zero temperature ($\zeta >1$) as well as the non-trivial
thermal rounding of depinning could be studied.


\section{Model and perturbation theory}

In this Section we discuss some general features of the field theory
of elastic manifolds in a random potential, both for the statics
and for the dynamics, driven or at zero applied force. Some issues
are indeed common to these three cases. At the end we
specialize to depinning.

\subsection{Static and Dynamical action and naive power counting}

The static, equilibrium problem, can be studied using replicas. The
replicated Hamiltonian corresponding to (\ref{ham}) is:
\begin{eqnarray}
 \frac{{\cal H}}{T} &=& \frac{1}{2T}   \int_x \sum_a [(\nabla u^a_x)^2
+ m^2 u^a_x] \nonumber \\ 
&& - \frac{1}{2 T^2} \int_x \sum_{a b} R(u^a_x - u^b_x)
\ ,
\end{eqnarray}
where, for now, we consider SR elasticity. $a$ runs from 1 to $n$. We
have added a small mass to provide an infrared cutoff, and we are
interested in the large scale limit $m \to 0$. The limit of zero
number of replicas $n=0$ is implicit everywhere. Terms with sums over three
replicas or more corresponding to third or higher cumulants of
disorder are generated in the perturbation expansion. These should in
principle be included, but as we will see below higher disorder
cumulants are not relevant for the $T=0$ depinning studied below.

The dynamics, corresponding to the equation of motion (\ref{eqmo1})
is studied using the dynamical action averaged over disorder:
\begin{eqnarray}
{\cal S}[\hat{u}, u] &=& \int_{xt} i \hat{u}_{xt} (\eta \partial_t
- \partial_x^2 + m^2) u_{xt} - \eta T  \int_{xt} i \hat{u}_{xt} i \hat{u}_{xt} \nonumber \\
&& - \frac{1}{2} \int_{xtt'} i \hat{u}_{xt} i \hat{u}_{xt'}
\Delta(u_{xt}- u_{xt'}) - \int_{xt} i \hat{u}_{xt} f_{xt}\ . \nonumber \\
&& \label{msr}  
\end{eqnarray} 
It generates disorder averaged correlations, e.g.  $\overline{\langle
A[u_{xt}]\rangle}= \langle A[u_{xt}]\rangle_S$ with $\langle
A\rangle_S=\int {\cal D}[u] {\cal D}[\hat{u}] A \rme^{-S}$ and
$\langle 1 \rangle_S=1$, and response functions $\overline{\delta
\langle A[u]\rangle/\delta f_{xt}}= \langle i \hat{u}_{xt} A[u]
\rangle _{S}$. The uniform driving force $f_{xt}=f>0$ (beyond
threshold at $T=0$) may produce a velocity $v = \overline{\partial_t
\langle u_{xt}\rangle}>0$, a situation which we study by going to the
comoving frame (where $\overline{\langle u_{xt}\rangle}=0$) shifting
$u_{xt} \to u_{xt} + v t$, resulting in $f \to f - \eta v$. This is
implied below. In general, for any value of $f$, we study the steady
state, which at finite temperature $T>0$ is expected to be unique and
time translational invariant (TTI) (all averages depend only on time
differences). In the zero temperature limit, one needs a priori to
distinguish the $T=0$ TTI theory as $\lim_{L \to \infty} \lim_{T \to
0}$ (e.g.\ the ground state in the static) and the $T=0^+$ theory as
$\lim_{T \to 0} \lim_{L \to \infty}$.

It is important to note that there are close connections, via the
fluctuation dissipation relations, between the dynamical formalism and
the statics.  Indeed, at equilibrium (for $f=0$ and when time
translation invariance is established) any equal time correlation
function computed with (\ref{msr}) is formally identical (e.g.\ to all
orders in perturbation theory) to the corresponding quantity computed in the
equilibrium theory (which is a single replica average).  Similarly,
the persistent parts, i.e.\ those $\propto \delta(\omega)$, of
dynamical correlations involving $p$ mutually very separated times,
are formally identical to the corresponding averages in the replica
theory involving $p$ replicas. The perturbative equilibrium
calculations in the statics can thus be indifferently performed either
with replicas or with (\ref{msr}). It is possible to generate all
dynamical graphs from static ones, a connection which, as will be
further explained below, also carries to some extent to the case $f>0$
at $T=0$.

\begin{figure}[tb]
\centerline{ \fig{0.45\textwidth}{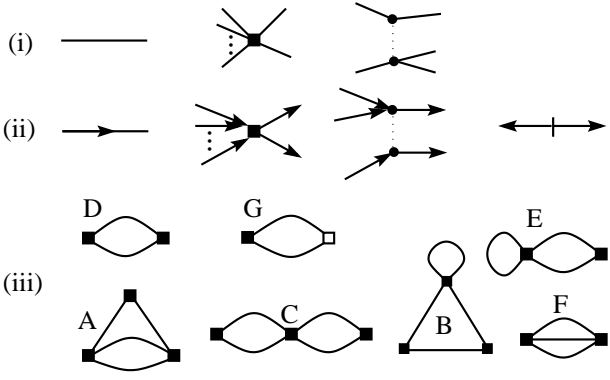} }
\caption{{
(i) diagrammatic rules for the statics: replica propagator $\langle u_a
u_b\rangle_0 \equiv T\delta_{ab}/q^{2}$,
unsplitted vertex, equivalent splitted vertex $-\sum_{ab} \frac{1}{2
T^2} R(u_a-u_b)$ and (ii) dynamics: response propagator $\langle \hat{u} u
\rangle_0 \equiv R_{q,t-t'}$,
unsplitted vertex, splitted vertex $\hat{u}_{xt} \hat{u}_{xt'}
\Delta(u_{xt}-u_{xt'})$
and temperature vertex. Arrows are along increasing time. An
arbitrary number of lines can enter these functional
vertices.
(iii) unsplitted diagrams to one loop D, one loop with inserted 1-loop
counter-term G and 2-loop A,B,C,E,F.}}
\label{fig1}
\end{figure}

We first study ``naive'' perturbation theory and power counting. The
quadratic part $S_0$ of the action (\ref{msr}) yields the free
response and correlation functions, used for perturbation theory in
the disorder. They read
\begin{eqnarray}
 \langle i \hat u_{q,t'} u_{-q,t} \rangle_0 &=& R_{q,t-t'}=
\frac{\theta(t-t')}{\eta} \rme^{-\frac{(t-t')}{\eta}
 (q^2+m^2)}
\nonumber
\\
\langle u_{q,t'} u_{-q,t} \rangle _0 &=& C_{q,t-t'}
\end{eqnarray}
respectively, with the FDT relation $T R_{q,\tau} = - \partial_\tau
C_{q,\tau}$ ($\tau>0$). Perturbation theory in $\Delta(u)$ yields a
disorder interaction vertex and at each (unsplitted) vertex there is
one conservation rule for momentum and two for frequency. It is thus
convenient to use splitted vertices, as represented in Fig \ref{fig1},
where the rules for the perturbation theory of the statics using
replica are also given. For the dynamics one can also focus on $T=0$
where graphs are made only with response functions and consider
temperature as an interaction vertex. The 1-loop and 2-loop diagrams
which correct the disorder at $T=0$ are shown in Fig. \ref{fig1} (unsplitted
vertices). There are three types of 2-loop graphs $A,B,C$. The graphs
$E$ and $F$ lead to corrections proportional to temperature.

At $T=0$ the model exhibits the property of dimensional reduction
\cite{EfetovLarkin1977,NattermannBookYoung,AharonyImryShangkeng1976,Grinstein1976,ParisiSourlas1979,Cardy1983}
(DR) both in the statics and dynamics.
Its ``naive'' perturbation theory,
obtained by taking for the disorder correlator $\Delta(u)$ an {\it
analytic function } of $u$ (or $R(u)$ for the statics), has a
triviality property.  As is easy to show using the above diagrammatic rules
the perturbative expansion of any correlation function $\langle \prod_i u_{x_i
t_i}\rangle _S$ in the derivatives $\Delta^{(n)}(0)$ yields to all
orders the same result as that obtained from the Gaussian theory
setting $\Delta(u) \equiv \Delta(0)$ (the so called Larkin random
force model).  The same result holds for the statics, for any
correlation $\langle \prod_i u^{a_i}_{x_i}\rangle _S$.  At $T=0$ these
correlations are independent of the replica indices $a_i$, their
dynamical equivalent being independent of the times $t_i$. The two
point function thus reads to all orders:
\begin{eqnarray}
\overline{\langle u_{q,t'} u_{-q,t} \rangle  }_{\mathrm{DR}} =
\frac{\Delta(0)}{(q^2 + m^2)^2}
\ .
\end{eqnarray}
This dimensional reduction results in a roughness exponent
$\zeta=(4-d)/2$ which is well known to be incorrect.  One physical
reason, in the statics, is that this amounts to solving the zero force
equation which, whenever more than one solution exists, is not
identical to finding the lowest energy configuration. Curing this
problem, within a field theory, is highly non-trivial. One way to do
that, as discussed later will be to consider a {\em non-analytic}
$\Delta(u)$.

It is important to note that despite the DR, dynamical averages
involving response fields remain non-trivial, even at zero
temperature.  Perturbation theory at finite temperature also remains
non-trivial. It is thus still useful to do power counting with an
analytic $\Delta(u)$, the modifications for a non-analytic $\Delta(u)$
being discussed in the following section.

Power counting at the Gaussian fixed point yields $t \sim x^2$ and
$\hat{u} u \sim x^{-d}$.  At $T=0$ nothing else fixes the dimensions
of $u$, since $u \to \lambda u$, $\hat{u} \to \lambda^{-1} \hat{u}$
leaves the $T=0$ action invariant. Denoting $u \sim x^{\zeta}$,
$\zeta$ is for now undetermined.  The disorder term then scales as
$x^{4-d+2\zeta}$. It becomes relevant for $d < 4$ provided $\zeta <
(4-d)/2$ which is physically expected (for instance in the random
periodic case, $\zeta=0$ is the only possible choice, and for other
cases $\zeta = O(\epsilon)$). With this power counting the temperature
term scales as $x^{-\theta}$ with $\theta = d - 2 + 2 \zeta$ and is
thus formally irrelevant near four dimension. In the end $\zeta$ will
be fixed by the disorder distribution at the fixed
point\cite{DSFisher1986}.

A more detailed study of divergences in the vertex functions allows to
identify all counter-terms needed to render the theory finite. We
denote by
\begin{eqnarray}
&& \Gamma_{\hat{u}..\hat{u};u ..u}(\hat{q}_i,\hat{\omega}_i,q_i,\omega_i) = \\
&&\qquad \qquad  \prod_{i=1}^{E_{u}} \frac{\delta}{\delta u_{q_i, \omega_i} }
\prod_{j=1}^{E_{\hat{u}}} \frac{\delta }{\delta \hat u_{\hat{q}_j,
\hat{\omega}_j} } \Gamma[u,\hat u] \ts _{u=\hat u=0} \nonumber
\end{eqnarray}
the irreducible vertex functions (IVF) with $E_{u}$ external fields
$u$ (at momenta $q_i,\omega_i$, $i=1,..E_{u}$) and $E_{\hat{u}}$
external fields (at momenta $\hat q_i,\hat \omega_i$, $i=1,..E_{\hat{u}}$).
Being the derivative of the effective action functional $\Gamma[u,\hat
u]$ they are the important objects since all averages of products of
$u$ and $\hat{u}$ fields are expressed as tree diagrams of the
IVF. Finiteness of the IVF thus imply finiteness of all such
averages. The present theory has the property of covariance under the
well known statistical tilt symmetry STS $u_{xt} \to u_{xt} + g_x$,
which yields that the two point vertex $\Gamma_{\hat{u} u}(\omega=0)$
remains uncorrected to all orders. This allows to fix the elastic
constant $c=1$ and shows that the mass term is uncorrected and can
thus safely be used as an IR cutoff.  It also implies that all higher
IVFs vanish when any of the $\omega_i$ is set to zero. The DR result
is a perturbative triviality statement about
$\Gamma_{\hat{u}..\hat{u}}(\hat{q}_i,\hat{\omega}_i)$ at $T=0$, all
other cases remain non-trivial. In a sense we will now expand around
dimensional reduction.  Similar replica IVFs can be defined for the
statics.

\begin{figure}[tb] \centerline{\fig{\columnwidth}{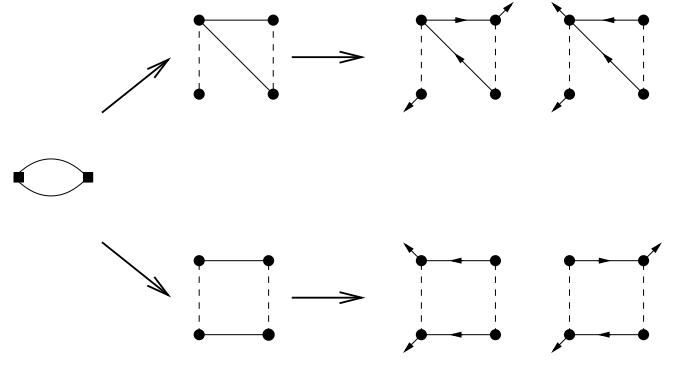}}
\caption{Construction of diagrams starting from an unsplitted static
diagram via two splitted static diagrams (2-replica component) to the
corresponding dynamical diagrams as explained in the text.}
\label{1loopexample}
\end{figure}

Perturbation expansion of a given IVF to any given order in the
disorder can be represented by a set of one particle irreducible (1PI)
graphs. As mentionned above there is a simple rule to generate the
dynamical graphs from the static ones. The static propagator being
diagonal in replicas, each static graph occurring in a $p$ replica IVF
contains $p$ connected components. At $T=0$ the rule is then to attach
one response field to each connected component of the static diagram,
each replica graph then generating one or more dynamical graphs. The
place where the response field is attached is the {\em root} of the
diagram. The direction of the remaining response functions is then
fixed unambiguously, always pointing towards the root. This procedure
to deduce the dynamical diagrams from the static ones is {\em unique}
and {\em exhaustive} and is illustrated in Fig.~\ref{1loopexample}. A
generalization exists at $T>0$ but is not needed here.

Any graph corresponding to a given dynamical IVF contains $p$
connected components (in the splitted diagrammatics) with $1 \le p \le
E_{\hat{u}}$ ($p=E_{\hat{u}}$ at $T=0$), each one leading to a
conservation rule between external frequencies, and thus one can write
symbolically:
\begin{eqnarray}
&&\!\! \Gamma_{\hat{u}..\hat{u};u
..u}(\hat{q}_i,\hat{\omega}_i,q_i,\omega_i) \nonumber \\ 
&&\qquad 
= \delta\left(\sum \hat{q} - \sum q\right) \prod_{i=1}^p
\delta\left(\sum \hat{\omega} -  \sum 
\omega \right) \tilde{\Gamma}
\ . \qquad \ \ 
\end{eqnarray}
Let us compute the superficial degree of UV divergence $\alpha$ of
such a graph with $v_\Delta$ disorder vertices and $v_T$ temperature factors
contributing to $\tilde{\Gamma} \sim \Lambda^\alpha$. Using momentum and
frequency
conservation laws at each vertex, and since there are only response
functions $E_{\hat{u}} + I = 2 ( v_\Delta + v_T)$ we obtain:
\begin{eqnarray}
\alpha= d + 2 p - d E_{\hat{u}} + (d-4) v_\Delta + (d-2) v_T
\label{superficial}
\ .\ \ 
\end{eqnarray}
At $T=0$ ($ v_T =0$, $p=E_{\hat{u}}$) at the critical dimension $d =4$
the only superficially UV divergent IVF are those with one external
$\hat{u}$ (quadratic divergence) or two external $\hat{u}$
(logarithmic divergence LD). The STS further restricts the possible
divergent diagrams. One sees that only three types of counter-term are
needed a priori. One counter-terms is needed for the $\Lambda^2$
divergence of $\Gamma_{\hat{u}}(q=0,\omega=0)$ (excess force $f - \eta
v$ in driven dynamics). This is analogous to the mass in the $\phi^4$
theory, i.e.\ the distance to criticality. If we are exactly at the
depinning critical point ($f=f_c$) we need not worry about this
divergence. Another counter-term is associated with the LD in $\eta$
and the last one with the LD in the second cumulant of disorder
$\Delta(u)$, i.e.\ a full function, which makes it different from the
conventional FT for critical phenomena (e.g.\ $\phi^4$). One notes
that higher cumulants are formally irrelevant, as they involve
$E_{\hat{u}} > 2$.

One sees from (\ref{superficial}) that each insertion of a temperature
vertex yields an additional quadratic divergence in $d=4$, more
generally a factor $T \Lambda^{d-2}$. Thus to obtain a theory where
observables are finite as $\Lambda \to \infty$ one must start from a
model where the initial temperature scales with the UV cutoff as
\begin{eqnarray}
T = \tilde{T} m^{- \theta} \left(\frac{m}{\Lambda}\right)^{d-2}
\label{rescaled}
\ .
\end{eqnarray}
This is similar to the $\phi^4$ theory where it is known that a
$\phi^6$ term can be present and yields a finite UV limit (i.e.\ does
not spoil renormalizability) only if it has the form $g_6
\phi^6/\Lambda^{d-2}$. It then produces only a finite shift to $g_4$
without changing universal properties \footnote{We thank E. Brezin for
a discussion on this point.}. Here each $\tilde{T}$ factor will thus
come with a $\Lambda^{2-d}$ factor which compensates the UV
divergence.  Computing the resulting shift in $\Delta(u)$ to order
$\Delta^2$ by resumming the diagrams $E$ and $F$ of Fig. \ref{fig1}
and all similar diagrams to any number of loops has not been attempted
here.

For convenience we have inserted factors of $m$ in the definition
of the rescaled temperature, using the freedom to
rescale $u$ by $m^{-\zeta}$ and $\hat u$ by $m^{\zeta}$. The disorder
term then reads as in (\ref{msr}) with $\Delta(u)$ replaced by $\Delta_0(u) =
m^{\epsilon - 2 \zeta} \tilde{\Delta}(u m^{\zeta})$ in terms of a dimensionless
rescaled function $\tilde{\Delta}$.

\subsection{Non-analytic field theory and depinning in the quasi-static limit}

From now on we study the zero temperature limit $T=0$. To escape the
DR triviality phenomenon, and since the fixed points found in 1-loop
studies exhibit a cusp at $u=0$, we must consider perturbation theory
in a {\em non-analytic} disorder correlator. In this section we show
how to develop perturbation theory and diagrammatics in a non-analytic
theory and what are the non-trivial issues which arise.

For now the considerations apply for zero or finite applied force.  In
usual diagrammatics, extracting a leg from a vertex corresponds to a
derivation. Here this can be done as usual with no ambiguity, provided
the corresponding vertex is evaluated at a generic $u$ (e.g.\ the
graphs in Fig~\ref{1loopexample}).  If the vertex is evaluated at
$u=0$ (here and in the following we call them {\em saturated vertices}) one
must go back to a careful application of Wick's rules. Any graph
containing such a vertex and which vanishes in the analytic theory is
called anomalous.  Let us  write the series expansion in powers
of $|u|$:
\begin{eqnarray}\label{dadatutu}
\Delta(u) = \Delta(0) + \Delta'(0^+) |u| + \frac{1}{2} \Delta''(0^+) u^2  + ...
\label{series}
\ .
\end{eqnarray}
Wick's rules can then be applied but usually end up in evaluating
non-trivial averages of e.g.\ sign or delta functions.

Let us consider as an example the following 2-loop 1PI diagram (noted
$\mathrm{e}_1$ in what follows) which is a correction to the effective
action of the form: \bea
&&\!\!\!\!\! \diagram{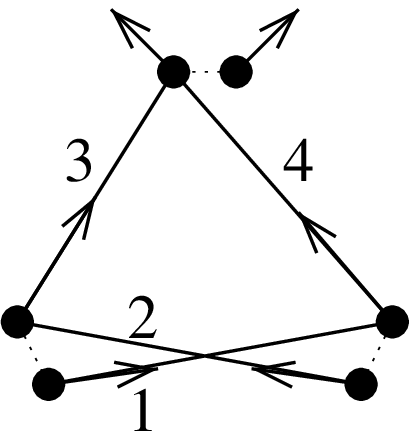} = \hat u_{r \tau}  \hat u_{r \sigma}
\Delta''(u_{r,\tau}-u_{r,\sigma}) \\ 
&& \times \int_{t_i>0,r_i}
R_{r_1 - r_2,t_1} R_{r_1 - r_2,t_2} R_{r - r_1,t_3} R_{r - r_2,t_4}
\nonumber \\ 
&& \times
\Delta'(u_{r_1,\tau-t_3}-u_{r_1,\tau-t_4-t_1})
\Delta'(u_{r_2,\tau-t_4}-u_{r_2,\tau-t_2-t_3})  \nonumber
\ .
\eea
Here four Wick contractions have been performed, as in any of the other
thirty 2-loop diagrams of the form A (studied in the next Section).
In an analytic theory performing the
local time expansion this would result
in a 2-loop correction to $\Delta(u)$ proportional to $\Delta''(u)$
but with a zero coefficient since the $\Delta'$ functions are evaluated
at zero argument. In the non-analytic theory, inserting the expansion
(\ref{series}) yields (upon some change of variables):
\begin{eqnarray}
 \mathrm{e}_1 &=& \Delta'(0^+)^2 \Delta''(u) \int_{t_i>0,r_i}
\!\!\!\!\!\!\!\!\!\!\!\!\!
R_{r_1,t_1} R_{r_1,t_2} R_{r_3-r_1,t_3} R_{r_3,t_4} F_{r_i,t_i} \nonumber \\
 F_{r_i,t_i}&=&\langle {\rm sgn}(X) {\rm sgn}(Y)\rangle \label{graphe} \\
X &=& u_{r_1,-t_3}-u_{r_1,-t_4-t_1} \nonumber \\
Y &=& u_{0,-t_4}-u_{0,-t_3-t_2} \nonumber
\end{eqnarray}
terms of higher order in (\ref{series}) do not contribute since we are
at $T=0$ and we have exhausted the number of $\hat u$ to contract (i.e
those terms would yield higher orders in $T$). The remaining average
in (\ref{graphe}) is evaluated with respect to a Gaussian measure, and
can thus be performed. It can be defined by using the $T >0 , v>0$
Gaussian measure ($u_{xt} \to v t + u_{xt}$) and taking the limit $T
\to 0, v \to 0$. The result is a continuous function of $v^2/T$ and
its value depends on how the limit is taken.

In the static theory one should take $T \to 0$ at $v=0$. This yields
\begin{eqnarray}
\langle {\rm sgn}(X) {\rm sgn}(Y)
\rangle&=&\frac{2}{\pi} {\rm asin}(\sigma) \label{asin} \\
\sigma&=&\frac{\langle XY \rangle}{\sqrt{\langle X^2\rangle}
\sqrt{\langle Y^2 \rangle}} \nonumber
\end{eqnarray}
i.e.\ the result for centered Gaussian variables. Expressing the
averages in (\ref{asin}) using correlation functions $C_{q,t}$ yields
a complicated $T=0$ expression for $e_{1}$. This expression will be
discussed in a companion paper on the statics
\cite{LeDoussalWieseChauve2002a}. A list of all anomalous diagrams is
presented in  Appendix \ref{appanomalous}.

The opposite limit $v \to 0$ at $T=0$ yields much simpler expressions:
\begin{eqnarray}
&&  \langle {\rm sgn}(X) {\rm sgn}(Y) \rangle  \to {\rm sgn}(t_4+t_1-t_3)
{\rm sgn}(t_3+t_2-t_4)  \nonumber
\ .
\end{eqnarray}
More generally this procedure corresponds to the substitution:
$\Delta^{(n)}(u_{r,t} - u_{r,t'}) \to \Delta^{(n)}(v(t - t'))$ in any
ambiguous vertex evaluated at $u=0$. That this is the correct
definition of the theory of the quasi-static depinning as the limit
$v=0^+$ is particularly clear here since it is well known (the no
passing property \cite{MiddletonFisher1991,MiddletonFisher1993}) that
the $u_{r,t}$ are increasing functions of time in the steady state. Of
course it remains to be shown that the procedure actually works and
does not produce singular terms such as $\delta(v t)$. It also remains
to be shown that it yields a renormalizable continuum theory where all
divergences can be removed by the appropriate counter-terms. This is
far from trivial and will be achieved below.

Let us comment again on the connections between dynamics and statics.
Consider a $T=0$ dynamical diagram with $p$ connected components
evaluated at zero external frequencies. All response functions can be
integrated over the times from the leaves towards the root on each
connected component.  Using the FDT relation this replaces response by
correlations and thus exactly reproduces a $p$ replica static diagram
except that it is differentiated once with respect to each replica
field (the sums over all possible positions of the response field
reproduces the derivation chain rule).  One simple way to establish
this rule is to consider the formal limit $\eta \to 0^+$ (equivalently
expansion of $R_{q,\omega}$ in powers of frequency), i.e.\ formally
replace $R_{q,t,t'} \to \delta_{tt'}/q^2$ (keeping track of
causality).  This reproduces exactly the zero frequencies dynamical
diagrams and treats ``replicas'' as ``times''.

Thus the $p$-th derivative of a $p$ replica static diagram gives a set
of dynamical diagrams with $p$ connected components. For $p=2$ this
ensures e.g.\ that the relation $\Delta(u) = - R''(u)$ remains
uncorrected to all orders. The flaw in this argument comes from the
anomalous diagrams (both in statics and dynamics).  In the analytic
theory the dynamical diagrams with response fields on a saturated
vertex vanish or cancel in pairs. This just expresses that taking a
derivative of a static saturated vertex gives zero and the rule still
works. But in the non-analytic theory the anomalous diagrams do not
vanish and contain an additional time dependence.  The above integration
of response functions from the leaves to the root cannot be performed
for these anomalous diagrams. As a result they can give non-trivial
contributions both in statics and dynamics which violate relations
such as $\Delta(u) = - R''(u)$, thus allowing to distinguish statics
from depinning.

To conclude this Section: The perturbative calculation of the
effective action and of the IVF vertices can also be performed in a
non-analytic theory. It can be expressed as sums of the same diagrams
one writes in the analytic theory, with the same graphical rules to
{\em draw} and generate the diagrams starting from the
statics. However the way to compute these diagrams and their {\em
values} is {\em different} from the analytic theory. The time ordering
of vertices comes in a non-trivial way and produces results which can
be different at depinning $f=f_c^+$ ($v=0^+$) and in the statics $f=0$,
as illustrated on the diagram $\mathrm{e}_1$ above. Thus we see the
principle  mechanism by which the statics and the depinning can
yield different field theories, which is a novel result. It remains to
perform the actual calculation of these non-analytic diagrams, which
is performed in the following sections.


\section{Renormalization program}
In this section we will compute the effective action to 2-loop order at $T=0$ for depinning.
From the above analysis we know that we  only need to compute the 1- and
2-loop corrections to $\Delta(u)$ and  $\eta$.

\subsection{Corrections to disorder}
We start by the corrections to the disorder, first at 1-loop and then
at 2-loop order.

\subsubsection{One loop}
\begin{figure}[h]
\centerline{\fig{4cm}{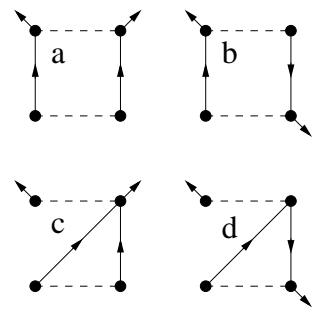}}
\caption{1-loop dynamical diagrams correcting $\Delta$}
\label{1loopDelta}
\end{figure}
At leading order, there are four diagrams, depicted on figure
\ref{1loopDelta}. Since
 diagram (d) is proportional to $\Delta'(u)
\Delta'(0)$, it is an odd function of $u$, and thus does not
contribute to the renormalization of $\Delta$. However its repeated
counter-term will appear at 2-loop order.
Diagram (a) is proportional to $-\Delta(u)\Delta''(u)$, diagram (b) to
$-\Delta'(u)^2$ and diagram (c) to $\Delta''(u)\Delta(0)$. All come
with a combinatorial factor of $1/2!$ from Taylor-expanding the
exponential function, $1/2$ from the action and $4$ from
combinatorics. Together, they add up to the 1-loop correction to disorder
\begin{eqnarray}
\delta^{1} \Delta(u) &=& \frac{4}{2!\, 2} \left[
- \Delta'(u)^2 - (\Delta(u) - \Delta(0)) \Delta''(u) \right] I_{1} \nn\\
 I_1& :=& \int_q \frac{1}{(q^2+m^2)^2} \label{I1}
\end{eqnarray}
with $I_1 = \int_q \rme^{-q^2} \Gamma ( 2 - \frac{d}{2}) = (4 \pi)^{-d/2} \Gamma \left( 2 - \frac{d}{2}\right)$.

\subsubsection{Two loops}
\begin{figure}[h]
\centerline{\fig{6cm}{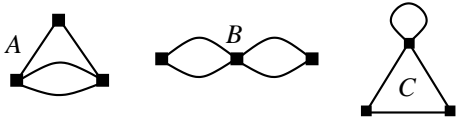}}
\caption{The 3 possible classes at second order correcting
disorder at $T=0$. Only classes A and B will contribute.}
\label{static-2-loop-DO-classes}
\end{figure}\begin{figure}[h] \centerline{\fig{6cm}{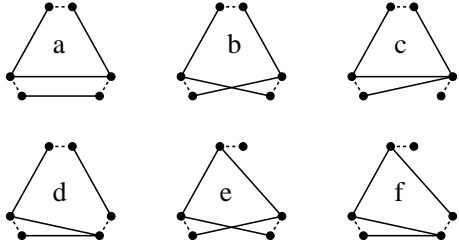}}
\caption{Static graphs at 2-loop order in the form of a hat (class A
in figure \ref{static-2-loop-DO-classes}) contributing to two replica
terms. Adding a response-field to each connected component leads to
the dynamic diagrams of figure \ref{dynamic-2-loop-DO}.}
\label{static-2-loop-DO}
\end{figure}
First, we have to find all diagrams correcting disorder at second
order. At $T=0$ they can be grouped in 3 classes A, B and C for the 3 possible
diagrams for unsplitted vertices. Class C does not contribute as is shown in
appendix \ref{app:DiagC}. We begin our analysis with class A.

\begin{figure}
\centerline{\fig{\columnwidth}{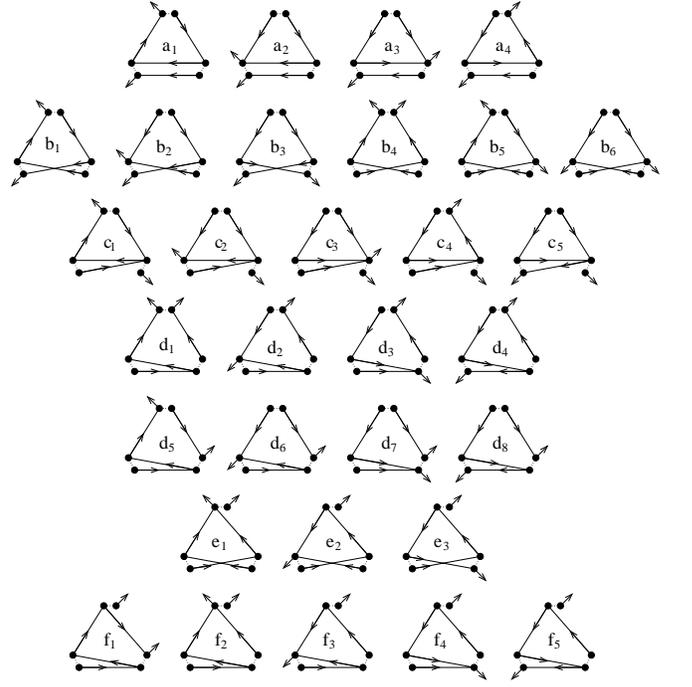}}
\caption{Dynamical diagrams at 2-loop order of type A with two
external response fields (two connected components) correcting the disorder; derived from the
two replica static diagrams of Fig. \protect\ref{static-2-loop-DO}.}
\label{dynamic-2-loop-DO}
\end{figure}

We now need to write all possible diagrams with splitted vertices of
type A. A systematic procedure is to start from all possible static
diagrams given in Fig.  \ref{static-2-loop-DO}. This relies on the
fact that dynamics and statics are related -- recall that in general a
dynamic formulation can be used to obtain the renormalization of the
statics. As mentionned in the previous Section, to go from the statics
to the dynamics, one attaches one response field to a root on each
connected component of the diagrams a to f in figure
\ref{static-2-loop-DO} and orient each component towards the root. The
result is presented on figure \ref{dynamic-2-loop-DO}.
The next step is to eliminate all diagrams which yield odd
functions of $u$ and thus do not contribute to the renormalized
disorder. The list is the following:
\begin{eqnarray}
\mathrm  a_1 &=&\mathrm  a_4 =\mathrm  c_3 = \mathrm d_1 =\mathrm  d_3 =\mathrm  d_5 =\mathrm  d_7 \nonumber \\
 &=&\mathrm  e_2 =\mathrm e_3 =\mathrm  f_1 =\mathrm  f_3 =\mathrm  f_4 =\mathrm  f_5
=0
\ .
\end{eqnarray}
Further simplifications come from diagrams, which mutually
cancel. Again this uses that $\Delta'(u)$ is an odd function. This gives:
\begin{equation}
\mathrm c_2 +\mathrm  c_5 = \mathrm d_2 + \mathrm d_4 =\mathrm  d_6 +\mathrm  d_8 = 0\ .
\end{equation}
In addition
\begin{eqnarray}
\mathrm c_4 = 0 \ ,
\end{eqnarray}
since $\int_{tt'} R_{xt} R_{xt'} \Delta'(t-t') =0$. This is explained
in more details in Appendix \ref{appanomalous} where the list of all anomalous (non-odd) graphs
is given together with their expressions in the non-analytic field theory.

Thus, the only non-zero graphs which we have to calculate are $\mathrm
a_2$,
$\mathrm a_3$, $\mathrm b_1,\dots,\mathrm  b_6$, $\mathrm c_1$,
$\mathrm e_1$ and $\mathrm f_2$. These calculations are
rather cumbersome, due to the appearance of theta-functions of sums or
differences of times as a result of the non-analyticity of the theory.
The correction to disorder is
\begin{eqnarray}
\delta^{2} \Delta(u) &=&  \frac{1}{3!} \frac{2}{2^3}
3 (2^3) \sum (\mathrm{a}_i + \mathrm{b}_i  + ..)  \nonumber  \\
&=& \sum (\mathrm a_i + \mathrm b_i + ..) \nonumber \ ,
\end{eqnarray}
where the combinatorial factors are: $1/3!$ from the Taylor-expansion
of the exponential function, $2/2^3$ from the explicit factors of
$1/2$ in the interaction, a factor of 3 to chose the vertex at the top
of the hat, and a factor of 2 for the possible two choices in each of
the vertices. Furthermore below some additional combinatorial factors
are given: a factor of $2$ for generic graphs
and $1$ if it has the mirror symmetry with respect to the vertical
axis: each diagram symbol ($a_i$..) denotes the diagram including the
symmetry factor.

We recall that we have defined {\em saturated} vertices as vertices
evaluated at $u=0$ while {\em unsaturated} vertices still contain $u$
explicitly. Diagrams with response-functions added to unsaturated vertices can
be obtained by deriving static diagrams:
\begin{eqnarray}
\mathrm  a_2 + \mathrm a_3 &=& \text{second derivative of the statics} \nonumber  \\
\mathrm b_1 + \mathrm  b_2 &+&\mathrm  b_3 +\mathrm  b_4 + \mathrm b_5 + \mathrm b_6\nonumber\\
  &=& \text{second derivative of the statics} \qquad
\ .
\end{eqnarray}

The graphs which contain external response-fields on {\em saturated}
vertices cannot be derivatives from static ones. For class A,  the
hat-diagrams,  the only non-zero such graph is $\mathrm c_1$.

Explicitly, this reads
\begin{equation}
\mathrm a_2 +\mathrm a_3 = - \partial_u^2 \left[ - R''(0) R'''(u)^2 \right] I_A
\ ,
\end{equation}
where (see (\ref{A.18}))
\begin{eqnarray}
I_A&:=&\int \frac{\rmd^d q_1}{(2\pi)^d} \frac{\rmd^d q_2}{(2\pi)^d}
\frac{1}{q_1^2+m^2}\frac{1}{q_2^2+m^2} \frac{1}{(
(q_1{+}q_2)^2+m^2)^2}\nn\\
&=& \left(\frac1{2\E^2} + \frac1{4\E} +O(\E^2) \right) (\E I_1)^2
\ .
\end{eqnarray}
Furthermore, we find
\begin{eqnarray}
\sum_{i=1}^6 \mathrm b_i = - \partial_u^2 \left[R''(u) R'''(u)^2 \right]I_A
\end{eqnarray}
and
\begin{eqnarray}
\mathrm c_1 = 2 \Delta'(0^+)^2 \Delta''(u) I_A
\ .
\end{eqnarray}
The diagram $\mathrm e_1$ is an explicit example for the appearance of
non-trivial sign-functions resulting from the monotonic increase of the
displacement. It was already discussed in the previous Section. In the
quasi-static depinning limit (\ref{graphe}) gives
(details are given in  appendix \ref{app:hat-diagrams}):
\begin{eqnarray}
 \mathrm e_1 &=&  \Delta'(0^+)^2 \Delta''(u)\nonumber \\
&& \times \int_{q_1,q_2}
\int_{t_1,t_2,t_3,t_4>0}\nn\\
&& \qquad \rme^{-[ (q_1^2+m^2)t_1 + (q_2^2+m^2)t_2+
((q_1+q_2)^2+m^2) (t_3 + t_4)}   \nonumber\\
&& \qquad \sgn(t_1 - t_3 + t_4) \, \sgn(t_2 - t_4 + t_3)\ . \qquad\qquad
\end{eqnarray}
The result of the explicit integration is:
\begin{eqnarray}
 \mathrm e_1 &=& \Delta'(0^+)^2 \Delta''(u) \left[ I_{l}
- I_A + \mbox{finite}\right]\\
I_{l} &:=& \int_{q_1,q_2}
\frac{1}{(q_1^2 + m^2)  (q_2^2 + m^2) (q_3^2 + m^2) ( q_1^2 + q_3^2 + 2 m^2)} \nn\\
&=&  \frac{\ln 2}{2 \epsilon}  ( \epsilon I_1)^2 + \text{finite}
\ .
\end{eqnarray}
The last diagram $f_2$ also involves a sign-function and reads:
\begin{eqnarray}
 \mathrm f_2 &=&  2 \Delta'(0^+)^2 \Delta''(u) \int_{q_1,q_2}
\int_{t_1,t_2,t_3,t_4>0}\sgn(t_4 {-} t_3 {-} t_2)\times\nonumber\\
&&\qquad\quad\  \times\rme^{- [(q_1+q_2)^2+m^2) (t_3 + t_4) + (q_1^2+m^2) t_1 + (q_2^2+m^2) t_2]} \nonumber \\
& =& -  \Delta'(0^+)^2 \Delta''(u)\, I_{l}
\ .
\end{eqnarray}
In appendix \ref{app:hat-diagrams} we show that (for any given elasticity)
the sum of $\mathrm e_{1}+\mathrm{f}_{2}$ only involves the integral
$I_{A}$, and that the combination takes the simpler form
\begin{equation}
\mathrm{e}_{1}+ \mathrm f_{2} =- \Delta' (0^{+})^{2} \Delta'' (u) I_{A}
\ .
\end{equation}

We now turn to graphs of type B (bubble-diagrams).
\begin{figure}[t]
\centerline{\Fig{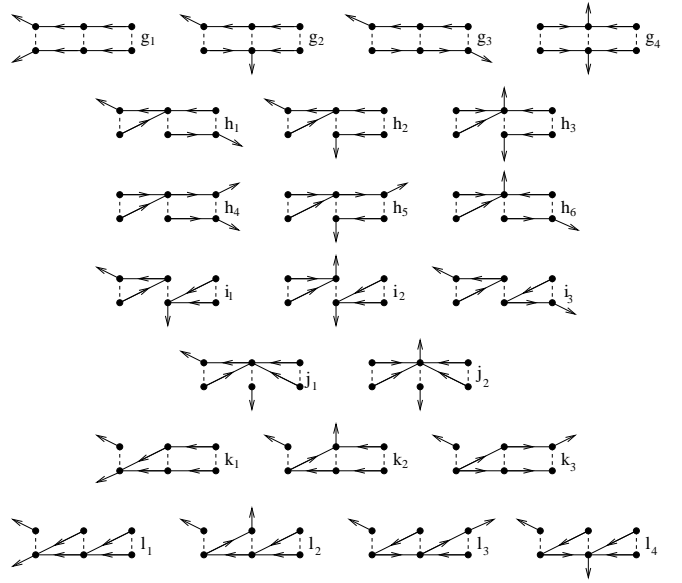}}
\caption{2-loop dynamical diagrams of type B (see figure \protect\ref{static-2-loop-DO-classes}).}
\end{figure}

Again diagrams, which are odd functions of $u$ vanish. These are:
\begin{equation}
\mathrm h_1=\mathrm h_2=\mathrm i_1=\mathrm j_1=\mathrm k_2=\mathrm k_3=\mathrm l_2=\mathrm l_3=\mathrm l_4=0
\ .
\end{equation}
Two other diagram mutually cancel:
\begin{equation}
\mathrm k_1 + \mathrm l_1 = 0
\ ,
\end{equation}
as discussed in Appendix \ref{appanomalous}.

The diagrams that are second derivative of the static have
all their response-fields  on their unsaturated vertices. These are:
\begin{eqnarray}
\mathrm g_1 + \mathrm g_2 + \mathrm g_3 + \mathrm g_4 &=& \partial^2_u \left[ \frac{1}{2} \Delta(u)^2
\Delta''(u)\right] I_1^2 \nonumber \\
\mathrm h_3 +\mathrm  h_4 + \mathrm h_5 + \mathrm h_6 &=& \partial^2_u \left[ - \Delta(0) \Delta(u) \Delta''(u)\rule{0mm}{2.2ex} \right]I_1^2 \nonumber  \\
\mathrm i_2 = \mathrm j_2 &=& \partial^2_u \left[\frac{1}{4} \Delta(0)^2  \Delta''(u)
\rule{0mm}{2.2ex} \right]I_1^2 \nonumber
\ .
\end{eqnarray}

The surprise is that $\mathrm i_3$, which is not the second derivative of a
static diagram, since it has both response fields on saturated
vertices, is non-trivial:
\begin{equation}
\mathrm i_3 = - \Delta'(0^+)^2 \Delta''(u) I_1^2
\ .
\end{equation}

To summarize, for the driven problem at $T=0$ in
perturbation  of $\Delta \equiv \Delta(u)$, the contributions to the
disorder to one and two loops, i.e.\ the corresponding terms in the
effective action $\Gamma[u,\hat u]$ are:
\begin{eqnarray}
 \delta^{1} \Delta (u) &=& - \left[{\Delta' (u)}^2 + (\Delta (u) - \Delta(0)) \Delta'' \right] I_1\qquad  \quad \label{d21} \\
\delta^{2} \Delta (u) &=& \left[(\Delta (u) - \Delta(0)) {\Delta' (u)}^2 \right]'' I_A  \nn  \\
&& + \frac{1}{2} \left[ (\Delta (u) - \Delta(0))^2 \Delta'' (u) \right]'' I_1^2\nn \\
&& + \Delta'(0^+)^2 \Delta'' (u) (I_A - I_1^2)   \label{d22}
\ .
\end{eqnarray}
Curiously, even though two diagrams contain contributions
proportional to $I_{l}\sim \ln 2$, these contributions cancel in the final
result for the corrections to the disorder.

\subsection{Corrections to the friction $\eta$}
\label{Main-text-calcul-eta}
\begin{figure}[h]
\centerline{\fig{1cm}{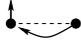}}
\caption{1-loop dynamical diagram correcting the friction.}
\label{fig:1loopabcd}
\end{figure}
We now calculate the divergent corrections to $\eta$, which will require a
counter-term proportional to $i \hat u \dot u$. Let us illustrate
their calculation at leading order. We start from the first order
expansion of the interaction $\rme^{- S_{\ind{int}}}$, which can be
written as
\be
\int_{t>t',x} i\hat u_{xt}\, \Delta(u_{xt}-u_{xt'})\, i \hat u_{xt'}
\ .
\ee
Contracting one $i \hat u_{xt'}$ leads to:
\be
\int_{t>t',x} i\hat u_{xt}\, \Delta'(u_{xt}-u_{xt'}) R_{r=0,t-t'} \ .
\ee
The response function contains a short-time divergence, which we deal
with in an operator product expansion.
Expanding $\Delta'(u_{xt}-u_{xt'})$ to the necessary order yields
\be
\int_{t>t',x} i\hat u_{xt}\, \left[ \Delta'(0^+)+
(u_{xt}{-}u_{xt'})\Delta''(0^+) + \dots \right] R_{r=0,t-t'} \ .
\label{III.20}
\ee
The first term of this expansion, proportional to $\Delta'(0^+)$, is
strongly UV-divergent and non-universal and gives the critical force
to lowest order in disorder. Since we tune $f$ to be exactly at the
depinning threshold we do not need to consider it. The second
contribution, proportional to
$\Delta''(0^+)$, corrects the friction: due to the short-range
singularity in the response-function, we can expand
$(u_{xt}{-}u_{xt'})$ in a Taylor-series, of which only the first
term contributes. (\ref{III.20}) becomes:
\be
\int_{t>t',x} i\hat u_{xt} \left[ (t-t') \dot u_{xt} + O(t-t')^2
\right] \Delta''(0^+)    R_{r=0,t-t'} \ .
\ee
The correction to friction at leading order thus is
\begin{eqnarray}
\delta \eta = -  \Delta''(0^{+}) \int_t t R_{r=0,t}
\ .
\end{eqnarray}
Here, the response-function is taken at spatial argument 0. In
momentum representation, the same expression reads
\begin{eqnarray}
\delta \eta &=& -  \Delta''(0^{+}) \int_t \int_q t R_{q,t} \nonumber\\
&=& -  \Delta''(0^{+}) \int_q t\, \rme^{-t (q^2 +m^2)} \nonumber\\
&=& - \Delta''(0^{+}) \int_q  \frac{1}{ (q^2 +m^2)^2} \nonumber\\
&=& - \Delta''(0^{+})\, I_1
\end{eqnarray}
with the already known integral $I_1$, equation (\ref{I1}).
\begin{figure}[b]
\centerline{\Fig{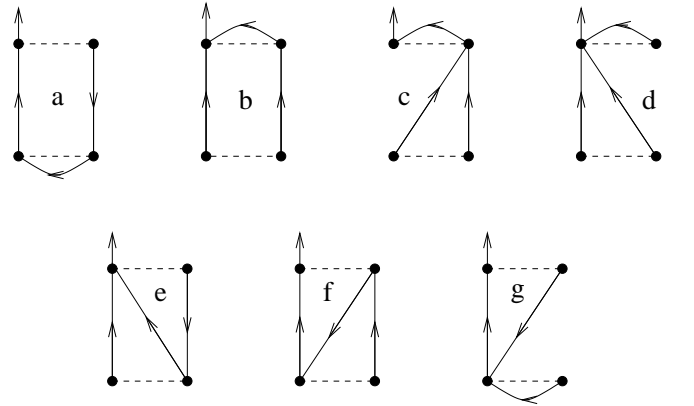}}
\caption{2-loop dynamical diagrams correcting the friction. They all
have multiplicity $8$ except (c) and (d) which have multiplicity $4$.}
\label{eta-2-loop}
\end{figure}%

We now turn to the 2-loop corrections. There are seven contributions,
drawn on figure \ref{eta-2-loop}. Their contribution to $\eta$ is
\begin{equation}
\delta \eta = - \frac{1}{8}  \times 4 \times 2 \left[\mathrm a +\mathrm  b + \mathrm  c+ \mathrm d + \mathrm e + \mathrm f + \mathrm g\right] \ .
\end{equation}
The combinatorial factor is $1/8$ from the interaction, 4 from the
time-ordering of the vertices, and an additional factor of 2 for the
symmetry of diagrams a, b, e, f and g. Details of the calculation of
diagrams a to g are given in appendix
\ref{app:eta}. Grouping diagrams, which partially cancel,
we find:
\begin{eqnarray}
\mathrm  a +\mathrm  g &=& - \Delta''(0^+)^2  I_1^2 \\
\mathrm  b +\mathrm  c + \mathrm d &=&  -\frac{1}{2} \Delta'''(0^+)
\Delta'(0^+) I_{1}^{2}\\
\mathrm{e} &=& -\Delta''' (0^{+}) \Delta' (0^{+}) I_{\eta }\\
\mathrm{f} &=& - 2 \Delta'''(0^+) \Delta'(0^+) I_{A} - 2
\Delta''(0^+)^2 I_A\qquad \quad
\ .
\end{eqnarray}
This involves the non-trivial diagram $I_{\eta }$
\begin{eqnarray}
I_{\eta }&:=&  \int\limits_{q_1,q_2}
\frac{1}{( q_1^2+m^{2}) ( q_2^2+m^{2})^{2} (q_2^2 + q_3^2+2m^{2})}\nn \\
&=& \left(\frac{1}{2\E^2}+\frac{1-2\ln 2}{4\E} \right) (\E I_1)^{2}
+\mbox{finite}
\end{eqnarray}
 calculated in appendix \ref{Ieta}.

\subsection{Renormalization program to two loops and calculation of
counter-terms}
\subsubsection{Renormalization of disorder}
\label{rgdis}
 Let us now discuss the strategy to renormalize the
present theory where the interaction  is not  a single
coupling-constant, but a whole function, the disorder-correlator
$\Delta(u)$. We denote by $\Delta_0$ the bare disorder --~this is
the object in which perturbation theory is carried out~-- i.e.\ one
consider the bare action (\ref{msr}) with $\Delta \to \Delta_0$.
We denote here by $\Delta$ the renormalized dimensionless disorder
i.e.\ the corresponding term in the effective action
$\Gamma[u,\hat u]$ is $m^{\epsilon} \Delta$.

We define the dimensionless bilinear 1-loop and trilinear 2-loop
symmetric functions  (see (\ref{d21}) and (\ref{d22})) such that:
\begin{eqnarray}
 \delta ^{(1)}(\Delta, \Delta) &=& m^{\epsilon} \delta^1 \Delta \\
 \delta ^{(2)}(\Delta, \Delta, \Delta) &=& m^{\epsilon} \delta^2 \Delta
\end{eqnarray}
thus extended to non-equal argument using $f(x,y) :=\frac12 \left[
f(x+y,x+y)-f(x,x)-f(y,y)\right]$ and a similar expression for the
trilinear function. Whenever possible we will use the shorthand notation
$\delta ^{(1)}(\Delta)=\delta ^{(1)}(\Delta, \Delta)$
and $\delta ^{(2)}(\Delta)=\delta ^{(2)}(\Delta, \Delta,\Delta)$.
The expression of $\Delta$ obtained perturbatively in powers of
$\Delta_0$  at 2-loop order reads:
\begin{equation}
\Delta = m^{-\epsilon} \Delta_0 + \delta ^{(1)}(m^{-\epsilon} \Delta_0)
+ \delta ^{(2)}(m^{-\epsilon} \Delta_0) +
O(\Delta_0^4)\ .
\end{equation}
It contains terms of order $1/\epsilon$ and
$1/\epsilon^2$. This is sufficient to calculate the RG-functions at
this order (In principle, one has to keep the finite part of the 1-loop
terms, but we will work in a scheme, where these terms are exactly 0,
by normalizing all diagrams by the 1-loop diagram).
Inverting this formula yields:
\begin{equation}
\Delta_0 = m^{\epsilon} \left[ \Delta - \delta ^{(1)}(\Delta) - \delta
^{(2)}(\Delta) + \delta^{(1,1)}(\Delta)+ \ldots \right]
\label{do}
\ ,
\end{equation}
where $\delta^{(1,1)} (\Delta)$ is the 1-loop repeated counter-term:
\begin{eqnarray}
\delta^{(1,1)}(\Delta) = 2 \delta ^{(1)}(\Delta, \delta ^{(1)}(\Delta, \Delta))
\ .
\end{eqnarray}
The $\beta$-function is by definition the derivative of $\Delta$ at
fixed $\Delta_0$. It reads:
\begin{eqnarray}
-m \partial_{m}\Delta\ts_{\Delta_0} &=& \epsilon  \Big[ m^{-\epsilon}\Delta_0
+ 2\delta ^{(1)} (m^{-\epsilon}\Delta_0 )\nn
\\
&& \ \ \ \ + 3 \delta ^{(2)} (m^{-\epsilon}\Delta_0) + \dots \Big]\qquad
 \label{betar0}
\ .
\end{eqnarray}
Using the inversion formula (\ref{do}), the  $\beta$-function can be written
in terms of the renormalized disorder $\Delta$:
\begin{eqnarray}
-m\partial_{m}\Delta\ts_{\Delta_{0}} &=&
\epsilon \Big[ \Delta+\delta ^{(1)} (\Delta)
\nonumber \\
&&\quad+2\delta ^{(2)} (\Delta)-\delta ^{(1,1)} (\Delta) + \dots \Big]
\qquad\qquad
\label{beta-renor-DO}
\ .
\end{eqnarray}
In order to proceed, let us calculate the repeated 1-loop counter-term
$ \delta^{1,1} (\Delta)$. We start from the 1-loop counter-term
(\ref{d21}), which has   the bilinear form
\bea
\delta^{(1)} (f,g) = - \frac12 \Big[ 2 f'(u) g'(u) &+& (f(u)-f(0))
g''(u) \nn\\ &+& (g(u)-g(0))f''(u) \Big] \tilde I_1  \nn\\
\eea
with the dimensionless integral $\tilde I_{1}:=I_{1}\ts _{m=1}$; we
will use the same convention for $\tilde I_{A}:=I_A\ts_{m=1}$.
Thus $ \delta^{1,1} (\Delta)$ reads
\begin{eqnarray}
 \delta^{(1,1)}( \Delta(u) )&=& 2
\delta^{(1)}\left(\Delta,\delta^{(1)}(\Delta)\right)\nn\\
&=& \Big[ (\Delta(u) - \Delta(0))^2 \Delta''(u)\nn\\
&&\  + (\Delta'(u)^2-\Delta'(0)^2) (\Delta(u) - \Delta(0))\Big]''
\tilde I_1^2 \nn \\
\label{1-loop-rep-CC}
\end{eqnarray} Note that this counter-term is non-ambiguos for $u\to
0$. 
Finally, as discussed at the end of the previous section
at any point we can rescale the fields $u$ by $m^{\zeta}$.
This amounts to write the $\beta $-function for the function
$\tilde{\Delta}(u) = m^{- 2 \zeta} \Delta(u m^{\zeta})$
which will be implicit in the following (in addition we will
drop the tilde superscript).

The 2-loop $\beta$-function (\ref{beta-renor-DO}) then becomes with the
help of (\ref{1-loop-rep-CC}) \begin{eqnarray}
-m \partial_m {\Delta}(u) &=&  (\epsilon - 2 \zeta) \Delta(u)
+ \zeta u \Delta'(u) \nn\\
&& - \frac{1}{2} \left[(\Delta(u) - \Delta(0))^2\right]'' (\epsilon
\tilde I_{1})  \nonumber \\&&
+  \left[ (\Delta(u) - \Delta(0)) \Delta'(u)^2 \right]'' \,\epsilon\!
\left( 2\tilde I_{A}-\tilde I_{1}^{2} \right)\nn\\
&& + \Delta'(0^+)^2 \Delta''(u) \, \epsilon\!
\left( 2\tilde I_{A}-\tilde I_{1}^{2} \right)
\label{rgdisorderunrescaled}\ .
\end{eqnarray}
One of our main results is now apparent: the $1/\E^2$-terms cancel in
the corrections to disorder. If it had not been the case it would lead
to a term of order $1/\E$ in the $\beta$-function and thus to
non-renormalizability.  Thus the $\beta$-function is finite to two
loops a hallmark of a renormalizable theory.  Note that this happened
in a rather non-trivial way since it required a consistent evaluation
of all anomalous non-analytic diagrams. Furthermore the precise type
of cancellation is unusual: usually the 2-loop bubble diagrams of type
B are simply the square of the 1-loop ones.  Here the easily missed
and non-trivial bubble diagram $i_3$ was crucial in achieving the
above cancellation. 

In order to simplify notations and further calculations, we absorb a
factor of $\E \tilde I_{1}$ in the definition of the renormalized
disorder (or equivalently in the normalization of momentum or space
integrals). With this, the $\beta $-function takes the simple form:
\begin{eqnarray}
-m \partial_m {\Delta}(u) &=&  (\epsilon - 2 \zeta) \Delta(u)
+ \zeta u \Delta'(u) \nn\\
&& - \frac{1}{2} \left[(\Delta(u) - \Delta(0))^2\right]'' \nonumber \\&&
+ \frac{1}{2} \left[ (\Delta(u) - \Delta(0)) \Delta'(u)^2 \right]'' \nn\\
&& + \frac{1}{2} \Delta'(0^+)^2 \Delta''(u)
\label{rgdisorder}\ .
\end{eqnarray}

Note several interesting features of this 2-loop
$\beta$-function. First it contains a non-trivial so called
``anomalous term'' (the last one) which is absent in an analytic
theory. Second, it can be shown to exhibit irreversibility, precisely
due to this term. Although, surprisingly, it can be formally be
integrated twice over $u$ the resulting flow equation for the double
primitive of $\Delta(u)$ does not, however, have the required property
for the flow of a potential function, i.e.\ a second cumulant of the
random potential in the static. This will be shown in details in the
next Section \ref{sec:fixedpoints} where we find that the fixed points
of the above equation are manifestly non-potential.  In
Ref. \cite{ChauveLeDoussalWiese2000a} we have obtained the corresponding beta function for
$R(u)$ in the statics.  The corresponding force force correlator
$\Delta_{\ind{stat}}(u)=-R''(u)$ obeys the same equation than
(\ref{rgdisorder}) but with the opposite sign for the anomalous term!
This shows that statics and depinning are indeed two different
theories at two loops.

\subsubsection{Renormalization of friction and dynamical exponent $z$}
In section \ref{Main-text-calcul-eta}, we have calculated the
effective (renormalized) friction coefficient $\eta_R$ as a function
of the bare
one $\eta_0$ and the bare disorder $\Delta_0$:
\begin{eqnarray}
\eta_R = \eta_0 Z[m^{-\epsilon} \Delta_0]^{-1}
\ .
\end{eqnarray}
This identifies the renormalization group $Z$-factor as
\begin{eqnarray}
 Z^{-1}[m^{- \epsilon}\Delta_0]  = 1& -&
 \Delta_0''(0^+)I_{1} \nn\\
&+&   \left[\Delta_0''(0^+)\right]^2 \left[
I_{1}^{2}+ 2   I_{A}\right]
\nn\\
&+&\Delta_0'''(0^+) \Delta_0'(0^+) \left[
\half I_1^{2}+2I_{A}+I_{\eta } \right]\ .\nn\\
\label{Z-factor}
\end{eqnarray}
The dynamical exponent $z$ is then given by
\be
z = 2 + m \frac{\rmd}{\rmd m} \ln Z(m^{-\E}\Delta_0)\ .
\ee
Equation (\ref{Z-factor}) yields
\begin{eqnarray}
\ln Z^{-1}  =  &-&  \Delta_0''(0^+) I_{1}\nn\\
&+&
\Delta_0''(0^+)^2 \left[\half I_1^2 +2 I_{A}\right]\nn \\
&+&  \Delta_0'''(0^+) \Delta_0'(0^+) \left[
\half I_1^{2}+2I_{A}+I_{\eta } \right]\qquad \quad
\end{eqnarray}
and thus (remind that $I_{1}\sim m^{-\E}$ and $I_{A}\sim I_{\eta }\sim
m^{{-2\E}}$)
\begin{eqnarray}
m \frac{\rmd }{\rmd m} \ln Z^{-1} &=& \Delta_0''(0^+) (\epsilon I_{1})
-
\Delta_0''(0^+)^2\, \E\! \left ( I_1^{2}+4 I_{A}\right) \nn \\
&& +\,
\Delta_0'''(0^+) \Delta_0'(0^+)\, \epsilon\! \left (
I_1^{2}+4I_{A}+2I_{\eta } \right)\nn\\
\end{eqnarray}
We now have to express $\Delta_0$ in terms of the renormalized
disorder $\Delta$ using (\ref{do}). For the second-order terms, this
relation is simply $\Delta_0=m^\E\Delta$. The non-trivial term is
$\Delta''(0^+)$.  Using (\ref{d21}), derived twice at $0^+$, we get
(with the factor of $(\E I_1)$ absorbed into the renormalized disorder
\bea
\lefteqn{ \Delta_0''(0^+) = (\E I_1)^{{-1}}\times}\nn\\
&&\qquad\times
 \left[\left( \Delta''(0^+)
+ \tilde I_{1} (4 \Delta'''(0^+) \Delta'(0^+) + 3
\Delta''(0^+)^2\right) \right]\nn\\
\eea
Putting everything together, the  result is
\begin{eqnarray} \label{III.49}
 m \frac{\rmd}{\rmd m} \ln Z^{-1} &=& \Delta''(0^+)
+\E\left(\frac{2}{\E^2}-\frac{4 I_{A}}{(\E I_{1})^{2}} \right)
\Delta''(0^+)^2\nn \\
&& +\epsilon
\left(\frac{3}{\E^2}-\frac{4I_{A}}{(\E I_1^2)}-\frac{2I_{\eta
}}{(\E I_1^2)} \right) \Delta'''(0^+) \Delta'(0^+)
 \nn\\
\end{eqnarray}
Again there is a non-trivial cancellation of the $1/\epsilon$ terms,
another manifestation of the renormalizability of the theory.
Inserting the values of the integrals $I_{A}$ and $I_{\eta }$,  the
dynamical exponent $z$ becomes:
\begin{equation} \label{III.50}
z = 2 - \Delta''(0^+) + \Delta''(0^+)^2 + \Delta'''(0^+) \Delta'(0^+)
\left[\frac{3}{2} - \ln 2\right]
\ .
\end{equation}

\subsection{Finiteness and scaling form of correlations and response
functions} \label{finite}

To complete the 2-loop renormalizability program one must 
check that all correlation and response functions are rendered
finite by the above counter-terms. In a more conventional theory
that would be more or less automatic. Here however there are
additional subtleties. The disorder counter-term is a full
function and is purely static.  This counter-term, and its
associated FRG equation (\ref{rgdisorder}) cannot be read at $u=0$
because of the non-analytic action (this point is further
explained in Appendix \ref{appanomalous}). Indeed, this equation
and the cancellation of divergent parts was established only for
$u \neq 0$. It remains to be checked that irreducible vertex
functions which are $u=0$ quantities are also rendered finite by
the above static $u \neq 0$ counter-terms.

We first examine the two point correlation function. We will first
show that it is {\em purely static}. Then, in appendix
\ref{appanomalous} we show that it is finite and perform its
calculation in the renormalized theory. One has
\begin{equation}
\langle u_{q \omega } u_{-q -\omega } \rangle = {\cal
R}_{q\omega } {\cal R}_{-q,-\omega } \Gamma _{i\hat u i\hat u}
(q\omega ) \ ,
\end{equation}
where ${\cal R}_{q\omega }$ is the (exact) response function. We will
thus only compute $\Gamma _{i\hat u i\hat u} (q t)$ (in time
variable). The 1-loop counter-term for $\eta $ is absent in this ${\cal
O} (\Delta ^{2})$ calculation of the proper vertex but it enters the
calculation of $\langle u _{q\omega }u _{-q-\omega } \rangle$ (it dresses the
external legs $R_{q\omega }$ into ${\cal R}_{q\omega }$). In fact
since we find that $\Gamma _{i\hat u i\hat u} (q t)$ is static
(independent of $t$) we will need only the exact response at zero
frequency, which is the bare one because of STS. 

To one loop, the proper vertex $\Gamma _{i\hat u i\hat u} (q\omega )$
is the sum of the graphs $\mathrm{a}$, $\mathrm{b}$, $\mathrm{c}$ and
$\mathrm{d}$ of Fig.~\ref{fig:1loopabcd} evaluated at finite frequency
and momentum, so we write $\Gamma _{i\hat u i\hat u} (q\omega )
=\mathrm{a}+\mathrm{b}+\mathrm{c}+\mathrm{d}$. The sum $\mathrm{a} +
\mathrm{b}$ yields after two Wick contractions and short distance
expansion a term proportional to
\begin{eqnarray}
&&\!\!\! \int_{k,t_i} i \hat u_t i \hat u_{t'} \Delta''(u_{t} - u_{t'}) \Delta(u_{t_1} - u_{t_2}) \nonumber \\
&&\qquad \times (R_{k,t'-t_2} - R_{k,t-t_2}) (R_{k,t-t_1} -
R_{k,t'-t_1}) \ ,\qquad \ \ \ 
\end{eqnarray}
where we have kept all times explicitly to resolve any
ambiguity. Expressing $\Delta$ in a series as in (\ref{series}), the
lowest order term is purely static (since one can integrate freely
over $t_1,t_2$), and proportional to $\Delta '' (0+)\Delta (0)
\int_{k} k^{-4}$, but vanishes from the cancellation between graphs
$\mathrm{a}$ and $\mathrm{b}$. As explained in detail in Appendix
\ref{appanomalous} there can be a priori another contribution coming
from $2 \Delta'(0^+)^2 \delta(u) u$ in the expansion of $\Delta''
\Delta$. It produces a term $\delta(v(t-t')) v |t_1 - t_2|$ which
vanishes when multiplied with the above response function combination
(since it vanishes at $t=t'$).

Thus the only contribution comes from $\mathrm{c}+\mathrm{d}$. There
the $\Delta'$ yields sign functions 
and there are no ambiguities. One finds: 
\begin{eqnarray}
\mathrm{d} &=&-2 \Delta ' (0+)^{2} \int_{\tau_{1},\tau_{2}>0}[\mbox{sgn}
(t -\tau _{2}) + 
\mbox{sgn} (-t -\tau _{2})] \nn \\
&&\qquad \times 
\int_{k}\rme^{-k^{2}\tau _{2}}\rme^{-(k+q)^{2}\tau _{1}} \nonumber \\
&=&-2 \Delta ' (0+)^{2}\int_{k}\frac{1}{k^{2} (k+q)^{2}}\rme^{-k^{2}|t|/\eta
} \nonumber\\
\mathrm{b} &=&\Delta ' (0+)^{2} \int_{\tau_{1},\tau_{2}>0}\mbox{sgn}
(\tau_{1}+t/\eta )\, \mbox{sgn} 
(\tau_{2}-t/\eta )\nn \\
&&\qquad \times \int_{k}\rme^{-k^{2}\tau _{2}}\rme^{- (k+q)^{2}\tau
_{1}}\nonumber \\ 
&=&\Delta ' (0+)^{2}\int_{k} \frac{1}{k^{2} (k+q)^{2}}\left(2
\rme^{-k^{2}|t|/\eta 
}-1 \right) \nonumber
\ , 
\end{eqnarray}
where we have accounted for the extra combinatoric factor of $2$ for graph $d$ and used
\begin{equation}
\int_{\tau >0}\rme^{-q^{2}\tau }\mbox{sgn} (\tau
-t)=\frac{1}{q^{2}}\left(\theta (t) (2\rme^{-q^{2}t}-1)+\theta (-t) \right)\ .
\end{equation}

We thus find that although each graph is time dependent, this
time-dependence  cancels in the sum. Thus we
find a static result:
\begin{equation}\label{tralala}
\Gamma _{i\hat u i\hat u} (q \omega) = \delta(\omega)  \left[ \Delta (0) -
\Delta ' (0+)^{2}\int_{k} \frac{1}{k^{2} (k+q)^{2}}  \right] \nonumber
\ .
\end{equation}
The static 1-loop counter-term should thus be sufficient to cancel the
divergence of (\ref{tralala}). This is further analyzed in Appendix
\ref{appcorrel} where the full correlation-function is  computed.

We have thus found the  commutation $\Gamma _{i\hat u i\hat
u}(u=0,q) = \Gamma _{i\hat u i\hat u}(u=0^+,q)$. Note that if all
correlation functions are purely static, i.e.\ strictly time-%
independent, it implies the commutation of the limits. Then it
also implies the finiteness since these static divergences have been
removed. We have not pushed  the analysis further but we found a simple
argument which indicates that all correlations are indeed static.  We
found that the time dependence in diagrams cancels by subsets, noting
\cite{ChauveLeDoussal2000} that graphs can be grouped in subsets
(e.g.\ pairs $\mathrm{a c}$, $\mathrm{b d}$, $\mathrm{e f}$ in Fig.\
\ref{static-2-loop-DO}) 
which vanish by shifting the endpoint of an internal line within a
splitted vertex.

Finally, let us note that our result that correlations at the quasi
static depinning are purely {\em static} for $v=0^+$ is at variance
with previous works
\cite{NattermannStepanowTangLeschhorn1992,LeschhornNattermannStepanowTang1997}.
Thus the only functions where the dynamical exponent comes in are
response function.

\subsection{Long range elasticity}

As was discussed in the Introduction there are physical systems where
the elastic energy does not scale with the square of the wave-vector
$q$ as $E_{\mathrm{elastic}}\sim q^{2}$ but as
$E_{\mathrm{elastic}}\sim q^{\alpha }$.  In this situation, the upper
critical dimension is $d_{c}=2 \alpha$ and we define:
\begin{equation}
\epsilon:= 2 \alpha - d \ .
\end{equation}
The most interesting case, a priori relevant to model wetting or crack-front
propagation is $\alpha =1$, thus $d_{c}=2$.

In order to proceed, we have again to specify a cut-off procedure. For
calculational convenience, we choose the elastic energy to be
\begin{equation}
E_{\mathrm{elastic}}\sim ( q^{2}+m^{2})^{\frac \alpha 2 }\ .
\end{equation}
This changes the response-function to
\begin{eqnarray}
R_{q,t}= \Theta(t)\, \rme^{- (q^2 + m^2)^{\frac{\alpha}{2}} t} \ .
\end{eqnarray}
Since contributions proportional to $I_{l}$, see (\ref{A.26}), cancel,
the only integrals which appear in the $\beta $-function are:
\begin{eqnarray}
 I_{1}^{(\alpha)} &=&
\int_{q} \frac{1}{(q^2 + m^2)^{\alpha}} = m^{-\epsilon}
\frac{\Gamma(\epsilon/2)}{\Gamma(\alpha)}  \int_q \rme^{- q^2}\\
I_A^{(\alpha) } &=&
\int_{q_1,q_2}
\frac{1}{(q_1^2 + m^2)^{\frac\alpha2} (q_2^2 + m^2)^{\alpha} ((q_1 +
q_2)^2 + m^2)^{\frac\alpha2}} \nonumber\\
\end{eqnarray}
The important combination is again $2I_{A}^{(\alpha) }- (I_{1}^{(\alpha
)})^{2}$. One finds (see appendix \ref{app:LongRange})
\begin{eqnarray}
 {X^{(\alpha) }} &:=& \frac{2\, \epsilon ( 2 I_A^{( \alpha) } - (
I_1^{( \alpha)})^2) }{ (\epsilon I_1^{(\alpha )})^2 }\nn \\
& =&
\int_{0}^{1}\frac{\rmd t}t\, \frac{1+t^{\frac{\alpha }{2}} -
(1+t)^{\frac{\alpha }{2}}}{(1+t)^{\frac{\alpha }{2}}}+
\frac{\Gamma ' (\alpha )}{\Gamma (\alpha )} -
\frac{\Gamma ' (\frac{a}{2})}{\Gamma (\frac {a}{2})}\nn \\
&& +\, O(\epsilon)\ .
\end{eqnarray}
Since this term is finite, the $\beta $-function is finite; this is of
course necessary for the theory to be renormalizable. For the
cases of interest $\alpha =1$ and $\alpha =2$, we find
\begin{eqnarray}
 X^{(2)} &=& 1  \\
 X^{(1)} &=& 4 \ln 2
\ .
\end{eqnarray}
Since there is only one non-trivial diagram at second order, all
2-loop terms in the $\beta $-function get multiplied by $X^{(\alpha )}$:
\begin{eqnarray}
 - m \partial_m \Delta (u) &=& (\epsilon - 2 \zeta)  \Delta (u) +
\zeta u \Delta' (u)  \nn \\
&&- \frac{1}{2} \left[(\Delta (u) - \Delta(0))^2 \right]'' \nonumber \\
&& + \frac{X^{(\alpha )}}{2}\left[
(\Delta (u) - \Delta(0)) \Delta' (u)^2\right]''\nonumber\\
&& +\frac{X^{(\alpha) }}2 \Delta'(0^+)^2 \Delta ''(u)  \ .
\label{genbetafunc}
\end{eqnarray}
The diagrams involved in the dynamics also change. Besides $I_{1}^{(1)}$ and
$I_{A} ^{(1)}$ given above we need
\begin{eqnarray}
I_{\eta }^{(1)} &:=& \int\limits_{q_1,q_2}
\frac{1}{(q_1^2 {+} m^2)^{\frac{1}{2}}\, (q_2^2 {+} m^2)\, [(q_2^2 {+}
m^2)^{\frac{1}{2}} +
(q_3^2 {+} m^2)^{\frac{1}{2}}]}  \nn \\
&=& \left(\frac{1}{2\E^2}+ \frac{\ln 2 -\frac{\pi }{4}}{\E}
\right) \big(\E I_1^{(1)}\big)^{2} + \mbox{finite}
\end{eqnarray}
calculated in appendix \ref{app:Ieta}.

Starting from (\ref{III.49}), the dynamical exponent $z$ is then in
straightforward generalization of (\ref{III.50}) given by
\begin{equation} \label{III.65}
z = \alpha  - \Delta''(0^+) + X^{(\alpha )}\Delta''(0^+)^2 +Y^{(\alpha )}
\Delta'''(0^+) \Delta'(0^+)
\end{equation}
with $X^{(\alpha )}$ given above and
\begin{eqnarray}
Y^{(\alpha )} &=& X^{(\alpha )} + \frac{2 I_{\eta }^{(\alpha )}- \left(
I^{(\alpha)}_{1}\right)^{2} }{\E \left( I_1^{(\alpha )}\right)^{2}} \\
Y^{(1)} &=& 6 \ln 2 -\frac{\pi }{2}\\
Y^{(2)} &=& \frac{3}{2} -\ln 2\ .
\end{eqnarray}
The case $\alpha =2$ reproduces  (\ref{III.50}).
Since both $X^{(1)}$ and $Y^{(1)}$ are finite, we have checked that
also in the case of long-range elasticity
the theory is renormalizable at second order.
\section{Analysis of fixed points and physical results}
\label{sec:fixedpoints}

The FRG-equation derived above describes several different
physical situations: periodic systems (such as charge density waves)
where the disorder correlator is
periodic and non-periodic systems (such as a
a domain-wall in a magnet). Within the latter, SR (random bond) and LR (random field)
disorder must
a priori be distinguished. In our analysis of the FRG-equations,
we have to study these situations separately.

Before we do so, let us mention an important property, valid under all
conditions: If $\Delta (u)$ is solution of (\ref{genbetafunc}), then
\begin{equation}\label{DeltaRescale}
\tilde{\Delta} (u):= \kappa ^{2} \Delta (u/\kappa )
\end{equation}
is also a solution.
We can use this property to fix $\Delta(0)$ in the case of non-periodic disorder.
(For periodic disorder the solution is unique, since the period is fixed.)

\subsection{Non-periodic systems}\label{non-period systems}

We now start our analysis with non-periodic systems, either with random field
disorder or any correlator decreasing faster than RF. Let us first recall
that at the level of the {\em bare} model the static RF obeys $R(u) \sim - \sigma |u|$
at large $|u|$ and thus $\int_0^{\infty} du \Delta(u) = R'(0^+) - R'(\infty)
= - \sigma$ ($\sigma$ is the amplitude of the random field)
while RB or any correlator decaying faster than RF satisfies $\int \Delta = 0$.

Let us first
integrate the disorder flow equation (\ref{genbetafunc}) from $u=0^+$ to $u=+\infty$. We obtain
\begin{equation}\label{FisherWrong}
- m \partial_m \int\limits_0^{\infty} \Delta (u)\,\rmd u
= (\epsilon - 3 \zeta) \int\limits_0^{\infty} \Delta (u)\,\rmd u
- X^{(\alpha )} \Delta'(0^+)^3
\ .
\end{equation}
The only assumption that we have made here is that $u \Delta(u)$ goes to zero at $u=+\infty$,
which is the case both for RB and RF.

Let us first recall the 1-loop analysis, where in the FRG equation
there is no distinction between statics and depinning. The last term
in (\ref{FisherWrong}) is then absent.  Thus one finds either fixed
points with $\int \Delta= \sigma >0$ with $\zeta = \epsilon/3$, the RF
universality class, or others with $\int \Delta=0$ for $\zeta <
\epsilon/3$ which corresponds to disorder with shorter range
correlations than than RF. This includes the RB fixed point with
exponentially decaying correlator and $\zeta_{RB} = 0.208
\epsilon$. It also includes a continuous family of intermediate power
law fixed points \cite{BalentsDSFisher1993,LeDoussalWiese2001} with
decay at large $u$ as $\Delta(u) = - R''(u) \sim (\alpha-2)
(\alpha-1) u^{- \alpha}$ with $\alpha^*>\alpha>1$. These have
$\zeta(\alpha) = \epsilon/(2 + \alpha)$ (from solving the linear part
of the FRG equation) and $\zeta(\alpha^*)=\zeta_{RB}$.

The last term in (\ref{FisherWrong}) shows that things work
differently to two loops at depinning. the condition $\int \Delta=0$
is no longer possible at the fixed point. Starting from RB one
develops a positive value for $\int \Delta$, i.e.\ a random field
component. The natural conclusion is then that all correlations
shorter range than RF flow to the RF universality class \footnote{If
one excludes a rather unnatural scenario with $\int \Delta =
O(\epsilon^2)$, where $\Delta(u) = \epsilon y_1(u) + \epsilon^2
y_2(u)$ with $y_1(u)$ corresponding to a 1-loop fixed point of range
shorter than RF.}.  Furthermore the fixed point condition
(\ref{genbetafunc}) equals $0$ implies a unique well defined value for
$\zeta$ identical for all ranges shorter than RF (including RB). This
value takes the form:
\begin{equation}\label{zetaAnsatz}
\zeta = \frac{\E}{3} - \frac{X^{(\alpha )} \Delta'(0^+)^3}{3
\int\limits_0^{\infty} \Delta} =\frac{\E}{3} + \zeta_{2}\E^2 + O
(\E^3)
\ ,
\end{equation}
where $\zeta_2>0$ can be obtained from the knowledge of the 1-loop
fixed point $\Delta \sim O(\epsilon)$ only.

Before we compute $\zeta_2$ and obtain the depinning fixed point to
two loop, let us note that in the static case
\cite{ChauveLeDoussalWiese2000a} the last term in (\ref{genbetafunc})
has the opposite sign and, integrating over $u$ one finds that there
is thus no term proportional to $\Delta' (0^{+})$ in
(\ref{FisherWrong}). Thus for the RF disorder $0 < \int \Delta < +
\infty$ one can again conclude that
\begin{equation}\label{zeta eq}
\zeta _{\mathrm{eq}}^{\mathrm{RF}} = \frac{\E}{3}
\end{equation}
to (at least) second order. In fact, as discussed in
\cite{ChauveLeDoussalWiese2000a} this is expected to be exact to all
orders due to the potentiality requirement of the static FRG equation,
which also implies that $\int \Delta=0$ holds to all orders at the
static RB fixed point. The corresponding value for
$\zeta_{\mathrm{eq}}^{\mathrm{RB}}$ is  given to order $\epsilon ^{2}$ in
\cite{ChauveLeDoussalWiese2000a}.

We now want to find the fixed-point function of equation
(\ref{genbetafunc}). Using the reparametrization-invariance
(\ref{DeltaRescale}), we set (with the factors $1/3$ and $1/18$ chosen
for later convenience)
\begin{eqnarray}\label{DeltaAnsatz}
\Delta(u) &=& \frac{\E}{3} y_{1} (u) + \frac{\E^2}{18} y_{2} (u) + O
(\E^3)\\
y_{1} (0)&=& 1 \label{y1Norm} \\
y_{2} (0)&=& 0  \label{y2Norm}
\ ,
\end{eqnarray}
where $y_{1} (u)$ is the 1-loop fixed point function for the RF case.
It was obtained in Ref. \cite{DSFisher1986} and
further studied in Ref. \cite{ChauveGiamarchiLeDoussal2000}. Let us recall
its properties. To lowest order in $\epsilon$ the 1-loop $\beta$-function
(\ref{genbetafunc}) reads:
\begin{equation}
\frac{\E}{3}\Delta (u) + \frac{\epsilon }{3}u \Delta'
(u)-\frac{1}{2}\left[\left(\Delta (u)-\Delta (0) \right)^{2} \right]'' =0
\end{equation}
inserting (\ref{DeltaAnsatz}) the function $y_1(u)$ must satisfy:
\begin{equation}
\left[u y_{1} (u) \right]' = \half \left[\left(y_{1} (u)-y_{1} (0)
\right)^{2} \right]''
\ ,
\end{equation}
which can be first integrated to
\begin{eqnarray}\label{IV.9}
 u y_{1} (u)  &=& \left(y_{1} (u)-1 \right) y_{1}' (u) \ ,
\end{eqnarray}
using (\ref{y1Norm}) in the last line. A second integration with
the boundary conditions implied by (\ref{y1Norm}) yields:
\begin{equation} \label{IV.12}
y_{1} (u) -\ln  y_{1} (u) = 1 + \half u^2\ .
\end{equation}
The derivatives of $y_{1} (u)$ at $u=0$ will be needed below. Deriving
(\ref{IV.9}) successively w.r.t.\ $u$, we find
\begin{eqnarray}
y_{1} (0)&=&1\ , \qquad y_{1}' (0^{+})= -1\nonumber \\
 y_{1}'' (0^{+})&=&
\frac{2}{3}\ , \qquad  y_{1}''' (0^{+})= -\frac{1}{6}\ .
\label{yDerivatives}\label{IV.13}
\end{eqnarray}
We have also determined the
fixed-point function at second order $y_2(u)$, which is given in appendix
\ref{app:fixedpointfunction}.
\begin{figure}[t]
\centerline{\Fig{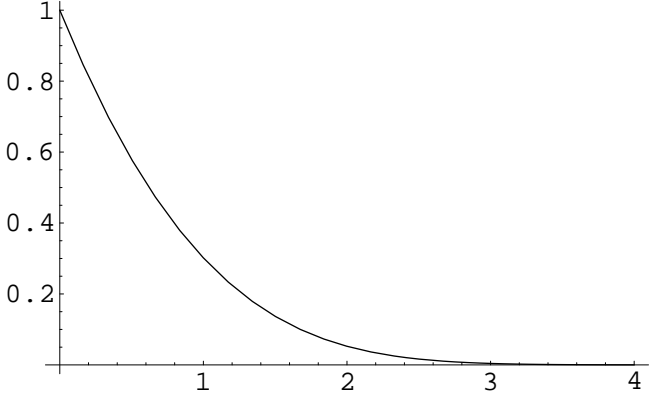}}
\caption{The fixed-point function $y_{1} (u)$ at 1-loop order for non-periodic
disorder.}
\label{fig:RFFP1loop}
\end{figure}%

In order to extract $\zeta $ from (\ref{FisherWrong}), we need
\begin{equation}\label{IV.14}
\sqrt{2} \gamma = \int_0^{\infty} y_1(u) \, \rmd u
\ ,
\end{equation}
which was computed in Ref. \cite{ChauveGiamarchiLeDoussal2000}. The method is
to convert (\ref{IV.14}) into an integral over $y:=y_{1}$:
\begin{eqnarray}\label{IV.15}
\int_{0}^{\infty } y\,  \rmd u &=& - \int _{0}^{1 }y\, \frac{\rmd u}{\rmd
y}\, \rmd y  =- \int _{0}^{1 } \frac{y}{y'}\, \rmd y \nn \\
&  =& - \int_{0}^{1} \rmd y \, \frac{y-1}{\sqrt{2}\sqrt{y- \ln
y -1}}\ ,
\end{eqnarray}
where in the last equality (\ref{IV.12}) has been used. Integrating by
parts, this yields
\begin{equation}\label{IV.16}
\gamma = \int_{0}^{1}
\rmd y \, \sqrt{y-\ln  y -1} \approx 0.5482228893..
\ .
\end{equation}
Combining the definition of $\Delta (u)$ in (\ref{DeltaAnsatz}) with
(\ref{IV.13}) and  (\ref{IV.16}), we find
\begin{equation}
\zeta _{2} = \frac{X^{(\alpha) }}{27 \sqrt{2} \gamma }
\end{equation}
and thus  for $\zeta$
\begin{equation} \label{zetaResult}
\zeta = \frac{\E}{3}+ \frac{X^{(\alpha )}\epsilon ^{2}}{27 \sqrt{2} \gamma }+ O
(\epsilon^{3} )  \ .
\end{equation}
This result violates the  conjecture of \cite{NarayanDSFisher1993a},
that $\zeta =\frac{\E}{3}$ to all orders in $\E$.
To compare (\ref{zetaResult}) with simulations, we have to specify to
the cases of interest: First, for short-range elasticity, i.e.\
$\alpha =2$, we find
\begin{equation}
\zeta = \frac{\epsilon}{3} \left(1 + \frac{\epsilon}{9 \sqrt{2} \gamma }\right)
= \frac{\epsilon}{3} \Big(1 + 0.143313\, \epsilon\Big)
\ .
\end{equation}
Our results are in excellent agreement with the numerical simulations,
see figure \ref{alpha=2 table}.%
\begin{figure}[t]
\begin{tabular}{|c|c|c|c|c|c|}
\hline
exponent & \,~dim~\, & 1-loop & 2-loop & estimate & simulation\\
\hline
\hline
        & $d=3$ & 0.33 & 0.38 & 0.38$\pm$0.02 & 0.34$\pm$0.01 \cite{NattermannStepanowTangLeschhorn1992,LeschhornNattermannStepanowTang1997} \\
\hline
$\zeta$ & $d=2$ & 0.67 & 0.86 & 0.82$\pm$0.1 & 0.75$\pm$0.02 \cite{Leschhorn1993} \\
\hline
        & $d=1$ & 1.00 & 1.43 & 1.2$\pm$0.2 &
\mbox{\begin{minipage}{1.5cm}
 1.25$\pm$0.01 \cite{TangUnpublished} \\
 1.25$\pm$0.05 \cite{Leschhorn1993}
\end{minipage}
}  \\
\hline
\hline
        & $d=3$ & 1.78 & 1.73 & 1.74 $\pm$0.02   & 1.75$\pm$0.15
\cite{NattermannStepanowTangLeschhorn1992,LeschhornNattermannStepanowTang1997} \\
\hline
$z$     & $d=2$ & 1.56 & 1.38 & 1.45$\pm$0.15   & 1.56$\pm$0.06
\cite{Leschhorn1993} \\
\hline
        & $d=1$ & 1.33 & 0.94 & 1.35$\pm$0.2  &
\begin{minipage}{1.8cm}
 1.42$\pm$0.04  \cite{Leschhorn1993} 1.54$\pm$0.05
\cite{TangUnpublished} \end{minipage}  \\
\hline
\hline
        & $d=3$ & 0.89 & 0.85 & 0.84$\pm$0.01 &~ 0.84$\pm$0.02
\cite{NattermannStepanowTangLeschhorn1992,LeschhornNattermannStepanowTang1997}~ \\
\hline
$\beta$ & $d=2$ & 0.78 & 0.62 & 0.53$\pm$0.15  &\begin{minipage}{1.8cm}
0.65$\pm$0.05 \cite{Leschhorn1993} 0.64$\pm$0.02
\cite{NattermannStepanowTangLeschhorn1992,LeschhornNattermannStepanowTang1997} 0.66$\pm$0.04  \cite{RotersHuchtLubeckNowakUsadel1999}
\end{minipage}
\\
\hline
        & $d=1$ & 0.67 & 0.31 & 0.2$\pm$0.2 &\begin{minipage}{1.8cm}
0.35$\pm$0.04 \cite{NowakUsadel1998}\\
$ 0.4  \pm0.05$ \cite{TangUnpublished}\\
 0.25$\pm$0.03 \cite{Leschhorn1993}\end{minipage} \\
\hline
\hline
        & $d=3$ & 0.58 & 0.61 &   0.62$\pm$0.01  & \\
\hline
$\nu$ & $d=2$ & 0.67 & 0.77 &  0.85$\pm$0.1   & 0.77$\pm$0.04 \cite{RotersHuchtLubeckNowakUsadel1999} \\
\hline
        & $d=1$ & 0.75 & 0.98 &   1.25$\pm$0.3
&\begin{minipage}{1.5cm} 1$\pm$0.05 \cite{NowakUsadel1998} 1.1$\pm$0.1
\cite{TangUnpublished} \end{minipage} \\
\hline
\end{tabular}
\caption{Depinning exponents, $\alpha =2$. First column: Exponents
obtained by setting $\epsilon =1$ in the 1-loop result. Second
column: Exponents obtained by setting $\epsilon=1$ in the two-loop
result. Third column: Conservative estimates based on three
Pad{\'e} estimates, scaling relations and common sense. Fourth
column: Results of numerical simulations obtained directly without
using scaling relations.} \label{alpha=2 table}
\end{figure}
For long-range elasticity, i.e.\ $\alpha =1$, equation
(\ref{zetaResult}) reads
\begin{equation}
\zeta = \frac{\epsilon}{3} \left(1 + \frac{ 4 \ln 2 }{9 \sqrt{2}
\gamma }\epsilon\right) 
= \frac{\epsilon}{3} \Big(1 + 0.39735\,\epsilon\Big) \ .
\end{equation}
This is in reasonable agreement with simulations, as shown on figure
\ref{alpha=1 table}.
\begin{figure}[b]
\begin{tabular}{|c|c|c|c|c|}
\hline
exponents & one loop & two loop & estimate & simulation \\
\hline
$\zeta $  & 0.33     &  0.47    &\mbox{ $0.5\pm 0.1$ }&\mbox{ $0.390\pm 0.002$\cite{RossoKrauth2001b}}\\
\hline
$z$ &0.78 &0.66 &$0.7\pm 0.1$ & $0.74 \pm  0.03$\cite{RamanathanFisher1998}  \\
\hline  
$\beta $ & 0.78 & 0.59 & $0.4\pm 0.2$ & ??? \\
\hline
$\nu $ & 1.33 & 1.58 & $2.\pm 0.4$ & ??? \\
\hline
\end{tabular}
\vspace{1mm}
\caption{\small Depinning exponents, $\alpha =1$. First column:
Exponents obtained by setting $\epsilon
=1$ in the 1-loop result. Second column: Exponents obtained by
setting $\epsilon=1$ in the two-loop result. Third column:
Conservative estimates based on three Pad{\'e} estimates together with
scaling relations between exponents.}
\label{alpha=1 table}
\end{figure}

We now turn to the calculation of the dynamical exponent $z$.
As can be seen from the general result of equation (\ref{III.65}), we
need $\Delta' (0^{+})$, $\Delta'' (0^{+})$ and $\Delta''' (0^{+})$ at
leading order, which can be inferred from (\ref{DeltaAnsatz}) and
(\ref{IV.13}). We further need $ \Delta''(0)$ at second order.
Expanding (\ref{genbetafunc}) to order $\E^3$, and Taylor-expanding to
second order in $u$, we can solve for $y_2'' (0)$, which yields
\begin{eqnarray}
 \Delta''(0)&=& \frac{\epsilon}{3}  y_1''(0) + \frac{\epsilon}{18} y_2''(0) \\
&=& \frac{2\epsilon}{9} + \frac{\epsilon ^{2}}{3}
\left(\frac{10X^{(\alpha )}}{27} - \zeta_2 \right)
\ .
\end{eqnarray}
Specializing to the SR-case ($\alpha =2$) yields with the help of
(\ref{III.50})
\begin{eqnarray}
z  &=& 2 - \Delta''(0^+) + \Delta''(0^+)^2 + \Delta'''(0^+) \Delta'(0^+)
\left(\frac{3}{2} - \ln 2\right)\!\! \nonumber \\
& =& 2 - \frac{2 \epsilon}{9} + \epsilon^2 \left( \frac{\zeta_2}{3} - \frac{\ln 2}{54}
- \frac{5}{108} \right)\nonumber  \\
& =& 2 - 0.222222 \epsilon - 0.0432087 \epsilon^2
\ .
\end{eqnarray}
The agreement with the numerical simulations given on figure
\ref{alpha=2 table} is again good.
Finally, the exponents $\beta $ and $\nu $ are obtained from scaling
relations. For $\alpha =2$ (SR) they read
\begin{eqnarray}
\beta &=& \frac{z - \zeta}{2 - \zeta} \nn \\
&=&
1 - \frac \epsilon 9 + \epsilon^2 \left(  \frac{\zeta_2}{6} - \frac{1}{24} - \frac{\ln 2}{108} \right)\nonumber  \\
&=&
1 - \frac \epsilon 9 + 0.040123 \epsilon^2  \\
\nu &=& \frac{1}{2 - \zeta} = \frac{1}{2} + \frac{\epsilon}{12} +
 \epsilon^2 \left(  \frac{\zeta_2}{4}  +  \frac{1}{72} \right) \nonumber\\
 &=&
\frac{1}{2} + \frac{\epsilon}{12} + 0.0258316\, \epsilon^2
\ .
\end{eqnarray}
We now turn to long range elasticity $\alpha \neq 2$. The general formula for $z$ reads
\begin{eqnarray}
z = \alpha - \frac{2}{9} \epsilon
+ \epsilon^2 \left( \frac{\zeta_2}{3} - \frac{2 X^{(\alpha )}}{27} +
\frac{Y^{(\alpha )}}{54} \right)
\ .
\end{eqnarray}
Specifying to  $\alpha=1$ yields
\begin{eqnarray}
z &=& 1 - \frac{2}{9} \epsilon + \epsilon^2 \left( \frac{\zeta_2}{3} - \frac{\pi + 20 \ln 2}{108}\right) \nn \\
&=& 1 - \frac{2}{9} \epsilon - 0.1132997 \epsilon^2
\ .
\end{eqnarray}
Again $\beta $ and $\nu $ are obtained from scaling as ($\alpha =1$)
\begin{eqnarray}
 \beta &=& \frac{z-\zeta}{1-\zeta} = 1 - \frac{2}{9} \epsilon
+ \epsilon^2 \left( \frac{\zeta_2}{3} - \frac{2}{27}
- \frac{\pi + 20 \ln 2}{108} \right) \nn \\
&=&  1 - \frac{2}{9} \epsilon - 0.1873737 \epsilon^2 \\
 \nu &=& 1 + \frac \epsilon 3 + \epsilon^2 \left(\frac{1}{9} + \zeta_2\right) \nn \\
&=& 1 + \frac{\epsilon}{3} + 0.24356 \epsilon^2\label{IV.29}
\ .
\end{eqnarray}
Numerical values are given on figure \ref{alpha=1 table}.

Note that to two loops at the RF fixed point there does not appear
to be any unstable direction. We thus conclude, as in
\cite{NarayanDSFisher1993a} that
\begin{eqnarray}
\nu_{\mathrm{FS}} &=& \nu
\ .
\end{eqnarray}
Finally, for depinning there should  also be a family of  fixed points
corresponding to correlations of the {\em force} which are  long range with
$\Delta(u) \sim u^{- \alpha}$ and $\alpha \geq 
\alpha^*  \geq 1$. The linear part of the FRG equation implies
that $\zeta(\alpha) = \epsilon/(2 + \alpha)$ and a crossover from
the RF fixed point occurs when $\zeta(\alpha^*) = \zeta_{\mathrm{RF}} =
\frac{\E}{3} + \zeta_{2}\E^2 + O (\E^3)$. We have not studied
these LR fixed points in details.

\subsection{Periodic systems}
\begin{figure}[bpt!h]%
\begin{tabular}{|c|c|c|c|c|c|}
\hline
exponent & dimension & 1-loop & 2-loop & estimate & simulation\\
\hline    & $d=3$ & 0.83 & 0.78 & 0.78$\pm$0.03 &
\mbox{\begin{minipage}{1.8cm}
0.81$\pm$0.03\rule{0mm}{2ex} \\
 0.84$\pm$0.05 \end{minipage} }
\\
\hline
$\beta$ & $d=2$ & 0.67 & 0.44 & 0.52$\pm$0.08
&\begin{minipage}{1.8cm}  0.63$\pm$0.06\rule{0mm}{2ex} \\
 0.68$\pm$0.07 \end{minipage}  \\
\hline
        & $d=1$ & 0.5 & 0. & 0.2$\pm$0.2 & \\
\hline
\end{tabular}
\caption{\small Depinning exponents for CDW. First column:
Exponents obtained by setting $\epsilon =1$ in the 1-loop result.
Second column: Exponents obtained by setting $\epsilon=1$ in the
two-loop result. Third column: Conservative estimates based on
three Pad{\'e} estimates, scaling relations and common sense.
Fourth column: Simulations from
\protect\cite{MiddletonFisher1993}. } \label{CDWtable}
\end{figure}%
For periodic $\Delta(u)$ as e.g.\ CDW depinning
\cite{NarayanDSFisher1992b,NarayanDSFisher1993a}, there is another
fixed point of (\ref{genbetafunc}). It is sufficient to study the case
where the period is set to unity, all other cases are easily obtained
using the reparametrization
invariance of equation (\ref{DeltaRescale}). This means, that no rescaling is possible in that direction,
and thus the rescaling factor is
\begin{equation}
\zeta =0\ .
\end{equation}
The fixed-point function is then periodic, and can in the
interval $\left[0,1 \right]$ be expanded in a Taylor-series in $u
(1-u)$. Evenmore, the ansatz
\begin{equation}
\Delta (u) = (a_{1}\epsilon +a_{2}\epsilon ^{2}+ \dots ) +
\left(b_{1}\epsilon +b_{2}\epsilon ^{2}+\dots  \right) u (1-u)
\end{equation}
allows to satisfy  the fixed-point equation
(\ref{genbetafunc}) to order $\E^2$, with coefficients\rechecked
\begin{equation}\label{DeltaPeriodicXalpha}
\Delta^*(u) = \frac{\epsilon}{36} + \frac{\epsilon^2 X^{(\alpha )}}{108}
- \left(\frac{\epsilon}{6} + \frac{\epsilon^2 X^{(\alpha )}}{9}
\right) u(1-u)\ .
\end{equation}
In the physically interesting situation of charge density
waves, the elasticity is short range, i.e.\ $\alpha=2$ and  $X^{(\alpha )}=1$ which yields:
\begin{equation}\label{DeltaPeriodic}
\Delta^*(u) = \frac{\epsilon}{36} + \frac{\epsilon^2 }{108}
- \left(\frac{\epsilon}{6} + \frac{\epsilon^2}{9}
\right) u(1-u)\ .
\end{equation}
This fixed point is manifestly non-potential, i.e.\ it describes a
force-force correlation-function, where the forces can not be derived from a
potential. In a potential environment, the integral of the force over
one period must vanish, and so must the force-force
correlation-function. In contrast we find here
\begin{equation}\label{IV.34}
\int_{0}^{1} \rmd u\, \Delta^* (u) = -\frac{\epsilon ^{2}X^{(\alpha
)}}{108}~ \stackrel{\alpha \to 2}{-\!\!\!-\!\!\!-\!\!\!\longrightarrow}~
-\frac{\epsilon ^{2}}{108} \ .
\end{equation}
Thus to two loop the fixed point correctly accounts for the
irreversibility in the driven system, which becomes manifest
beyond the Larkin-length. This was not apparent to one loop.

An important feature of the periodic case is that the fixed point is
{\em unstable}. The direction of instability is simply adding a constant
to $\Delta(u)$ and its eigenvalue is trivial equal to $\epsilon$ to two loops
and presumably to all orders. The full stability analysis is performed in
appendix \ref{appstability}   but it can be seen already from:
\begin{equation}\label{FisherWrong2}
- m \partial_m \int\limits_0^{1} \Delta (u)\,\rmd u
= \epsilon  \int\limits_0^{1} \Delta (u)\,\rmd u
- 2 X^{(\alpha )} \Delta'(0^+)^3
\end{equation}
obtained by integration of the 2-loop FRG equation on the
interval $[0^{+} , 1^{-}]$. One sees that $\int \Delta$ flows
away if it does not coincide with its fixed point value (\ref{IV.34}).

Thus the asymptotic flow as the dimensional parameter
$m \to 0$ takes the simple form
\begin{eqnarray}
 \Delta_m(u) &=& \Delta^*(u) +  c m^{- \epsilon} \\
 c &=& m_0^{\epsilon} \int\limits_0^{1} (\Delta_{m_0} - \Delta^*)
\end{eqnarray}
i.e.\ it takes the fixed point form shifted by a growing constant. In the statics
$\int_{0}^{1} \rmd u\, \Delta_{m} = \int_{0}^{1} \rmd u\, \Delta_{*} =0$
from potentiality ( the last term in (\ref{FisherWrong2}) is absent )
and thus $c=0$.
At depinning $c$ is non-zero at 2-loop order ($c \approx -
\int_0^{1} \Delta^* >0)$ using that the bare disorder has zero
integral) and this has several consequences.  First one obtains the
static deformations as the sum
\begin{eqnarray}
\overline{(u_x-u_0)^2} =
B_{\mathrm{nl}}(x) + B_{\mathrm{RF}}(x)
\end{eqnarray}
of a universal logarithmic growth-term
\begin{eqnarray}
 B_{\mathrm{nl}}(x) &=& A_d \ln|x| \\
 A_d &=& \frac1 {18} \epsilon + \frac{2 X^{(\alpha)} - 3 }{108}
\epsilon^2 \label{amp}
\end{eqnarray}
(the calculation of $A_d$ is presented in Appendix
\ref{appcorrel} as an example of an explicit calculation of a
correlation function in the renormalized theory); and of the
contribution of the generated ``random force'' of the Larkin type
\begin{eqnarray}
B_{\ind{RF}}(x) \sim c |x|^{4 - d} \ ,
\end{eqnarray}
which completely decouples from the other one.  This is very similar
to what was found in other driven systems where a random force is
generated \cite{GiamarchiLeDoussal1996,LeDoussalGiamarchi1997}. In
particular this implies that the true roughness-exponent at depinning
is not $\zeta=0$ but
\begin{eqnarray}
\zeta_\mathrm{dep} = \frac{4-d}{2}
\ .
\end{eqnarray}

Another consequence is that the the two exponents $\nu$ and
$\nu_{\mathrm{FS}}$ are different. We find:
\begin{eqnarray}
 \nu &=& \frac{1}{2 - \zeta} = \frac{1}{2} \\
 \nu_{\mathrm{FS}} &=& \frac{1}{2 - \zeta_{\mathrm{dep}}} = \frac{2}{d}
\end{eqnarray}
and given the generality of the above argument this should holds
to all orders. Note then that the CCFFS-bound
\cite{ChayesChayesFisherSpencer1986}   for $\nu_{\mathrm{FS}}$ is
saturated. This is very different to the case of interfaces
(saturation of the bound there would lead to the incorrect result
$\zeta=\epsilon/3$).

The dynamical exponent $z$ is \rechecked
\begin{equation}\label{RPz}
z = \alpha  - \frac{\epsilon}{3}  - \frac{\epsilon^{2} X^{(\alpha )}}{9}~
\stackrel{\alpha \to
2}{-\!\!\!-\!\!\!-\!\!\!\longrightarrow}~ 2 - \frac{\epsilon}{3}  -
\frac{ \epsilon^2}{9}
\ .
\end{equation}
Curiously, it does not depend on the diagram $I_{\eta }$ or
equivalently $Y^{(\alpha )}$.

In CDW depinning, the best observable quantity is
$\beta $. From the scaling relation
\cite{NattermannStepanowTangLeschhorn1992,%
LeschhornNattermannStepanowTang1997,NarayanDSFisher1992b,%
NarayanDSFisher1993a}
$\beta=(z-\zeta)/(2-\zeta)$, and $\zeta =0$,  we find
$\beta =z/2$
and thus for CDW ($\alpha =2$)
\begin{equation}
\beta  = 1 - \frac{\epsilon}{6}  -
\frac{ \epsilon^2}{18} \ .
\end{equation}
This expansion is however ill-behaved, at least at large $\epsilon
$. It therefore seems advisable, to use one of the Pade-variants. The
only one which respects common sense down to $d=1$ and even beyond, is
the Pade (0,2), reading
\begin{equation}
\beta = \frac{1}{\textstyle   1+\frac{\epsilon }{6}+ \frac{\epsilon
^{2}}{12}} \ .
\end{equation}
Again simulations are in reasonable good agreement with our
theoretical predictions, as can be seen on table
\ref{CDWtable}. Further simulations would  be welcome.


\section{Conclusion}

To conclude we have constructed a consistent field theory of isotropic
depinning at zero temperature to 2-loop order. While the 1-loop
flow-equations for statics and driven dynamics are identical, our 
2-loop equations distinguish these physically different situations,
yielding different universal predictions for both cases. This is an
encouraging progress. The non-analytic field theory that we have
developed here will be discussed  in companion studies
\cite{LeDoussalWieseChauve2002a,LeDoussalWiese2002b} for the 
static theory to two and three loops. 

A lot remains to be done and understood. If universality is to
hold at depinning then a  renormalizable theory 
should exist to any number of loops. We have not attempted a proof
to all orders here, and the mechanism in which the 
$1/\epsilon$-divergences cancel is non-trivial. We have however
checked the applicability of formal constructions like the subtraction
operator {\bf R} on sample diagrams. This  could further be tested in a
3-loop calculation. Although short time singularities ($\delta(v t)$
terms) did not appear their role to any order remains to be
clarified. Next,  effects of temperature have not been included
here. One expects that although at $T=0$ the statics and the
depinning should be two distinct field theories, this distinction
becomes blurred at finite temperature. How this will work out is not
yet elucidated. Some efforts in that direction are reported in
\footnote{L. Balents, P. Le Doussal in preparation.}. Similarly it
would be quite interesting to understand how to describe
$f=f_c^{-}$, i.e.\ the approach to the threshold from below. From
the considerations here this appears to be  quite non-trivial.

Extension of the present method to systems with $N>1$ is also far
from trivial. The monotonous increase of $u_{t}$ does not apply to all
components, which  leads to complications. The large-$N$ limit
of the static FRG was solved exactly recently
\cite{LeDoussalWiese2001}and it would be interesting to extend it
to the dynamics. Finally the
threshold dynamics of other systems, such as random field spin
models, which can be described by the FRG, is of
interest.

From the point of view of simulations our results together with
recent more powerful algorithms offer hope that more precise
comparisons could be made, not only for  exponents but
also for other universal quantities which  offer stronger tests
such as scaling functions, amplitudes or finite size effects. The
exponent $\nu_{\mathrm{FS}}$ should be measured independently. We encourage
further precise numerical studies on both manifolds and CDW with a
comparison to theory in mind. Agreement between numerics and
theory would allow to rule out or to accept elastic models for the
description of more complex experimental situations.

\appendix

\section{Corrections to disorder: Diagrams of type A}
\label{app:hat-diagrams}
In the following, we give explicit expressions for the diagrams
contributing to the renormalization of disorder. To simplify
notations, we have introduced $q_3:=q_1-q_2$ and set the mass $m$ to
zero. The mass-dependence can easily be reconstructed by replacing
$q_i^2$ by $q_i^2+m^2$. We start with the diagrams of class A, given
on figure \ref{dynamic-2-loop-DO}. For illustration, we show the
complete calculation of the first non-vanishing diagram $\mathrm{a}_2$:
\bea
\diagram{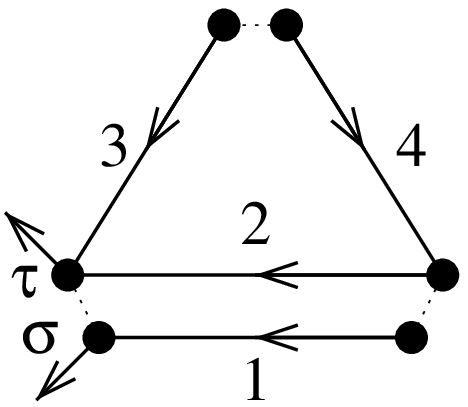} &=&- \int_{q_1,q_2} \int_{t_1,t_2,t_3,t_4>0} \rme^{-q_1^2
t_1 -q_2^2 t_2  -q_3^2(t_3+t_4)}\nn\\
&& \qquad \times  \Delta'''(u_\tau-u_\sigma)
 \Delta'(u_{\tau-t_2}-u_{\sigma-t_1})\nn\\
&& \qquad \times \Delta(u_{\tau-t_3}-u_{\tau-t_2-t_4})
\ .
\eea
For the field $u$, we have given the time-arguments, but suppressed
the spatial arguments, since the result is taken at constant
background field. We also do not write explicitly the two
response-fields. The given configuration is for $\tau-\sigma$ large,
and we can set $u:=u_{\tau'}-u_{\sigma'}$ for all $\tau'=\tau \pm
\mbox{ some } t_i$, $\sigma'=\sigma\pm \mbox{ some } t_i$ since the
$t_i$ are small (due to the exponentially suppressing factors)
compared to
the difference of $\tau -\sigma$. Finally, since $\Delta$ is
continuous, $\Delta(u_{\tau-t_3}-u_{\tau-t_2-t_4})$ can be replaced by
$\Delta(0)$. Integrating over all times leads to
\be
\mathrm a_2 =  - 2 \Delta(0) \Delta'(u) \Delta'''(u) \int_{q_1,q_2}
\frac{1}{q_1^2 q_2^2 q_3^4} \ .
\ee
Similarly, we find
\begin{eqnarray}
\mathrm a_3 &=&  - 2 \Delta(0) \Delta''(u)^2 \int_{q_1,q_2}
\frac{1}{q_1^2 q_2^2 q_3^4}
\\
 \mathrm a_2 + \mathrm a_3 &=& - \partial_u^2 \left[  \Delta(0) \Delta'(u)^2
\int_{q_1,q_2}  \frac{1}{q_1^2 q_2^2 q_3^4} \right] \qquad \\
  \mathrm b_1 &=& 2 \Delta'(u)^2  \Delta''(u) \int_{q_1,q_2}  \frac{1}{q_1^2 q_2^2 q_3^4} \\
 \mathrm  b_2 &=& 2 \Delta(u) \Delta'(u)  \Delta'''(u) \int_{q_1,q_2}  \frac{1}{q_1^2 q_2^2 q_3^4} \\
  \mathrm b_3 &=&  \Delta(u) \Delta''(u)^2  \int_{q_1,q_2}  \frac{1}{q_1^2 q_2^2 q_3^4} \\
  \mathrm b_4 &=&  \Delta''(u) \Delta'(u)^2 \int_{q_1,q_2}  \frac{1}{q_1^2 q_2^2 q_3^4} \\
  \mathrm b_5 &=& 2 \Delta''(u) \Delta'(u)^2  \int_{q_1,q_2}  \frac{1}{q_1^2 q_2^2 q_3^4} \\
  \mathrm b_6 &=&  \Delta(u) \Delta''(u)^2  \int_{q_1,q_2}  \frac{1}{q_1^2 q_2^2 q_3^4}
\ .
\end{eqnarray}
The contribution of the $ \mathrm b_i$'s can be summed as
\be
\sum_{i=1}^6  \mathrm b_i = \partial_u^2 \left[ \Delta(u) \Delta'(u)^2
\int_{q_1,q_2}  \frac{1}{q_1^2 q_2^2 q_3^4}\right] \ .
\ee
Diagram $ \mathrm c_1$ is
\begin{eqnarray}
 \mathrm c_1 = 2 \Delta'(0^+)^2 \Delta''(u) \int_{q_1,q_2}  \frac{1}{q_1^2
q_2^2 q_3^4} \ .
\end{eqnarray}
\begin{widetext}
All these diagrams contain the hat-diagram known from the statics and
$\phi^4$ theory. It can be calculated as follows:
\begin{eqnarray}
\parbox{0.7cm}{\fig{0.7cm}{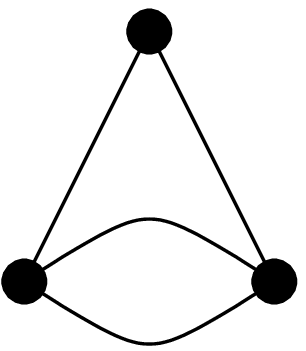}} &=& \int_{q_1,q_2} \frac{1}{(q_1^2 + m^2) (q_2^2 + m^2)^2 ((q_1+ q_2)^2+ m^2)}
= \int_{\alpha, \beta, \gamma >0}
\beta \rme^{- \alpha (q_1^2 + m^2) - \beta (q_2^2 + m^2) - \gamma
((q_1+ q_2)^2+ m^2)} \nn \\
& =& \left(\int_{q} \rme^{-q^2}\right)^2
\int_{\alpha>0,\beta>0,\gamma>0} \beta
\rme^{- m^2 ( \alpha + \beta +  \gamma)}
\left[ \text{Det}
\left(
\begin{array} {cc} \alpha + \gamma & \gamma \\
                                    \gamma     & \beta + \gamma \end{array}
\right) \right]^{-d/2}\nn \\
& =&
\left(\int_{q} \rme^{-q^2}\right)^2
\int_{\alpha>0,\beta>0,\gamma>0} \beta \gamma^{3-d}
\rme^{- m^2 \gamma ( \alpha + \beta + 1 )}
(\alpha + \beta + \alpha \beta)^{-d/2} \nn \\
& =& \left(\int_{q} \rme^{-q^2}\right)^2 \Gamma ( 4-d) m^{- 2 \epsilon} J
\ ,
\end{eqnarray}
where we split the divergent integral $J$ in pieces, which are
either finite or where the divergence can be calculated analytically:
\begin{eqnarray}
  J &=& \int_0^{\infty} \rmd \alpha \int_0^{\infty} \rmd \beta\,
\frac{\beta}{(\alpha + \beta + \alpha \beta)^2}
\frac{(\alpha + \beta + \alpha \beta)^{\frac{\epsilon}{2}}}{
( \alpha + \beta + 1 )^\epsilon} = J_{1} + J_{2} + J_{3} \\
J_{1} &=& \int_0^{\infty} \rmd \alpha \int_0^{1} \rmd \beta \,
\frac{\beta}{(\alpha + \beta + \alpha \beta)^2}
\frac{(\alpha + \beta + \alpha \beta)^{\frac{\epsilon}{2}}}{
( \alpha + \beta + 1 )^\epsilon}
= \ln 2 + O(\epsilon)
\\
 J_{2} &=& \int_0^{\infty} \rmd \alpha \int_1^{\infty} \rmd \beta
\left( \frac{\beta}{(\alpha + \beta + \alpha \beta)^2}
\frac{(\alpha + \beta + \alpha \beta)^{\frac{\epsilon}{2}}}{
( \alpha + \beta + 1 )^\epsilon}
- \frac{1}{(1+\alpha )^{2-\frac{\epsilon}{2}} \beta^{1 + \frac{\epsilon}{2}}} \right)
= - \ln 2 + O(\epsilon) \\
J_3 &=& \int_0^{\infty} \rmd \alpha \int_1^{\infty} \rmd \beta\,
\frac{1}{(1+\alpha )^{2-\frac{\epsilon}{2}} \beta^{1 + \frac{\epsilon}{2}}}
= \frac{2}{\epsilon} + 1 + O(\epsilon)
\end{eqnarray}
\end{widetext}
This gives the final result for the hat-diagram
\begin{eqnarray} \label{IA}
I_{A}= \parbox{0.7cm}{\fig{0.7cm}{LH}} &=&
\left(\int_{q} \rme^{-q^2}\right)^{\!2}\, \Gamma(4-d)\, m^{- 2 \epsilon}
\left(\frac{2}{\epsilon} + 1 + O(\epsilon)\right) \nn \\
&=&
\left(\frac{1}{2 \epsilon^2} + \frac{1}{4 \epsilon}\right)
\left(\epsilon I_1^{(\alpha )}\right)^2 \label{A.18}
\ .
\end{eqnarray}
We now turn to the non-trivial diagram $ \mathrm e_1$. At  finite velocity $v$,
the diagram is
\bea
\diagram{e1}&=&\int_{q_1,q_2}\int_{t_1,\ldots,t_4>0} \Delta''(u)
\Delta'(v(t_1+t_4-t_3)) \nn\\&& \times\Delta'(v(t_2+t_3-t_4))
\rme^{-t_1q_1^2-t_2q_2^2-(t_3+t_4) q_3^3} \nn\\
\eea
In the limit of vanishing velocity $v\to 0$,  we can replace  %
$\Delta'(v(t_2+t_3-t_4))$ by $\Delta(0^+) \sgn(t_2+t_3-t_4)$ a.s.o. Let us
stress that this replacement is correct both before and after reaching
the Larkin length. Its result is
\begin{eqnarray}
&&\!\! \mathrm  e_1 =  \Delta'(0^+)^2 \Delta''(u) \nn\\
&& \qquad \times \int_{q_1,q_2}
\int_{t_3,t_4>0} \rme^{- q_3^2 (t_3 + t_4)}
I(t_3,t_4,q_1,q_2) \label{A.20}\\
&&\!\! I(t_3,t_4,q_1,q_2) = \int_{t_1,t_2} \theta(t_1)\theta(t_2)
 \sgn(t_1+ t_4 - t_3 )
 \nn \\
&&\hphantom{I(t_3,t_4,q_1,q_2) = \int} \times \sgn(t_2  + t_3- t_4 )
\rme^{- (q_1^2 t_1 + q_2^2 t_2)}\ .\nn\\
\end{eqnarray}
Using that $\rme^{-(q_1^2t_1 + q_2^2 t_2)} =\frac {1}{q_1^2 q_2^2}
\partial_{t_1} \partial_{t_2} \rme^{-(q_1^2t_1 + q_2^2 t_2)}$ and
integrating $I$ by parts in $t_1$ and $t_2$ yields
\begin{eqnarray}
&& I(t_3,t_4,q_1,q_2) = \frac{1}{q_1^2 q_2^2}\times \nn\\
&& \times  \Big[ 2 \theta(t_3-t_4)
\rme^{- q_1^2 (t_3-t_4)}+
2 \theta(t_4-t_3) \rme^{- q_2^2 (t_4-t_3)} - 1) \Big] \nn\\
\end{eqnarray}
The integral over the two remaining times $t_3$ and $t_4$  in
(\ref{A.20}) gives
\begin{eqnarray}
\lefteqn{ \int_{t_3,t_4>0}
I(t_3,t_4,q_1,q_2)=}&& \nn\\
&&\qquad \frac{1}{q_1^2 q_2^2} \left(
\frac{1}{q_3^2 (q_1^2 + q_3^2)} +
\frac{1}{q_3^2 (q_2^2 + q_3^2)} - \frac{1}{q_3^4} \right)\qquad  \nn\\
\end{eqnarray}
and thus
\begin{eqnarray}
 \mathrm e_1 &=&  \Delta'(0^+)^2 \Delta''(u)\times\nn\\
&& \times \int_{q_1,q_2}
 \frac{1}{q_1^2 q_2^2 q_3^2}
\left( \frac{1}{q_1^2 + q_3^2} + \frac{1}{q_2^2 + q_3^2}
- \frac{1}{q_3^2} \right)\qquad\ \
\end{eqnarray}
with $q_3=q_1-q_2$. In presence of a mass this reads:
\begin{eqnarray}
  \mathrm e_1 &=&  \Delta'(0^+)^2 \Delta''(u)\times\nn\\
&& \times \int_{q_1,q_2}
\frac{1}{(q_1^2 + m^2)  (q_2^2 + m^2) (q_3^2 + m^2) }\nn \\
&&~~~~~\left(
\frac{1}{q_1^2 + q_3^2 + 2 m^2}
+ \frac{1}{q_2^2 + q_3^2 + 2 m^2}
- \frac{1}{q_3^2 + m^2} \right)\!\nn\\
\end{eqnarray}
We now calculate the new integral. It is relatively simple, since it has
only a single pole in $1/\E$: 
\begin{widetext}
\begin{eqnarray}
I_{l} &:=& \int_{q_1,q_2}
\frac{1}{(q_1^2 + m^2)  (q_2^2 + m^2) (q_3^2 + m^2) ( q_1^2 + q_3^2 + 2 m^2)} \nn\\
& =& \int_{q_1,q_3}
\int_{\alpha>0,\beta>0,\gamma>0,\delta >0}
\rme^{- \alpha (q_1^2 + m^2)
- \beta((q_1+q_3)^2 + m^2) - \gamma (q_3^2 +  m^2) - \delta (q_1^2 + q_3^2 + 2 m^2)}  \nn\\
&=&
\left(\int_{q} \rme^{-q^2}\right)^2
\int_{\alpha>0,\beta>0,\gamma>0,\delta >0}
\rme^{- m^2 ( \alpha + \beta + \gamma + 2 \delta)}
\left[ \text{Det}
\left(
\begin{array} {cc} \alpha + \beta + \delta & \beta \\
   \beta  & \beta + \delta +
\gamma \end{array}
\right) \right]^{-d/2} \nn \\
&=&
\left(\int_{q} \rme^{-q^2}\right)^2
\int_{\alpha>0,\beta>0,\gamma>0,\delta >0}
\rme^{- m^2 \delta ( \alpha + \beta + \gamma + 2)} \delta^{3-d}
(1 + \gamma + \alpha + 2 \beta +  \alpha \beta + \alpha \gamma + \beta
\gamma)^{-d/2} \nn \\
&=&  \Gamma(4-d)
\left(\int_{q} \rme^{-q^2}\right)^2
\int_{\alpha>0,\beta>0,\gamma>0}
(1 + \gamma + \alpha + 2 \beta +  \alpha \beta + \alpha \gamma + \beta \gamma)^{-d/2}
\left[m^2 ( \alpha + \beta + \gamma + 2) \right]^{d-4} \nn \\
&= &  \left(\int_{q} \rme^{-q^2}\right)^2 \frac{1}{\epsilon} m^{- 2 \epsilon}
\int_{\alpha>0,\beta>0,\gamma>0}
(1 + \gamma + \alpha + 2 \beta +  \alpha \beta + \alpha \gamma + \beta
\gamma)^{-2} + \mbox{finite}\nn \\
&=& 2 \ln 2 \left(\int_{q} \rme^{-q^2}\right)^2 \frac{1}{\epsilon} m^{- 2 \epsilon}
+\mbox{ finite}\nn \\
&=&  \frac{ \ln 2}{2 \epsilon} \, ( \epsilon I_1)^2 + \text{finite}
\label{A.26}\label{Il}
\end{eqnarray}
\end{widetext}
This gives the final result for $\rme_1$
\be
\mathrm{e}_1 = \Delta'(0^+)^2 \Delta''(u) \left(2 I_{l}
 -I_{A}\right)
\ .
\ee
The last non-vanishing diagram is $f_{2}$:\rechecked
\begin{eqnarray}
\mathrm{f}_2 &=&  2 \Delta'(0^+)^2 \Delta''(u) \int_{q_1,q_2}
\rme^{- q_3^2 (t_3 + t_4) - (q_1^2 t_1 + q_2^2 t_2)} \times\nn\\
&& \hphantom{ 2 \Delta'(0^+)^2 \Delta''(u) \int_{q_1,q_2}  }
\times \sgn(t_4 - t_3 - t_2)
\ .\qquad \qquad 
\end{eqnarray}
Integrating first over $t_{4}$ and then over the remaining times gives
\rechecked
\begin{equation}
\mathrm{f}_2= - 2 \Delta'(0^+)^2 \Delta''(u)
\int_{q_1,q_2} \frac{1}{q_1^2 q_2^2 q_3^2 (q_2^2 + q_3^2)}\ .
\end{equation}
The integral has already been calculated in (\ref{A.26}), yielding the
result
\begin{equation}
\mathrm{f}_2 = - 2 \Delta'(0^+)^2 \Delta''(u)  I_{l}
\ .
\end{equation}
Note that the non-trivial integrals in $\mathrm{e}_{1}$ and
$\mathrm{f}_{2}$ are in fact identical and cancel:
\begin{equation}
\mathrm{e}_{1}+\mathrm{f}_{2} = -\Delta'(0^+)^2 \Delta''(u) I_{A}
\ .
\end{equation}


\section{Corrections to disorder: Diagrams of type B}
In this appendix, we calculate diagrams of type B (the bubble-chains).

The diagrams which are odd functions of $u$ are:
\be
\mathrm h_1=\mathrm h_2=\mathrm i_1=\mathrm j_1=\mathrm k_2=\mathrm
k_3=\mathrm l_2=\mathrm l_3=\mathrm l_4=0
\ .
\ee
The diagrams that are second derivative of  static ones have
all their response-fields on their unsaturated vertices. These are:
\begin{eqnarray}
\mathrm g_1&=& \Delta''(u)^2 \Delta I_1^2 \\
\mathrm g_2& =& 2 \Delta'(u) \Delta'''(u) \Delta(u) I_1^2 \\
\mathrm g_3 &=& \Delta'(u)^2 \Delta''(u) I_1^2 \\
\mathrm g_4 &=& \frac{1}{2} \Delta(u)^2 \Delta''''(u) I_1^2 \\
\mathrm g_1 + \mathrm g_2 + \mathrm g_3 + \mathrm g_4 &=& \partial^2_u
\left[ \frac{1}{2} \Delta(u)^2 \Delta''(u)\right]  I_1^2 \\
\mathrm  h_3 &=& - \Delta(0) \Delta''''(u) \Delta (u) I_1^2 \\
\mathrm  h_4 &=& - \Delta(0) \Delta''(u)^2 I_1^2 \\
\mathrm  h_5 = \mathrm h_6 &=& - \Delta(0) \Delta'''(u) \Delta'(u) I_1^2 \\
\mathrm  h_3 + \mathrm h_4 + \mathrm h_5 + \mathrm h_6 &=&
\partial^2_u \left[ - \Delta(0) \Delta(u) \Delta''(u) \right] I_1^2 \qquad \\
\mathrm i_2 = \mathrm j_2 &=& \frac{1}{4} \Delta(0)^2 \Delta''''(u) I_1^2 \\
\mathrm  k_1 =  - \mathrm l_1 &=& -  \Delta(u)'' \Delta''(0^+) \Delta(0)  I_1^2
\ .
\end{eqnarray}
The surprise is that $\mathrm{i}_3$, which is not the second derivative of a
static diagram (since it has both $\hat{u}$ on saturated vertices)
is non-trivial:
\be
\mathrm i_3 = - \Delta'(0^+)^2 \Delta''(u) I_1^2\ .
\ee
This diagram  is necessary to ensure renormalizability.

\section{Corrections to disorder: Diagrams of type C}
\label{app:DiagC}%
\begin{figure}[h]
\centerline{\Fig{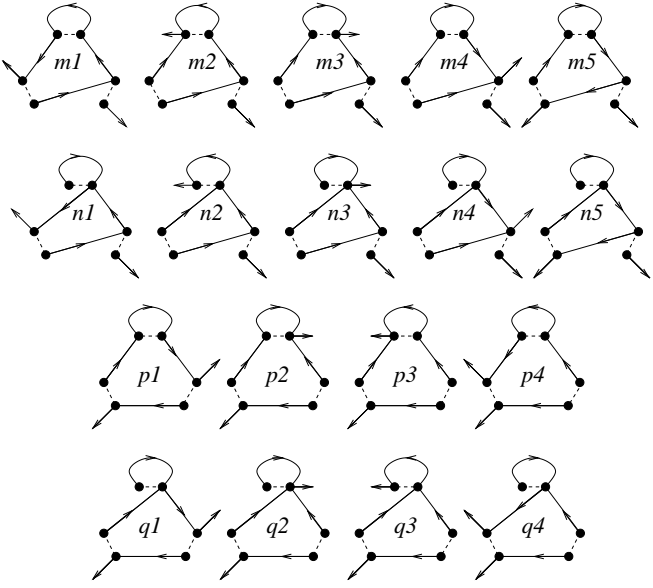}}
\caption{2-loop diagrams of class C}
\end{figure}%
\noindent
In this appendix,
we show that diagrams of type C do not contribute
to the renormalization of disorder. This is fortunate, since they
involve a strongly diverging diagram (the tadpole), which would render
perturbation theory non-universal.

The diagrams which are odd functions of $u$ are
\begin{eqnarray}
\mathrm m_1&=&\mathrm m_3=\mathrm m_4= \mathrm m_5=\mathrm n_1=\mathrm n_3 \nn \\
& =&\mathrm n_4=\mathrm n_5=\mathrm p_2=\mathrm p_3=\mathrm q_2=\mathrm q_3=0
\ .
\end{eqnarray}
The following diagrams cancel:
\begin{eqnarray}
\mathrm m_2+\mathrm n_2 &=&0\\
\mathrm p_1+\mathrm q_1&=&0\\
\mathrm p_4+q_4&=&0
\ .
\end{eqnarray}
No contribution remains.

\section{Corrections to $\eta$: 2-loop diagrams}
\label{app:eta}
In this appendix, we give  all diagrams contributing to the
correction of $\eta$ at second order. For simplicity of notation, we again drop
the explicit mass-dependence.  We group together those diagrams
which partially cancel. We demonstrate explicitly how to calculate the
first diagram $\mathrm{a}$ from the very beginning.
\begin{eqnarray}
\mathrm{a}&=&\diagram{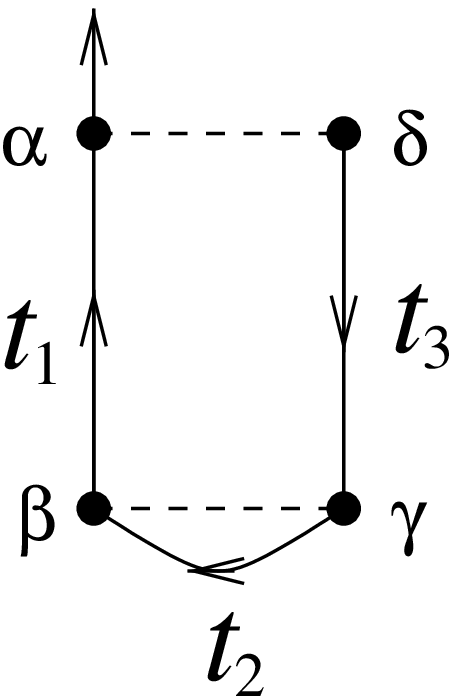} \\
& =& \int\limits_{q_1,q_2}\,
\int\limits_{t_1,t_2,t_3} R_{q_{1}t_{1}} R_{q_{2}t_{2}}
R_{q_{1}t_{3}} \times \nn \\
&& \hphantom{ \int\limits_{q_1,q_2}\, \int\limits_{t_1,t_2,t_3}~~}
 \Delta' (u_{t=0}-u_{-t_{1}-t_{2}-t_{3}}) (-\Delta'' (u_{-t_{1}}))\nn
\end{eqnarray}We have drawn three response functions. We have chosen
to start counting time at 0 for
vertex $\alpha $, such that vertex $\beta $ is at time $-t_{1}$, vertex
$\gamma $ at time $-t_{1}-t_{2}$ and vertex $\delta $ at time
$-t_{1}-t_{2}-t_{3}$. This gives the times for the arguments  of
$\Delta$. The upper $\Delta$ in diagram $\mathrm{a}$ has one time
derivative, the lower vertex two,
resulting in $\Delta'$ and $-\Delta''$ respectively (the minus-sign
is a consequence of the both response-functions entering at different
``ends'' of $\Delta$). We have suppressed
the space-arguments in the fields $u$, since all diagrams correcting
$\eta $ are calculated at a spatially constant background
field. Inserting the response-functions $R_{qt}=\Theta
(t)\rme^{-q^2t}$, we arrive at
\rechecked
\begin{eqnarray}
\mathrm{a} &=&-  \int\limits_{q_1,q_2}\, \int\limits_{t_1,t_2,t_3>0}
\rme^{- q_1^2 (t_1 + t_3) - q_2^2 t_2}  \times\nn \\
&& \hphantom{ \int\limits_{q_1,q_2}\, \int\limits_{t_1,t_2,t_3}~~}
\Delta'(u_0 - u_{-t_1-t_2-t_3})  \Delta''(u_{-t_1} - u_{-t_1-t_2}) \nonumber \\
\end{eqnarray}
The crucial point is now that this diagram corrects the critical force
and $\eta $. The correction to the critical force is obtained by
setting the arguments of the $\Delta$'s to $0^{+}$ (unique here due to
the time-arguments). This contribution is non-universal and we shall
not calculate it in the following. The {\em universal} correction to
$\eta $ is obtained by Taylor-expanding the argument of e.g.\ $\Delta'
(u_{0}-u_{-t_{1}-t_{2}-t_{3}})$ as
\begin{equation}
u_{0}-u_{-t_{1}-t_{2}-t_{3}} \approx (v + \dot u_{0} ) (t_{1}+t_{2}+t_{3})
\end{equation}
and thus
\begin{equation}
\Delta'(u_{0}-u_{-t_{1}-t_{2}-t_{3}}) \approx \Delta'' (0^{+}) \dot
u_{0} (t_{1}+t_{2}+t_{3}) \ ,
\end{equation}
which naturally leads to the generation of a correction to friction.
For our diagram, this is (sloppily dropping $\dot{u}_{0} $ and the
response-field for simplicity of notation)\rechecked
\begin{eqnarray}
\mathrm{a}& =&-
\int\limits_{q_1,q_2}\, \int\limits_{t_1,t_2,t_3>>0} \rme^{- q_1^2 (t_1 +
t_3) - q_2^2 t_2} \times\nonumber\\
&&\hphantom{ \int\limits_{q_1,q_2}\, \int}
\left[\Delta''(0^+)^2 (t_1 + t_2 + t_3) + \Delta'(0^+) \Delta'''(0^+)
t_2 \right] \nonumber  \\
& =&
-\! \int\limits_{q_1,q_2}\!
\Delta''(0^+)^2 \left(\frac{2}{q_1^6 q_2^2} {+} \frac{1}{q_1^4 q_2^4} \right)
+ \Delta'(0^+) \Delta'''(0^+) \frac{1}{q_1^4 q_2^4}\nn \\
&&
\end{eqnarray}
where in the last line we have explicitly performed the
time-integrations.

Diagram $\mathrm{g}$ is \rechecked
\begin{eqnarray}
\mathrm{g} &=& \int\limits_{q_1,q_2} \int\limits_{t_1,t_2,t_3>0}
\rme^{- q_1^2 (t_1 + t_3) - q_2^2 t_2}\times \nn\\
&& \hphantom{ \int\limits_{q_1,q_2} \int\limits_{t_1,t_2,t_3>0}}
\Delta'(u_0 - u_{-t_1-t_3}) \Delta''(u_{-t_1} - u_{-t_1-t_2})\nn  \\
&=&
\int\limits_{q_1,q_2}\, \int\limits_{t_1,t_2,t_3>0} \rme^{- q_1^2 (t_1 +
t_3) - q_2^2 t_2} \times \nn\\
&& \hphantom{ \int\limits_{q_1,q_2} \int\limits_{t_1,t_2,t_3>0}}
\left(\Delta''(0^+)^2 (t_1 + t_3) + \Delta'(0^+) \Delta'''(0^+)
t_2\right)\nn  \\
&=&
\int_{q_1,q_2}
\Delta''(0^+)^2 \frac{2}{q_1^6 q_2^2}
+ \Delta'(0^+) \Delta'''(0^+) \frac{1}{q_1^4 q_2^4}
\ .
\end{eqnarray}
Thus
\begin{eqnarray}
\mathrm{a}+\mathrm{g}  = - \int\limits_{q_1,q_2}
\Delta''(0^+)^2 \frac{1}{q_1^4 q_2^4} = - \Delta''(0^+)^2 I_1^2 \qquad
\ .
\end{eqnarray}
Note that both diagrams $\mathrm{a}$ and $\mathrm{g}$ contain a
tadpole-like sub-divergence, which is  canceled by a counter-term
for the critical force. However their sum does not involve such a
term and thus there is no need to specify it.

Graphs b, c and d:
\begin{eqnarray}
\mathrm{b} &=& - \int\limits_{q_1,q_2} \int\limits_{t_1,t_2,t_3>0}
\rme^{- q_1^2 (t_1 + t_3) - q_2^2 t_2}\times \nn\\
&& \hphantom{ \int\limits_{q_1,q_2} \int\limits_{t_1,t_2,t_3>0}}
\Delta'''(u_0 - u_{-t_2}) \Delta(u_{-t_3} - u_{-t_1-t_2})\nn  \\
& =&
- \int_{q_1,q_2} \int_{t_1,t_2,t_3} \rme^{- q_1^2 (t_1 + t_3) - q_2^2
t_2}\times \nn\\
&& \hphantom{ -\int}
\Big[\Delta''''(0^+) \Delta(0) t_2 + \Delta'''(0^+) \Delta'(0^+)|t_1 +
t_2 - t_3|\Big]\nn \\
\end{eqnarray}
while
\begin{eqnarray}
\mathrm{c} &=& \mathrm{d} = \frac{1}{2}\int\limits_{q_1,q_2}
\int\limits_{t_1,t_2,t_3>0}
\rme^{- q_1^2 (t_1 + t_3) - q_2^2 t_2}\times \nn \\
&&\qquad
\Big[\Delta''''(0^+) \Delta(0) t_2 + \Delta'''(0^+) \Delta'(0^+)|t_1 -
t_3|\Big]\ . \nn \\
\end{eqnarray}
Note the factor $1/2$ for the symmetry in the  lower vertex.
Together they are \rechecked
\begin{eqnarray}
&\mathrm{b}&+\,\mathrm{c}+\mathrm{d}= \Delta'''(0^+) \Delta'(0^+)\nn \\
&& \times  \int\limits_{q_1,q_2}\,  \int\limits_{t_1,t_2,t_3}
\rme^{- q_1^2 (t_1 + t_3) - q_2^2 t_2}
 ( |t_1 - t_3| - |t_1 + t_2 - t_3|)\nn  \\\label{D.10}
\end{eqnarray}
Changing variables to $u=t_{1}-t_{3}$ and $s=\frac{1}{2}
(t_{1}+t_{3})$, the integral $\int_{t_{1},t_{3}>0}$ becomes
$\int_{0}^{\infty }\rmd s \int_{-2s}^{2s} \rmd u$. The integral over
$u $ can be performed, but  for fixed $s$ the second term in
(\ref{D.10}) depends on  the value of
$t_{2}$. Distinguishing the both cases, we obtain \rechecked
\begin{widetext}
\begin{eqnarray}
\mathrm{b}+\mathrm{c}+\mathrm{d} &=& \Delta'''(0^+) \Delta'(0^+) \int_{q_1,q_2}
\int_{s>0} \rme^{- 2 q_1^2 s} \left[ \int\limits_0^{\infty}\rmd t_2\,
\rme^{- q_2^2 t_2} 4 s^2 - \int\limits_0^{2 s}\rmd t_2\, \rme^{- q_2^2
t_2} (t_2^2 + 4 s^2)
- \int\limits_{2 s}^{\infty}\rmd t_2\, \rme^{- q_2^2 t_2} 4 s t_2
 \right] \nn \\
&=&  - \Delta'''(0^+) \Delta'(0^+) \int_{q_1,q_2}
\frac{1}{q_1^2 q_2^4 (q_1^2 + q_2^2)}\label{D.11}
\ .
\end{eqnarray}
This integral can be simplified through symmetrization.
Using that
\begin{eqnarray}
&\displaystyle \int\limits_{q_{1},q_{2}}\Bigg[& \frac{1}{q_{1}^{2}
q_{2}^{4} (q_{1}^{2}+q_{2}^{2})} +  \frac{1}{q_{1}^{4} q_{2}^{2}
(q_{1}^{2}+q_{2}^{2})}\Bigg] = \int\limits_{q_{1},q_{2}}
\frac{1}{q_{1}^{4}
q_{2}^{4}} =I_{1}^{2}\ ,
\end{eqnarray}
we obtain
\begin{equation}
\mathrm  b +\mathrm  c + \mathrm d =  - \frac{1}{2} \Delta'''(0^+)
\Delta'(0^+) I_{1}^{2}
\ .
\end{equation}
The next diagram is $\mathrm e$:
\begin{eqnarray}
\mathrm{e} &=& \int_{q_1,q_2} \int_{t_1,t_2,t_3} \rme^{- q_1^2 t_1 -
q_2^2 t_2 - q_3^2 t_3}
\Delta''(u_0 - u_{-t_2- t_1})) \Delta'(u_{-t_2} - u_{-t_3}) \nonumber  \\
&=&  \int_{q_1,q_2} \int_{t_1,t_2,t_3} \rme^{- q_1^2 t_1 - q_2^2 t_2 -
q_3^2 t_3}
\left[ \Delta'''(0^+) \Delta'(0^+) (t_2 + t_1)\, \sgn(t_3-t_2)
+  \Delta''(0^+)^2 (t_3-t_2) \right]
\ .
\end{eqnarray}
By symmetrizing  in ($2\leftrightarrow 3$), the term proportional to
$t_{1}$ and the term proportional to $(t_{3}-t_{2})$ vanish.
The remaining term can be written as
\begin{eqnarray}
\mathrm{e}
&=& -\frac12  \int\limits_{q_1,q_2} \int\limits_{t_1,t_2,t_3} \rme^{-
q_1^2 t_1 -
q_2^2 t_2 - q_3^2 t_3}  \Delta'''(0^+) \Delta'(0^+) |t_{2}-t_{3}|
\ .
\end{eqnarray}
Making the same change of variables to $u$ and $s$ as for
(\ref{D.11}), the integration over $u$, $s$ and $t_{1}$ can be
performed in this order, distinguishing the cases $u<0$ and
$u>0$. Both cases give the same result for a total of
\begin{equation}
 \mathrm{e} = - \Delta'''(0^+) \Delta'(0^+) \int_{q_1,q_2}
\frac{1}{q_1^2 q_2^4 (q_2^2 + q_3^2)}
\ .
\end{equation}
This contains the new integral (given regularized)
\begin{equation}\label{D17}
I_{\eta }:= \int_{q_{1},q_{2}} \frac{1}{(q_{1}^{2}+m^{2})
(q_{2}^{2}+m^{2})^{2} (q_{2}^{2}+q_{3}^{2}+2 m^{2})}\ .
\end{equation}
It is related to $I_l$, see (\ref{Il}) and  $I_{A}$, see (\ref{IA}):
\begin{equation}
I_{\eta }+ I_{l} = I_{A} \ .
\end{equation}
It is calculated in appendix \ref{Ieta}.
The last diagram to be calculated is $\mathrm{f}$:
\begin{eqnarray}
\mathrm{f}&=& \int_{q_1,q_2} \int_{t_1,t_2,t_3} \rme^{- q_1^2 t_1 -
q_2^2 t_2 - q_3^2 t_3}
\Delta''(u_0 - u_{-t_1- t_2}) \Delta'(u_{-t_1 - t_2 - t_3} -
u_{-t_2})\nonumber  \\
& =&- \int_{q_1,q_2} \int_{t_1,t_2,t_3} \rme^{- q_1^2 t_1 - q_2^2 t_2 - q_3^2 t_3}
\left[ \Delta'''(0^+) \Delta'(0^+) (t_2 + t_1)
+  \Delta''(0^+)^2 (t_1+ t_3)  \right]\nonumber  \\
&=&
-2 \Delta'''(0^+) \Delta'(0^+)I_{A} -2 \Delta''(0^+)^2  I_{A}
\ .
\end{eqnarray}

\section{The integral $I_{\eta }$}\label{Ieta}
We have to calculate the integral $I_{\eta }$ defined as
\begin{eqnarray}\label{Ietadef}
I_{\eta }:= \int_{q_1,q_2} \frac{1}{(q_1^2 + m^2)
(q_2^2 + m^2)^2 (q_2^2 + q_3^2 + 2 m^2)}
\ .
\end{eqnarray}
This is done as follows:\rechecked
\begin{eqnarray}
I_{\eta } & =& \int_{q_{1},q_{2}}\int_{\alpha ,\beta ,\gamma >0} \beta
\,\rme^{-\alpha (q_{1}^{2}+m^{2})-\beta (q_{2}^{2}+ m^{2})-\gamma
(q_{2}^{2}+q_{3}^{2}+2 m^{2}) } \nonumber \\
&=&
\left(\int_{q} \rme^{-q^2} \right)^2
\int_{\alpha,\beta,\gamma>0} \beta\,
\rme^{- m^2 ( \alpha + \beta + 2 \gamma)}
\left[ \text{Det}
\left(
\begin{array} {cc} \alpha + \gamma & \gamma \\
     \gamma     & \beta + 2 \gamma \end{array}
\right) \right]^{-d/2}\nonumber  \\
& =&
\left(\int_{q} \rme^{-q^2} \right)^2
\int_{\alpha,\beta,\gamma>0} \beta\, \gamma^{3-d}
\rme^{- m^2 \gamma ( \alpha + \beta + 2 )}
(1+2 \alpha + \beta + \alpha \beta)^{-d/2}\nonumber  \\
& =& \left(\int_{q} \rme^{-q^2}\right)^2 \Gamma ( 4-d)\, m^{- 2 \epsilon}\, J
\end{eqnarray}
with \rechecked
\begin{eqnarray}
J &:=& \int_0^{\infty} \rmd \alpha \int_0^{\infty} \rmd \beta\,
\frac{\beta}{(1+2 \alpha + \beta + \alpha \beta)^2}
\frac{(1+2 \alpha + \beta + \alpha \beta)^{\frac{\epsilon}{2}}}{
( \alpha + \beta + 2 )^\epsilon} = J_1 + J_2 + J_3 \\
 J_1 &:=& \int_0^{\infty} \rmd \alpha \int_0^{1} \rmd \beta \,
\frac{\beta}{(1+2 \alpha + \beta + \alpha \beta)^2}
\frac{(1+2 \alpha + \beta + \alpha \beta)^{\frac{\epsilon}{2}}}{
( \alpha + \beta + 2 )^\epsilon}
= 2 \ln 3 - 3 \ln 2 + O(\epsilon)
\\
 J_2 &:=& \int_0^{\infty} \rmd \alpha \int_1^{\infty} \rmd \beta \,
\left[ \frac{\beta}{(1+2 \alpha + \beta + \alpha \beta)^2}
\frac{(1+2 \alpha + \beta + \alpha \beta)^{\frac{\epsilon}{2}}}{
( \alpha + \beta + 2 )^\epsilon}
- \frac{1}{(1+\alpha )^{2-\frac{\epsilon}{2}} \beta^{1 +
\frac{\epsilon}{2}}} \right]
= \ln 2 - 2 \ln 3 + O(\epsilon)\qquad  \\
 J_3 &:=& \int_0^{\infty} \rmd \alpha \int_1^{\infty} \rmd \beta\,
\frac{1}{(1+\alpha )^{2-\frac{\epsilon}{2}} \beta^{1 + \frac{\epsilon}{2}}}
= \frac{2}{\epsilon} + 1 + O(\epsilon)
\ .
\end{eqnarray}
Thus \rechecked
\begin{equation}\label{IetaInt}
 I_{\eta }:= \int_{q_1,q_2} \frac{1}{(q_1^2 + m^2)
(q_2^2 + m^2)^2 (q_2^2 + q_3^2 + 2 m^2)} = \left(
\frac{1}{2 \E^{2}}+\frac{1-2\ln 2}{4 \E}\right) (\E I_1)^{2} + \mbox{finite}
\ .
\end{equation}

\end{widetext}

\section{Integrals in long range elasticity calculation}
\label{app:LongRange}
In the long-range case, there are two integrals which contribute to
the renormalization of the disorder. At 1-loop order this is
\begin{eqnarray}
I_{1}^{(\alpha )} &:=&
\int_{q} \frac{1}{(q^2 + m^2)^{\alpha}} =\frac{1}{\Gamma (\alpha
)}\int_0^{\infty }\frac{\rmd s}{s}s^{{\alpha } }  \int_{q} \rme^{-s
(q^{2}+ m^{2})}\nonumber \\
&=&  m^{-\epsilon}
\frac{\Gamma(\frac{\epsilon }{2})}{\Gamma(\alpha)} \left(\int_{q}\rme^{-q^2}
\right) \ .
\end{eqnarray}
At 2-loop order, this is
\begin{eqnarray}
 I_A^{(\alpha )} &:=& \int\limits_{q_1,q_2} \frac{1}{(q_1^2 +
m^2)^{\frac{\alpha }{2}} (q_2^2 + m^2)^{\alpha} ((q_1 + q_2)^2 +
m^2)^{\frac{\alpha }{2}}}\nonumber \\
\end{eqnarray}
This is evaluated as follows
\begin{eqnarray}
I_{A}^{(\alpha )} &=&
\int_{0}^{\infty }\frac{\rmd s}{s} \frac{s^{\frac{\alpha }{2} }}{\Gamma
(\frac{\alpha }{2})} \int_{0}^{\infty }\frac{\rmd t}{t}  \frac{
t^{\alpha }}{\Gamma (\alpha
)}\int_{0}^{\infty }
 \frac{\rmd u}{u} \frac{u^{{\frac{\alpha
}{2}}} }{\Gamma (\frac{\alpha }{2})} \nonumber \\
&&\times \int_{q_{1},q_{2}} \rme^{- s ( q_{1}^{2}+m^{2})-t (
q_{2}^{2}+m^{2})-u ( ( q_{1}+q_{2})^{2}+m^{2})  } \nonumber \\
&&\hspace{-0.7 cm}= \left(\int_{q}\rme^
{-q^{2}}\right)^{\!\!2}
\int_{0}^{\infty }\frac{\rmd s}{s} \frac{s^{\frac{\alpha }{2} }}{\Gamma
(\frac{\alpha }{2})} \int_{0}^{\infty }\frac{\rmd t}{t}  \frac{
t^{\alpha }}{\Gamma (\alpha
)}\int_{0}^{\infty }
 \frac{\rmd u}{u} \frac{u^{{\frac{\alpha
}{2}}} }{\Gamma (\frac{\alpha }{2})} \nonumber \\
&&\qquad \times \left[ \mbox{det}\left(\begin{array}{cc}
s+u & u \\
u & t+u
\end{array} \right)\right]^{-\frac{d}{2}} \rme^{-(s+t+u) m^2}\qquad \quad
\ .
\end{eqnarray}
Making the replacement $s\to s u$ and $t\to t u$ and integrating over
$u$, we obtain
\begin{eqnarray}
I_{A}^{(\alpha )} &=& \left(\int_{q}\rme^
{-q^{2}}\right)^{\!\!2}
 m^{-2 \epsilon}  \frac{\Gamma(\epsilon) }{\Gamma(\alpha)
\Gamma(\frac{\alpha }{2})^2} \nonumber \\
&& \times
\int_{s,t>0}
\frac{s^{\frac{\alpha }{2}-1} t^{\alpha-1}}{ (st + s+ t)^\alpha } (s t + s + t)^{\frac{\epsilon }{2}} (1+s+t)^{-\epsilon}\nonumber  \\
& =&\left(\int_{q}\rme^ {-q^{2}}\right)^{\!\!2}
 m^{-2 \epsilon} \frac{\Gamma(\epsilon) }{\Gamma(\alpha)
\Gamma(\frac{\alpha }{2})^2}
\left(J_{1} + J_{2} + J_{3}+O (\epsilon ) \right)\nonumber \\
&&  \\
J_{1} &=& \int_0^1 \rmd t \int_0^{\infty} \rmd s\,  \frac{s^{\frac{\alpha }{2}-1} t^{\alpha-1}}{ (st + s+ t)^\alpha } \\
J_{2} &=&
\int_1^{\infty} \rmd t \int_0^{\infty} \rmd s\,
s^{\frac{\alpha }{2}-1} t^{\alpha-1}\times\nonumber \\
&&\hphantom{\int_1^{\infty} \rmd t \int_0^{\infty} }
\left[ (st + s+ t)^{-\alpha} - (1+s)^{-\alpha} t^{-\alpha} \right]  \\
J_{3} &=&
\int_1^{\infty} \rmd t \int_0^{\infty} \rmd s\,
s^{\frac{\alpha }{2}-1}  (1+s)^{\frac{\epsilon }{2}-\alpha} t^{-1-\frac{\epsilon }{2}}\rule{0mm}{5ex}\nonumber \\
&=& \frac{2}{\epsilon} \frac{\Gamma(\frac{\alpha }{2})
\Gamma(\frac{\alpha-\epsilon}2) }{
\Gamma(\alpha - \frac \epsilon 2)}
\end{eqnarray}
$J_{1}$ and $J_{2}$ are now both integrated over $s$. Changing in
$J_{2}$ the integration over $t$ to that over $1/t$, we obtain
\begin{eqnarray}
J_{1} &+& J_{2} =
2^{1-\alpha}\, \sqrt{\pi}\, \frac{\Gamma(\frac{\alpha }{2})}{\Gamma
(\frac{1+\alpha }{2})} \int_0^1 \rmd t\, \frac{1 + t^{\frac{\alpha }{2}} -
(1+t)^{\frac{\alpha }{2}}}{ t (1+t)^{\frac{\alpha }{2}}}\nonumber \\
&&\hphantom{ J_{2}} \\
  ( J_{1}&+&J_{2}) \ts_{\alpha=1}  = 0  \\
  ( J_{1}&+&J_{2})\ts_{\alpha=1}  = 2 \pi  \ln 2
\ .
\end{eqnarray}
Putting everything together, the final result is
\begin{eqnarray}
I_{A}^{(\alpha )} &=& \left[\frac{1}{2\E^2}+\frac{1}{4\epsilon}
\left(\int_0^1 \rmd t\, \frac{1 + t^{\frac{\alpha }{2}} -
(1+t)^{\frac{\alpha }{2}}}{ t (1+t)^{\frac{\alpha }{2}}}
\right.\right.\\
&&
\left.\left.\hphantom{\frac{1}{2\E^2}+\frac{1}{4\epsilon} \quad }  +
\frac{\Gamma ' (a)}{\Gamma (a)}-\frac{\Gamma '
(\frac{a}{2})}{\Gamma
(\frac{a}{2})} \right) \right]\rule{0mm}{4.8ex}\!
\left(\epsilon I_1
^{(\alpha )}\right)^{2} +O (\epsilon ^{0})\nonumber \\
I_{A}^{(1)} &=& \left[\frac{1}{2\E^2}+\frac{\ln 2}{\E }
\right]\left(\epsilon I_1
^{(\alpha )}\right)^{2} +O (\epsilon ^{0}) \\
I_{A}^{(2)} &=& \left[\frac{1}{2\E^2}+\frac{1}{4\E }
\right]\left(\epsilon I_1
^{(\alpha )}\right)^{2} +O (\epsilon ^{0})
\ .
\end{eqnarray}

\section{Calculation of the  integral $I_{\eta}^{\alpha } $}\label{app:Ieta}
The calculations for the corrections to friction are the same as with
short-range elasticity, except that the integrals $I_{1}$, $I_{A}$ and
$I_{\eta }$ change. The first two have already been calculated in
appendix \ref{app:LongRange}. We now attack the masterpiece, $I_{\eta }^{(\alpha)
}$. For simplicity, we restrict ourselves to $\alpha =1$.
\begin{eqnarray}
I_{\eta }^{(1)}\! :=\! \int\limits_{q_1,q_2}
\frac{1}{(q_1^2 {+} m^2)^{\frac{1}{2}}\, (q_2^2 {+} m^2)\, [(q_2^2 {+}
m^2)^{\frac{1}{2}} +
(q_3^2 {+} m^2)^{\frac{1}{2}}]}  \nn \\
\ .
\end{eqnarray}
Using:
\begin{eqnarray}
\rme^{- \sqrt{x}} = \frac{1}{2 \sqrt{\pi}} \int_0^{\infty} \rmd s\, s^{-3/2} \rme^{- \frac{1}{4 s}} \rme^{-s x}
\end{eqnarray}
we have
\begin{eqnarray}
\frac{1}{\sqrt{a} + \sqrt{b}} &=& \int_{t_{3}>0}\rme^{-t_3
(\sqrt{a}+\sqrt{b}) }
\nn \\
&=&
 \frac{1}{4 \pi} \int\limits_{t_3,s_1,s_2>0} (s_1 s_2)^{-\frac{3}{2}}
\rme^{- \frac{1}{4 s_1} - \frac{1}{4 s_2} } \rme^{- s_1 t_3^2 a - s_2
t_3^2 b}\nn \\  \label{E.3}
\end{eqnarray}
With the help of (\ref{E.3}), we can write $I_{\eta }^{(1)}$ as 
\begin{widetext}
\begin{eqnarray}
 I_{\eta }^{(1)} &=& \frac{1}{4 \pi}
\frac{1}{\Gamma (\frac{1}{2}) }\int\limits_{{q_{1},q_{2}}}\
\int\limits_{t_1,t_2,t_3,s_1,s_2 >0}
t_1^{-\frac{1}{2}}  (s_1 s_2)^{-\frac{3}{2}}
\, \rme^{- t_1 (q_1^2 + m^2) - (t_2 + s_1 t_3^2 ) (q_2^2 + m^2) - s_2
t_3^2 (q_3^2 + m^2)} \, \rme^{- \frac{1}{4 s_1} - \frac{1}{4 s_2} }
 \nonumber \\
& =& \frac{1}{4 \pi} \frac{1}{\Gamma ( \frac{1}{2}) } \left(\int_q
\rme^{-q^2}\right)^2 \int\limits_{t_1,t_2,t_3,s_1,s_2 >0} \frac{
t_1^{-\frac{1}{2}}
(s_1 s_2)^{-\frac{3}{2}} \,
\rme^{- m^2 ( t_1 + t_2 + (s_1 + s_2) t_3^2) } \, \rme^{- \frac{1}{4 s_1}
- \frac{1}{4 s_2} }}{
(t_1 t_2 + s_1 s_2 t_3^4 + (s_1 + s_2) t_1 t_3^2 + s_2 t_2 t_3^2 )^{1
- \frac{\epsilon}{2}}  }
 \nonumber \\
& =& \frac{1}{4 \pi} \frac{1}{\Gamma (\frac{1}{2}) }  \left(\int_q
\rme^{-q^2}\right)^2
\int\limits_{t_1,t_2,t_3,s_1,s_2 >0} t_3^{-1 + 2 \epsilon   }\, \frac{
t_1^{-\frac{1}{2}} (s_1 s_2)^{-\frac{3}{2}}  \,
\rme^{- m^2 t_3^2 ( t_1 + t_2 + s_1 + s_2) }\,
 \rme^{- \frac{1}{4 s_1} - \frac{1}{4 s_2} }}
{(t_1 t_2 + s_1 s_2  + (s_1 + s_2) t_1 + s_2 t_2 )^{1 - \frac{\epsilon}{2}}}
\nonumber \\
& =&  \frac{1}{4 \pi} \frac{1}{\Gamma (\frac{1}{2}) }  \left(\int_q
\rme^{-q^2}\right)^2  \frac{ \Gamma ( \epsilon)}2 \,  m^{- 2 \epsilon }
\int\limits_{t_1,t_2,s_1,s_2 >0}\, \frac{
t_1^{-\frac{1}{2}} (s_1 s_2)^{-\frac{3}{2}} \,
 \rme^{- \frac{1}{4 s_1} - \frac{1}{4 s_2} }}
{(t_1 t_2 + s_1 s_2  + (s_1 + s_2) t_1 + s_2 t_2 )^{1 -
\frac{\epsilon}{2}} (t_{1}+t_{2}+s_{1}+s_{2})^{\E}}\nonumber   \\
&=& \frac{1}{4 \pi} \frac{1 }{\Gamma (\frac{1}{2})}  \left(\int_q
\rme^{-q^2}\right)^2  \frac{\Gamma(\epsilon)}2 \, m^{- 2 \epsilon }
J \ .
\end{eqnarray}
In  the third line we have made the replacement  $t_1 \to t_3^2
t_1$ and $t_2 \to t_3^2 t_2$. In the fourth line
we have integrated over $t_3$. The integral $J$ is again decomposed in
converging parts (which can be evaluated at $\E=0$) and parts that can
be integrated analytically:
\begin{eqnarray}
 J &=& J_1 + J_2 + J_3 + O (\E)\\
 J_1 &=&
\int_{t_1,s_1,s_2 >0}  t_1^{-\frac{1}{2}} (s_1 s_2)^{-\frac{3}{2}} \rme^{- \frac{1}{4 s_1} - \frac{1}{4 s_2} }
\int_{t_2>1} \, \frac1 {t_1 t_2 + s_1 s_2  + (s_1 + s_2) t_1 + s_2 t_2}
- \frac{1}{t_2 (s_2 + t_1)}  \\
J_2 &=&
\int_{t_1,s_1,s_2 >0}  t_1^{-\frac{1}{2}} (s_1 s_2)^{-\frac{3}{2}}
\rme^{- \frac{1}{4 s_1} - \frac{1}{4 s_2} }
\int_{0<t_2<1}\, \frac{1}{t_1 t_2 + s_1 s_2  + (s_1 + s_2) t_1 + s_2 t_2 }  \\
 J_3 &=& \int_{t_1,s_1,s_2 >0} \,  t_1^{-\frac{1}{2}} (s_1
s_2)^{-\frac{3}{2}} \rme^{- \frac{1}{4 s_1} - \frac{1}{4 s_2} }
\int_{t_2>1}
t_2^{-1 - \frac{\epsilon}{2}} (s_2 + t_1)^{-1 + \frac{\epsilon}{2}}
\ .
\end{eqnarray}
Integrating in $J_{3}$ over $t_{2}$, $t_{1}$, $s_{1}$ and $s_{2}$ (in
this order) we find \rechecked
\begin{equation}\label{E.9}
J_3 =  2^{-\epsilon}  \frac{16 \pi}{\epsilon}\,
\Gamma\left(\frac{1-\epsilon}{2} \right)
\ .
\end{equation}
In order to calculate $J_{1}$ and $J_{2}$, it is convenient to do the
integration over $t_{2}$ in both integrals first. Taking the sum, some
terms cancel: \rechecked
\begin{equation}
 J_1 + J_2 =
\int_{t_1,s_1,s_2 >0}\,
t_1^{-\frac{1}{2}}  (s_1 s_2)^{-\frac{3}{2}} \rme^{- \frac{1}{4 s_1} -
\frac{1}{4 s_2} } \,
\frac{ \ln ( s_2 + t_1) - \ln(s_1 s_2 + t_1 (s_1 + s_2)) }{ s_2 + t_1}
\ .
\end{equation}
The logarithms have to  be written as derivatives: \rechecked
\begin{equation}
 J_1 + J_2 =\left. \frac{\partial }{\partial b}\right|_{b=0}
\int_{t_1,s_1,s_2 >0}\, t_1^{-\frac{1}{2}} (s_1 s_2)^{-\frac{3}{2}}
\rme^{- \frac{1}{4 s_1} - \frac{1}{4 s_2} }
 \Big( (s_2 + t_1)^{-1 + b} - (s_2 + t_1)^{-1} (s_1 s_2 + t_1 (s_1 +
s_2))^b \Big)
\ .
\end{equation}
Making the change of variables  $s_1 \to 1/s_1$, $s_2 \to 1/s_2$, $t_1
\to t_1/s_2$ and $s_1 \to s_1 s_2$ (in this order), we obtain \rechecked
\begin{equation}\label{E.12}
 J_1 + J_2 =\left. \frac{\partial }{\partial b}\right|_{b=0}
\int_{t_1,s_1,s_2 >0} s_1^{-\frac{1}{2}} s_2^{\frac{1}{2}-b}
t_1^{-\frac{1}{2}} \Big(
(1+t_1)^{-1+b} - (1+t_1)^{-1} s_1^{-b} s_2^{-b} (1 + t_1 (1+s_1))^b \Big)\,
\rme^{- \frac{1}{4} s_2 (1+s_1)}
\ .
\end{equation}
The integration over $s_{2}$ can now be done analytically: \rechecked
\begin{equation}\label{E.13}
J_{1}+J_2=\left. \frac{\partial }{\partial b}\right|_{b=0}
\int_{t_1,s_1}
 s_1^{-\frac{1}{2}}
t_1^{-\frac{1}{2}}  \bigg[ \frac{\Gamma\left( {\textstyle \frac{3}{2}} - b\right)}
{(1+t_1)^{1-b}}
\left(\frac{1+s_1}{4}\right)^{b-\frac{3}{2}}
-\frac{ \Gamma\left({\textstyle \frac{3}{2}} - 2 b \right) (1 + t_1
(1+s_1))^b }{ (1+t_1) s_1^{b} }
\left(\frac{1+s_1}{4}\right)^{2 b-\frac{3}{2}} \bigg]\nn \\
\ .
\end{equation}
In order to proceed, we split these integrals as follows \rechecked
\begin{eqnarray}
J_{1}+J_2&=&K_{1}+K_{2}+K_{3}\\
K_1&=&\left. \frac{\partial }{\partial b}\right|_{b=0}
\int_{t_1,s_1}
 s_1^{-\frac{1}{2}}
t_1^{-\frac{1}{2}}  \frac{\Gamma\left( {\textstyle \frac{3}{2}} - b\right)}
{(1+t_1)^{1-b}}
\left(\frac{1+s_1}{4}\right)^{b-\frac{3}{2}} =  \left. \frac{\partial
}{\partial b}\right|_{b=0} 8\, \pi\, 4^{-b}\,\Gamma
\left({\textstyle\frac{1}{2}-b } \right)  \\
K_{2}&=& - \left. \frac{\partial }{\partial b}\right|_{b=0}
\int_{t_1,s_1}
 s_1^{-\frac{1}{2}}
t_1^{-\frac{1}{2}} \frac{ \Gamma\left({\textstyle \frac{3}{2}} - 2 b
\right)  }{ (1+t_1) s_1^{b} }
\left(\frac{1+s_1}{4}\right)^{2 b-\frac{3}{2}} =- \left. \frac{\partial
}{\partial b}\right|_{b=0} 8\,\pi ^{\frac{3}{2}} \, 4^{-b}\,\Gamma \left(1-2b
\right) \\
K_{3}&=& - \left. \frac{\partial }{\partial b}\right|_{b=0}
\int_{t_1,s_1}
 s_1^{-\frac{1}{2}}
t_1^{-\frac{1}{2}} \frac{ \Gamma\left({\textstyle \frac{3}{2}}  \right) (1 + t_1
(1+s_1))^b }{ (1+t_1) }
\left(\frac{1+s_1}{4}\right)^{-\frac{3}{2}}
\end{eqnarray}
\end{widetext}
To evaluate $K_{3}$ one first has to take the derivative:\rechecked
\begin{equation}
K_{3} = - 8 \, \Gamma \left({\textstyle \frac{3}{2}} \right)  \int_{t_1,s_1}
 \frac{\ln  (1 + t_1 (1+s_1)) }{\sqrt{s_{1}}\, \sqrt{t_{1}}\,
(1+t_1)\left({1+s_1}\right)^{\frac{3}{2}} }
\ .
\end{equation}
Integrating first over $t_{1}$ and then $s_{1}$ gives \rechecked
\begin{eqnarray}
K_{3} &=& - 4 \pi ^{\frac{3}{2}} \int_{s_1}
 \frac{ 2\, \mbox{atanh}  (1/\sqrt{1+s_{1}})+\ln (s_{1}) }{\sqrt{s_{1}}\,
\left({1+s_1}\right)^{\frac{3}{2}} } \nonumber  \\
&=&  8 \pi ^{\frac{3}{2}} \left(2 \ln - 2\pi \right)
\ .
\end{eqnarray}
Putting everything together his gives finally
\begin{equation}
I_{\eta}^{(1)} = \left(\frac{1}{2\E^2}+ \frac{\ln 2 -\frac{\pi }{4}}{\E}
\right) (\E I_1^{(1)})^{2} + \mbox{finite}\ .
\end{equation}

\section{Fixed-point function at second order}
\label{app:fixedpointfunction}
\begin{figure}[tbp!h]
\centerline{\Fig{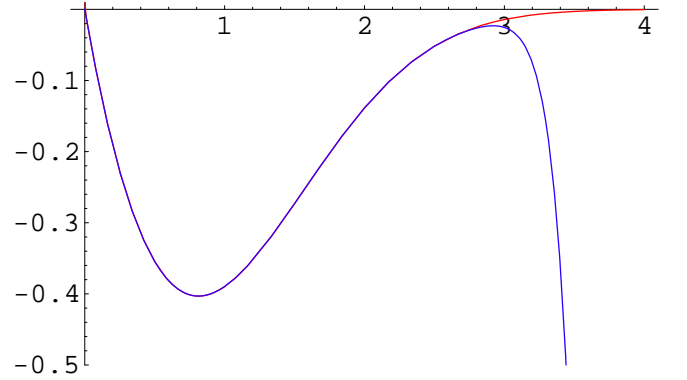}}
\caption{The fixed-point function of the RG-flow
$y_{2} (u)$ at second
order in $\E$. Upper curve: Numerical integration. Lower curve: Taylor-expansion
about 0.}
\label{fig:RFFP2loop}
\end{figure}

In this appendix, we show how to obtain the fixed-point function
for $\Delta (u)$ at second order. We restrict the discussion to
$\alpha =2$.  We use the  notations of equation
(\ref{DeltaAnsatz}). First, one needs the 1-loop function $y_{1} (u)$
both by solving (\ref{IV.12}) numerically and as a Taylor-series about 0.
The latter is obtained by deriving the 1-loop $\beta $-function at the
origin and fitting the coefficients as
\begin{eqnarray}
y_{1} (u)&=&
1-u+\frac{{u^2}}{3}-\frac{{u^3}}{36}-\frac{{u^4}}{270}-\frac{{u^5}}{4320}+\frac{{u^6}}{17010}
\nn  \\ &&+
 \frac{139
{u^7}}{5443200}+\frac{{u^8}}{204120}+\frac{571
{u^9}}{2351462400} \nn  \\ && -
 \frac{281 {u^{10}}}{1515591000}-\frac{163879
{u^{11}}}{2172751257600} \nn  \\&& -
 \frac{5221
{u^{12}}}{354648294000}-\frac{5246819
{u^{13}}}{10168475885568000}\nn   \\ &&+
 \frac{5459
{u^{14}}}{7447614174000}+\frac{534703531
{u^{15}}}{1830325659402240000}\dots \nn \\
\end{eqnarray}
The $\beta $-function at second order yields a linear differential
equation for $y_{2} (u)$. It is numerically singular at small
$u$. Therefore one has to expand it in a Taylor-series about 0. Using
the above information and the knowledge of $\zeta _{2}$, one finds
\begin{eqnarray}
y_{2} (u)&=& -1.14012\,u + 0.967798\,u^2 - 0.202495\,u^3\nn \\
&& -
  0.019299\,u^4 + 0.00259234\,u^5 + 0.0015302\,u^6\nn \\
&& +
  0.000286423\,u^7 - 6.25533\,{10}^{-6}\,u^8\nn \\
&& -
  0.0000206648\,u^9 - 6.48801\,{10}^{-6}\,u^{10}\nn \\
&& -
  7.85669\,{10}^{-7}\,u^{11} + 1.88404\,{10}^{-7}\,u^{12}\nn \\
&& +
  1.24668\,{10}^{-7}\,u^{13} + 3.13093\,{10}^{-8}\,u^{14} + \dots\nn \\
\end{eqnarray}
The differential equation for $y_{2} (u)$ is then solved numerically,
starting at $u\approx 0.5$. By integrating from that point both
towards 0 and towards infinity, one verifies that Taylor-expansion and
numerically obtained curve coincide in their respective domain of
validity. This is shown on figure \ref{fig:RFFP2loop}. One also
verifies that the numerically obtained function converges to 0 for
large u, thus the exponent $\zeta _{2}$ obtained above is correct.

It is a good question to ask for how large $\E$ the fixed-point
function $\Delta (u)=\frac{\E}{3}y_{1} (u)+\frac{\epsilon
^{2}}{18}y_{2} (u)$ might be a good approximation for the true
disorder correlator. Let us note that if one demands that $\Delta
(u)>0$, thus that forces be never anti-correlated, this is only
satisfied if
\begin{equation}
\epsilon < \epsilon _{c} \approx 1.6  \ .
\end{equation}

\section{Stability of the fixed points}
\label{appstability}
We now consider the stability of the periodic fixed point given in
(\ref{DeltaPeriodic}). Define $K [f]$ as
\begin{equation}
K [f]:=\lim_{\kappa \to 0} \frac{1}{\kappa }\left[\beta (\Delta^{*}
(u)+\kappa f (u))-\beta (\Delta^{*} (u)) \right] \ .
\end{equation}
The eigenfunctions and eigenvalues are
\begin{equation}
K[f] = \lambda f \ .
\end{equation}
We find the following solutions (with $x = u (1-u) $ and normalized to
$f (0)=1$)
\begin{widetext}
\begin{eqnarray}\label{}
\begin{array}[b]{rclrcl}
\lambda_{1}&=&\epsilon\ , &\quad  f_{1} &=& 1\\
\lambda_{2} &=& -\epsilon -\frac{7}{3}\epsilon ^{2} \ , &\quad
f_{2} &=&   1-(6+4\epsilon ) x \\
\lambda_{3} &=& -4\epsilon - 5 \epsilon ^{2} \ , &\quad f_{3} &=&
1-(15+20\epsilon)x+(45+85{\epsilon})x^2\\\lambda_{4} &=& \frac
{-25{\epsilon}}{3}-\frac{140{{\epsilon}}^2}{9}\ , &\quad f_{4} &=&
1-(28+\frac{238{\epsilon}}{3})x+(\frac{616}{3}+\frac{23548{\epsilon}}{27})x^
2
-(\frac{4004}{9}+\frac{185402{\epsilon}}{81})x^3\\
\lambda_{5} &=&-14{\epsilon}-35{{\epsilon}}^2 \ , &\quad f_{5} &=&
1-(45+225{\epsilon})x+(585+4500{\epsilon})x^2-(2925+\frac{110475{\epsilon}
}{4})x^3 +(\frac{9945}{2}+\frac{424755{\epsilon}}{8})x^4\\
\lambda_{6} &=&-21{\epsilon}-66{{\epsilon}}^2
\ , &\quad f_{6} &=& 1-(66+517{\epsilon})x+(1320+16148{\epsilon})x^2-
(11220+169928{\epsilon})x^3+(42636+\frac{3672944{\epsilon}}{5})x^4\\
&&&&&\hphantom{1}-(\frac{298452}{5}+\frac{28055588 \epsilon }{25})x^5
\end{array} 
\end{eqnarray}
\end{widetext}
This shows that apart from the constant mode (the shift) discussed in the
text, the fixed point is stable.

\section{Calculation of correlation functions}
\label{appcorrel}
In this Appendix we show how to compute a correlation function in
the renormalized theory. As an example we study the periodic case,
i.e.\ we compute the amplitude $A_d$ in (\ref{amp}). To do that we
 assume that we are exactly at the fixed point.

The correlation function is time-independent, as was shown in
Section (\ref{finite}), and takes the scaling form:
\begin{eqnarray}
\langle u_q u_{-q} \rangle _{\mathrm{nl}} = \frac{1}{\epsilon \tilde{I}_1}
\Delta^*(0) m^{-d} F_d\left(\frac{q}{m}\right)
\ ,
\end{eqnarray}
where we have restored the factor previously absorbed in $\Delta$. The
scaling function is universal and satisfies $F(0)=1$ since our
calculation was performed at zero external momenta in presence of a
mass and $F(z) \sim B/z^d$ at large $z$.  In $d=4$ one has $F_4(z) =
1/(1 + z^2)^2$. We want to obtain the scaling function to the next order;
in particular to compute $A_d$ we need $B=1 + b \epsilon +
O(\epsilon^2)$. The universal amplitude reads:
\begin{eqnarray}
A_d &=& \frac{2 S_d}{(2 \pi)^d \epsilon \tilde{I}_1} B \Delta^*(0) \\
&=& (1+ b \epsilon) (2 + \epsilon) \left(\frac{\epsilon}{36} +
\frac{\epsilon^2 X^{(\alpha )}}{108}\right) + O(\epsilon^3) \nonumber \ ,
\end{eqnarray}
which yields:
\begin{equation}
 A_d = \frac1 {18} \epsilon + \frac{2 X^{(\alpha)} + 3 + 6 b}{108} \epsilon^2
\ .
\end{equation}
Computing $b$ requires computing diagrams with external
momentum which we do now.
Let us use straight perturbation theory with $\Delta_0$, as
in Section (\ref{rgdis}). One has
\begin{eqnarray}
&&(q^2 + m^2)^2 \langle u_q u_{-q} \rangle  = \Delta_0(0) -
\Delta_0'(0^+)^2 I(q) \qquad \\
&&I(q)=\int_p \frac1{(p^2 + m^2)((p+q)^2 + m^2)}
\ .
\end{eqnarray}
Let us reexpress this by the renormalized dimensionless
disorder given in (\ref{do}) and (\ref{d21}) at $u=0$:
\begin{eqnarray}
\Delta_0(0) = m^\epsilon ( \Delta(0) + \Delta'(0^+)^2 m^\epsilon
I(0))
\ .
\end{eqnarray}
This gives:
\begin{eqnarray}
&&\!\!\! (q^2 + m^2)^2 \langle u_q u_{-q} \rangle  \nn \\
&&\qquad  = m^\epsilon ( \Delta(0) -
\Delta'(0^+)^2 m^\epsilon (I(q) - I(0)) )\nonumber  \\
&&\qquad  = m^\epsilon \frac{1}{\epsilon \tilde{I}_1} \Delta^*(0) ( 1 -
\epsilon \frac{1}{\epsilon \tilde{I}_1} m^\epsilon (I(q) - I(0)) )
\ ,\qquad 
\end{eqnarray}
where we have reestablished the factor $\Delta(u) = \frac{1}{\epsilon
\tilde{I}_1} \Delta^*(u)$ and used the fixed point condition
$\Delta^{* \prime}(0^+)^2 = \epsilon \Delta^{*}(0)$ This substitution
acts as a counter-term which exactly subtract the divergence as it
should. The result is finite. Using that:
\begin{eqnarray}
 I(q)
&=& \int_p \int_{s,t>0}  \rme^{ - s (p+q/2)^2 - t
(p-q/2)^2 - (s+t) m^2} \nonumber \\ 
&=& \int_p \rme^{-p^2}
\int_{s,t>0} (s+t)^{-d/2} \rme^{  - q^2 \frac{s t}{s+t}
- (s+t) m^2}\nonumber  \\
&=& \int_p \rme^{-p^2} m^{-\epsilon} \Gamma \left( 2-\frac{d}{2}\right)\nn \\
&&\times \int_{t>0} (1+t)^{-d/2} \left[(1+t) + \frac{t}{1+t}
\frac{q^2}{m^2}\right]^{-\epsilon/2} \ .\qquad
\end{eqnarray}
One obtains the scaling function in the form
($z=|q|/m$):\begin{widetext}
\begin{eqnarray}
F_d(z) &=& \frac1{(1 + z^2)^2} \left\{ 1 - \int_0^{\infty} \rmd t\,
\frac{1}{(1+t)^{2}} \left[  \left(1+ \frac{t
z^2}{(1+t)^2}\right)^{-\epsilon/2} - 1\right]  \right\}  \\ 
& =& \frac1{(1 + z^2)^2} \left\{ 1 + \frac{\epsilon}{2} \int_0^{1}
\rmd s
 \ln(1+ z^2 (s - s^2) )  \right\} + O(\epsilon^2) \nonumber \\
& =& \frac1{(1 + z^2)^2} \left\{ 1 + \frac{\epsilon}{2}\! \left[ - 2 +
\sqrt{ \frac{ 4+z^{2}}{z^{2}} } \left( \ln 2 - \ln\left( 2 + z^{2} - z
\sqrt{4+z^{2}}
\right) \right) 
\right] \right\} + O(\epsilon^2)\nn \\
&&\!\!\!\!\! \stackrel{\zeta \to \infty
}{-\!\!\!\longrightarrow} 
\frac{1}{z^{4}}\left\{1+ \frac{\epsilon }{2}\left[-2 + 2 \ln z \right] \right\}
 + {\cal O} (\epsilon ^{2}) \nn 
\end{eqnarray}
\end{widetext}
We want to match at large $z$:
\begin{equation}
 F_d(z) = \frac1{z^4} (1 + b \epsilon) z^\epsilon
= \frac1{z^4} (1 + \epsilon (\ln z + b ) + O(\epsilon^2) )
\end{equation}
The above result  yields
\begin{equation}
 b =  - 1\ .
\end{equation}

\section{Anomalous and non-odd graphs}
\label{appanomalous}
In this Appendix we write all anomalous
2-loop  graphs contributing to the correction of a non-%
analytic disorder. In a first step we make no assumption and give
their general expressions: Already at that stage some cancellations
are apparent. In a second step we consider the limit $v\to 0^+$ at
$T=0$. We check all cancellations given in the text and show that no
additional singularities occur. The multiplicity factors are included
in the given expressions. Of course since we want only corrections to
disorder we will give only the expressions when the separations of the
times between the two external response fields are much larger than
the separations within each connected component.  If this were not the
case, as is needed e.g.\ in the calculation of a 2-point
correlation function to order $\Delta^3$, the above expressions should
be
reexamined separately. Equivalently, the expressions given here are
correct only for $u>0$ and may become incorrect at $u=0$.

Graphs which are odd need not be considered (see main text). Each
remaining graph, e.g.\ $\mathrm{c}_i$ is written in the shorthand notation
form:
\begin{eqnarray}
\text{graph} ~~ \mathrm{c}_i = \int_{x,y} \int_{t_i>0} F_{ \mathrm{c}}
\mathrm{c}_i 
\ .
\end{eqnarray}
The only anomalous non-vanishing graphs of class A are:
\begin{eqnarray}
 F_{\mathrm{c}} &=& R_{yt_1}  R_{yt_2} R_{x-yt_3} R_{x t_4} \\
  \mathrm{c}_1 &=&
2 \langle  \Delta'(u^x_0 - u^x_{-t_3-t_2-t_4}) \Delta'(u^y_{-t_3} -
u^y_{-t_3-t_2-t_1}) \rangle    \Delta''(u)\nonumber \\
 \\
 \mathrm{c}_2 &=&
2 \langle  \Delta(u^x_{-t_4} - u^x_{-t_2-t_3}) \Delta''(u^y_{0} -
u^y_{-t_2-t_1}) \rangle    \Delta''(u)\\
 \mathrm{c}_4 &=&
2  \Delta'(u^x_{0} - u^x_{-t_2-t_3-t_4}) \Delta'(u^y_{-t_4-t_2} -
u^y_{-t_4-t_1})   \Delta''(u) \nonumber \\
\\
  \mathrm{c}_5 &=&
- 2  \Delta(u^x_{-t_1-t_2-t_3} - u^x_{-t_1-t_4}) \Delta''(u^y_{0} -
u^y_{-t_2-t_1})   \Delta''(u)\nonumber \\
 \\ 
F_{ \mathrm{e}} &=& F_{ \mathrm{f}} = R_{x-yt_1}  R_{x-yt_2} R_{xt_3}
R_{y t_4} \\ 
  \mathrm{e}_1 &=& \Delta''(u)  \langle  \Delta'(u^x_{-t_3} - u^x_{-t_1-t_4})
\Delta'(u^y_{-t_4} - u^y_{-t_3-t_2}) \rangle  \\   
 \mathrm{f}_2 &=& 2  \Delta''(u) \langle  \Delta'(u^x_{-t_3} -
u^x_{-t_3-t_2-t_1}) 
\Delta'(u^y_{-t_2-t_3} - u^y_{-t_4}) \rangle  
\ .\nonumber \\
\end{eqnarray}
For the graphs d one easily sees that the following relations are
exact (with no other assumption than $u \neq 0$):
\begin{eqnarray}
 F_{\mathrm{c}} = F_{\mathrm{d}} &=& R_{yt_1}  R_{yt_2} R_{x-yt_3} R_{x t_4} \\
  \mathrm{d}_2 +  \mathrm{d}_4 &=& 0 \\
  \mathrm{d}_6 +  \mathrm{d}_8 &=& 0
\ .
\end{eqnarray}
The only anomalous non-vanishing graphs of class B are:
\begin{eqnarray}
 F_{ \mathrm{k}} &=& F_{ \mathrm{l}} = R_{xt_1}  R_{xt_2} R_{y-xt_3}
R_{y-x t_4} \\ 
  \mathrm{k}_1 &=& c \Delta''(u)
\langle  \Delta''(u^x_{-t_2} - u^x_{-t_1}) \Delta(u^y_{-t_2-t_4} -
u^y_{-t_1-t_3}) \rangle \nonumber \\
 \\
  \mathrm{l}_1 &=&  - c \Delta''(u)
\langle  \Delta''(u^x_{-t_2} - u^x_{-t_1}) \Delta(u^y_{-t_1-t_4} -
u^y_{-t_1-t_3}) \rangle\nonumber \\
  \\
 F_{ \mathrm{i}} &=& R_{xt_1} R_{xt_2} R_{yt_3} R_{y t_4} \\
  \mathrm{i}_3 &=& - \Delta''(u) \langle  \Delta'(u^x_0 - u^x_{-t_1 -t_2})
\Delta'(u^y_0 - u^y_{-t_3 -t_4}) \rangle
\ .\nonumber \\
\end{eqnarray}
All graphs of class C exactly vanish. For instance:
\begin{eqnarray}
 \mathrm{m}_2 &=& \Delta(0) \Delta'''(u_0 - u_{-t_3}) \Delta'(u) \\
\mathrm{n}_2 &=& - \Delta(0) \Delta'''(u_0 - u_{-t_3}) \Delta'(u)
\ .
\end{eqnarray}
We now evaluate these graphs in the quasi-static depinning limit,
substituting $\Delta(u)$ by its power series as a function of $u$, as
explained in the main text. We need in addition to (\ref{dadatutu}):
\begin{eqnarray}
 \Delta'(u) &=& \Delta'(0^+) \sgn(u) + \Delta''(0^+) u + \dots\nonumber   \\
 \Delta''(u) &=& 2 \Delta'(0^+) \delta(u) + \Delta''(0^+)  + \dots 
\label{series2}
\ .
\end{eqnarray}
In $\Delta''(u)$ evaluated at zero we have written the
$\delta$-function which may in principle be needed. If this were the
case that would pose two unpleasant problems: Firstly a different
viewpoint were to argue that $\Delta''(u)$ should simply be continued
to zero which does not pose any problem since it is pair. Second it
would open the possibility to problematic singular terms ($\delta(v)$
or $1/v$) as $v \to 0^+$. Fortunately, in all our 2-loop calculations
this never happens: these $\delta$-functions, if put by hand,
cancel. This confirms that, at least to this order, no pathology arises.

Let us start with the sum $c_2 + c_5$. Using (\ref{dadatutu}) and
(\ref{series2}) one sees that the term proportional to $\Delta(0)
\Delta''(0^+)$ cancels. Let us test the $\delta$-function.  Then one
needs to go one order further in the expansion of the $\Delta$ term since
averages of the type $\delta(u_1) u_2$ have dimension one, similar to
$\langle \sgn ( u_1) \sgn ( u_2) \rangle $, and can thus yield a non-zero
result at zero temperature (higher order terms yielding dimensions as
positive powers of the field are not needed as they vanish at zero
$T$). This yields 
\begin{widetext}
\begin{equation}
 \mathrm{c}_2 + \mathrm{c}_5 = 4 \Delta'(0^+)^2 \Delta''(u) \langle
(|u^x_{-t_4} - 
u^x_{-t_2-t_3}| - |u^x_{-t_1-t_2-t_3} - u^x_{-t_1-t_4}|)
\delta(u^y_{0} - u^y_{-t_2-t_1}) \rangle
\ ,
\end{equation}
which strictly vanish upon the replacement $u^x_{t} - u^x_{t'} \to v
(t-t')$. This is fortunate since this term would have led to a $1/v$
singularity. Note that all diagrams $a-g$ in the 2-loop correction to
$\eta$ could a priori suffer from the same problem since $\Delta''$
functions must be expanded. However one notes that their arguments are
always strictly positive in the depinning limit, which avoids, as it
did here, the problem. Similarly one has
\begin{equation}
 c_4 =   2  \Delta'(0^+)^2 \Delta''(u)
 \langle  \sgn(u^x_{0} - u^x_{-t_2-t_3-t_4}) \sgn(u^y_{-t_4-t_2} -
u^y_{-t_4-t_1}) \rangle \ .
\end{equation}\end{widetext}
Performing the replacement $u^x_{t} - u^x_{t'} \to v (t-t')$, since
the $t_i>0$ and because $F_{c}$ is symmetric in $t_1 \leftrightarrow  t_2$ one
finds that
\begin{equation}
 \mathrm{c}_2 + \mathrm{c}_5 = \mathrm{c}_4 = 0
\end{equation}
at depinning. Note that these cancellations do not happen any longer,
if the field is not a monotonic function, a question which will be
discussed in Ref.\ \cite{LeDoussalWieseChauve2002a}.

A similar calculation shows that at depinning one has also:
\begin{equation}
 \mathrm{k}_1 + \mathrm{l}_1 = 0
\ .
\end{equation}
There, in the singular part, the $\delta$-function implies that $t_1 =
t_2$ yielding the cancellation via a slightly different mechanism than
above.

Finally we are left with the only non-zero anomalous non-trivial
graphs $c_1$, $e_1$, $f_2$ and $i_1$ to compute, which is done in the
text.
\begin{acknowledgements}
It is a pleasure to thank E.~Bouchaud, E.~Br\'ezin, J.~Ferr\'e, W.~Krauth,
S.~Lemerle, E.~Rolley and A.~Rosso for stimulating discussions.
\end{acknowledgements}

\vfill  


\end{document}